\documentclass{article}

\usepackage{PRIMEarxiv}

\usepackage[utf8]{inputenc} 
\usepackage[T1]{fontenc}    
\usepackage{hyperref}       
\usepackage{url}            
\usepackage{booktabs}       
\usepackage{amsfonts}       
\usepackage{nicefrac}       
\usepackage{microtype}      
\usepackage{lipsum}
\usepackage{fancyhdr}       
\usepackage{graphicx}       
\graphicspath{{media/}}     
\usepackage{authblk}
\usepackage{array}
\usepackage{threeparttable}
\usepackage{tabularx}
\usepackage{amsmath}
\usepackage[table]{xcolor}
\usepackage{rotating}
\title{Toward Robust, Reproducible, and Widely Accessible Intracranial Speech Brain-Computer Interfaces: A Comprehensive Narrative Review of Neural Mechanisms, Hardware, Algorithms, Evaluation, Clinical Pathways and Future Directions}

\author[1,2]{Dongyi He}
\author[1,*]{Wai Ting Siok}
\author[1,*]{Nizhuan Wang}

\affil[1]{Department of Language Science and Technology, The Hong Kong Polytechnic University, Hung Hom, Hong Kong SAR, China}
\affil[2]{School of Artificial Intelligence, Chongqing University of Technology, Chongqing 401135, China}
\affil[*]{Correspondence: wai-ting.siok@polyu.edu.hk (W.T.S.); wangnizhuan1120@gmail.com (N.W.)}

\begin{document}
\maketitle

\begin{abstract}
Intracranial speech brain-computer interfaces (BCIs) offer a promising route for restoring communication in individuals with severe, chronic, and otherwise irreversible motor and speech impairments who cannot communicate adequately using conventional assistive technologies, but clinical translation remains limited by fragmented and heterogeneous evidence, as well as unresolved design trade-offs across neuroscience, hardware, algorithms, validation methods, and clinical integration. This narrative review synthesizes recent progress across four key domains in intracranial speech neuroprosthetics: i) the neural mechanisms underlying overt, attempted (including silent miming and attempts without observable movement), and imagined speech (speech imagery without an attempt to articulate); ii) decision-oriented hardware comparisons of surgically implanted recording modalities, including microelectrode array (MEA), electrocorticography (ECoG), and stereotactic electroencephalography (SEEG); iii) experimental strategies for cross-subject generalization; and iv) advances in neural decoding, including sequence models, attention-based architectures (e.g., transformers), articulatory intermediate representations, and language-prior-assisted frameworks. We highlight persistent bottlenecks, including limited transfer without individual calibration, long-term non-stationarity and recalibration burden, heterogeneous and non-comparable evaluation practices, limited naturalistic expressivity, and the variable signal-to-noise ratio (SNR) and limited temporal alignment of neural activity in imagined speech decoding. Our contributions are threefold: (1) an end-to-end, decision-oriented synthesis that links neural representations to recording choices, experimental design, decoding model architectures, and translational constraints; (2) a structured framework organized around five coupled design questions, accompanied by a unified evaluation framework and a conceptual cross-linguistic, cross-task benchmark template that integrates objective, perceptual, expressive, conversational, and longitudinal metrics; and (3) user-centered translational guidance that includes agency-preserving shared control, verifiable performance priorities, and scenario-specific minimum viable product (MVP) profiles for differentiating between reliability-first home communication and fidelity-first conversational speech restoration. We conclude with a call for larger multilingual, multi-center longitudinal datasets; harmonized benchmarks; adaptive yet interpretable decoders; prospective clinical validation; and transparent data-sharing and reporting practices with robust ethical safeguards. These efforts are essential to accelerate the safe and equitable deployment of speech neuroprostheses.
\end{abstract}


\section{Introduction}
Speech is the most natural and efficient channel for human communication, yet it is among the first abilities compromised in a range of devastating neurological conditions~\cite{willett2023high, 
chaudhary2022spelling, masrori2020amyotrophic}. Individuals with locked-in syndrome (LIS), advanced amyotrophic lateral sclerosis (ALS), brainstem stroke, or severe paralysis may remain cognitively intact while unable to produce intelligible speech or purposeful movement~\cite{pasinelli2006molecular,tiryaki2014and, masrori2020amyotrophic,chio2014neuroimaging}. For these patients, speech BCIs, systems that decode neural activity associated with speech into text or synthesized speech (and ideally expressive prosody), offer a pathway to restore communication with the potential for improved speed, privacy, and user agency when conventional assistive technologies are inadequate~\cite{willett2023high,luo2022brain,silva2024speech,makin2020machine,card2024accurate}. However, intracranial speech BCIs are not intended as first-line interventions for all individuals with communication disorders. Their most realistic near-term therapeutic candidates are individuals with severe, chronic, and otherwise irreversible loss of intelligible speech or reliable motor output, particularly those with late-stage ALS accompanied by anarthria or profound dysarthria, brainstem stroke or LIS, and selected individuals with severe paralysis or tetraplegia whose speech and conventional access channels are inadequate~\cite{moses2021neuroprosthesis,chaudhary2022spelling,willett2023high,luo2023stable,card2024accurate,angrick2024online}. Suitable candidates should retain sufficient cognition, language comprehension, communicative intent, and the ability to formulate intended linguistic content; they should also derive limited benefit from conventional augmentative and alternative communication (AAC) and have an acceptable surgical risk--benefit profile~\cite{silva2024speech,stavisky2025restoring,jhilal2025implantable,sankaran2023recommendations}. Patients undergoing temporary Electrocorticography (ECoG) or Stereoelectroencephalography (SEEG) implantation for epilepsy monitoring remain an important research population for mechanistic investigation and short-term decoder validation, but they are distinct from the intended population for chronic therapeutic implantation~\cite{makin2020machine,verwoert2022dataset,wu2024speech}. For broader populations, particularly individuals with milder, potentially reversible, or less disabling impairments, conventional AAC remains a practical communication option where usable, while non-invasive approaches based on scalp Electroencephalography (EEG), Magnetoencephalography (MEG), functional near-infrared spectroscopy (fNIRS), or functional magnetic resonance imaging (fMRI) may support research, functional assessment, rehabilitation, or specific BCI applications, according to their evidence and deployment requirements~\cite{panachakel2021decoding,su2025systematic,dash2020decoding,lu2020comparison,tang2023semantic}.

\begin{figure}
\centering
\includegraphics[width=\textwidth]{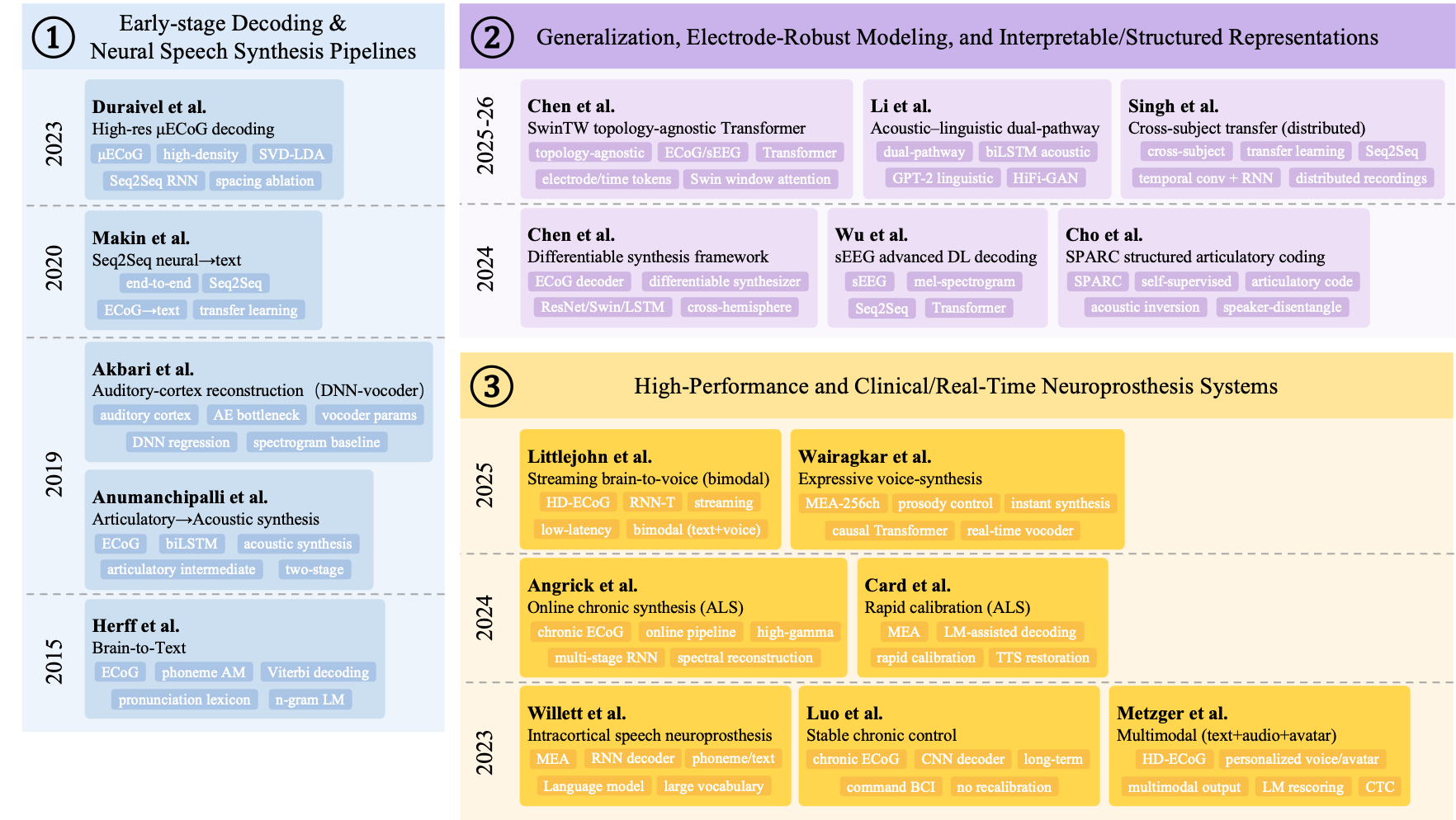}
\caption{Lane-based timeline of representative intracranial speech-decoding frameworks (selected representatives from Tables~\ref{tab:decoding-method-review-early}--\ref{tab:decoding-method-review-generalization}). The frameworks are grouped into three categories and ordered by publication year. Each card includes the representative studies with a short framework label and technical tags summarizing key methods.}
\label{fig:timeline}
\end{figure}

For these carefully selected populations, intracranial neural recordings have catalyzed rapid progress in speech BCI research over the past decade. Compared with non-invasive modalities~\cite{d2025towards, zhao2025rgftslanet,lu2020comparison, jiang2026comprehensive,li2025tale,he2025spec2volcamu}, intracranial approaches provide higher SNR and access to high-frequency activity that tightly tracks local cortical computations~\cite{he2026non,cao2022virtual}, enabling decoding at the levels of phonemes, words, and continuous speech~\cite{leuthardt2004brain,leuthardt2011using,willett2023high}. In parallel, the field has shifted from proof-of-concept demonstrations toward system-level designs that must operate reliably in real-world settings: across days and months, across electrode configurations, across languages and speaking styles, and across overt speech, attempted speech (including mimed articulation), and imagined speech. Attempted speech denotes an effort to articulate, whether or not movement or sound results; imagined speech denotes speech imagery without an articulation attempt. Each paradigm is characterized by its residual movement and vocalization, cueing, and feedback conditions~\cite{luo2023stable,angrick2024online}. As summarized jointly in the lane-based timeline in Fig.~\ref{fig:timeline} and the corresponding evidence tables in Tables~\ref{tab:decoding-method-review-early}--\ref{tab:decoding-method-review-generalization}, this evolution represents not only a chronology of decoder improvements but a broader shift toward chronic, online systems and frameworks that explicitly address cross-subject transfer, electrode heterogeneity, and deployment constraints. The figure and tables organize the same representative frameworks by chronology, theme, and evidential weight: multi-participant cohorts establish within-study consistency, single-participant studies establish participant-level feasibility, and explicitly marked same-participant extensions establish longitudinal reproducibility. Numerical word error rate (WER) / phoneme error rate (PER) replication requires matched datasets, vocabulary or sentence sets, and tasks, and independent-group agreement across non-matched settings supplies convergent evidence for representations or system feasibility, according to the study design. Advancing from laboratory demonstrations to these evidence-weighted, chronic, and transferable systems is more demanding and necessitates the tight integration of neuroscience, hardware, machine learning, evaluation methodology, and clinical translation.

At the mechanistic level, speech production and perception recruit distributed cortical and subcortical circuits, spanning ventral sensorimotor cortex, premotor and supplementary motor regions, inferior frontal and temporal language areas, and higher-order association networks. Intracranial recordings have revealed rich articulatory representations and population dynamics that support rapid sequencing, coarticulation, and context-dependent planning, and have clarified how neural signatures differ across overt, mimed and imagined speech~\cite{silva2024speech,musch2020transformation,castellucci2022speech,wang2023distributed,proix2022imagined,de2024imagined}. For speech BCIs, these observations motivate the choice of decoding targets (e.g., acoustics, phonemes, articulatory features, or semantic units), the selection of temporal context, and the design of intermediate representations that may improve robustness and interpretability~\cite{anumanchipalli2019speech,chartier2018encoding,bouchard2013functional,stavisky2019neural}.

On the technology side, intracranial speech BCIs sit at the intersection of multiple recording modalities, each with fundamentally different trade-offs. Intracortical MEAs can capture high-dimensional spiking activity with fine spatial granularity but typically cover limited cortical territory and face chronic stability challenges~\cite{willett2023high,card2024accurate}. Macro-ECoG provides broad, contiguous coverage of the cortical surface, whereas SEEG enables sparse, distributed three-dimensional sampling of cortical and subcortical structures. Both rely on clinically established implantation approaches but differ in spatial scale and noise characteristics~\cite{leuthardt2004brain,leuthardt2011using,chen2024neural}. The ``best'' modality therefore depends on the intended use case (research vs.\ home use), surgical feasibility, target neural features, and acceptable calibration burden. Experiment design is another critical consideration. The field has evolved from small, single-subject datasets with limited linguistic diversity toward larger, multi-subject datasets that enable cross-subject learning and benchmarking. However, challenges remain in standardizing data collection protocols, ensuring multilingual coverage, and producing rich annotations that capture not only lexical content but also prosody, emphasis, and affective features~\cite{verwoert2022dataset,pescatore2025decoding,feng2025acoustic,li2025high,wairagkar2025instantaneous,jhilal2025implantable}.

Algorithmically, speech BCIs have evolved from feature-engineered pipelines and frame-wise classifiers to Seq2Seq models that jointly learn neural feature extraction, alignment, and decoding~\cite{makin2020machine}. Advances such as transformers, convolutional temporal models, and hybrid architectures have enabled larger vocabularies, streaming or online decoding, and improved utilization of long-range context. Additionally, physically interpretable articulatory intermediates provide structural constraints that can enhance data efficiency and transfer under the evaluated adaptation protocols~\cite{chen2024neural,chen2025transformer}. More recently, language models and other linguistic priors have been integrated to reduce error rates and increase output coherence, raising new questions about controllability, bias, latency, and attribution of ``who said what'' in shared-control systems~\cite{willett2023high,card2024accurate}. These trends underscore the importance of evaluating models not only in terms of accuracy but also considering calibration requirements, real-time constraints, robustness to neural non-stationarity, and interpretability.

Despite these advances, several core bottlenecks continue to hinder clinical translation of intracranial speech BCIs. First, generalization without individual calibration remains limited: many decoders remain subject-specific, do not transfer across electrode layouts, and cannot fully leverage data across participants or centers~\cite{chen2024neural, chen2025transformer, makin2020machine}. Second, long-term stability of both neural signals and end-to-end performance remains a central challenge, with MEAs often requiring frequent recalibration and ECoG/SEEG systems still necessitating rigorous chronic validation under naturalistic use~\cite{luo2023stable,leuthardt2011using,leuthardt2004brain,angrick2024online,card2024accurate}. Third, evaluation practices remain heterogeneous: inconsistent use of WER, PER, latency, information-transfer measures, and human perceptual scoring makes cross-task comparisons difficult and can obscure clinically meaningful differences~\cite{defossez2023decoding,wu2025improved,chen2025neural,stavisky2025restoring}. Fourth, limited naturalistic expressivity restricts many systems to lexical content, while paralinguistic features such as prosody, intonation, emphasis, and affect are less consistently modeled. This challenge is amplified for tonal languages (e.g., Mandarin and Cantonese), in which pitch contours and syllable structure are integral to lexical meaning~\cite{feng2025acoustic,burnham2022seeing,chen2022cantonese}. However, cross-linguistic evidence in intracranial speech BCIs remains limited, with current work dominated by English and only initial decoding studies in Mandarin~\cite{feng2025acoustic}. Building on these early advances, systematic multilingual validation can extend intracranial decoding across diverse phonological systems and provide a rigorous test of decoder generalizability. Finally, imagined-speech decoding poses unique difficulties, as neural signals are often less discriminable and less time-locked than overt-speech signals, with no directly observable acoustic ground truth~\cite{proix2022imagined,kunz2025inner}. In addition, intracranial speech BCI research remains fragmented across neuroscience, hardware development, algorithmic design, evaluation methods, and clinical translation. This fragmentation complicates cross-study comparisons, obscures clinically meaningful trade-offs, and limits the development of actionable design principles. Although recent reviews have provided valuable complementary perspectives, their principal emphases differ. For example, Luo et al.~\cite{luo2022brain} focus on speech-decoding and synthesis pipelines, Silva et al.~\cite{silva2024speech} on cortical speech representations, decoding approaches, and evaluation practices, Stavisky~\cite{stavisky2025restoring} on restorative speech BCIs, and Jhilal et al.~\cite{jhilal2025implantable} on implantable decoders and clinical translation. Collectively, these reviews cover many central components of the field, but they do not explicitly organize them within a single end-to-end, decision-oriented framework that jointly links neural mechanisms, recording choices, experimental design, cross-subject and electrode-robust modeling, evaluation, and user-centered deployment.

To address this gap, this review moves beyond a component-wise summary of the literature to provide an end-to-end, decision-oriented framework for intracranial speech BCI development. Its distinctive contribution lies not simply in covering neural mechanisms, recording technologies, experimental design, decoding algorithms, evaluation, and clinical translation, but in explicitly examining how decisions at one stage constrain and reshape choices at the others. We therefore organize the evidence around coupled design trade-offs—from the selection of neural targets and recording modalities to cross-subject modeling, longitudinal validation, and user-centered deployment—and translate this synthesis into actionable design principles, a unified evaluation framework, and scenario-specific translational priorities. This narrative synthesis focuses on speech restoration within the broader field of language BCIs. Within this review, the term ``speech neuroprosthesis'' denotes an assistive speech-BCI implementation intended to restore communicative output, whereas ``language BCI'' denotes the broader research domain concerned with lexical-semantic, conceptual, and compositional representations of intended meaning. Perception and semantic studies are drawn on to characterize neural representations and inform decoding methods, rather than treated as direct demonstrations of intended-speech communication. Recent reviews and landmark primary studies were used to define the evidence base, with emphasis on work published from 2015 through early 2026 and earlier seminal reports retained where necessary for historical or mechanistic context. Primary intracranial studies were prioritized when they introduced substantive advances in recording, decoding, generalization, online or chronic performance, or clinical relevance, while reviews, preprints, and non-invasive studies were included to provide conceptual framing, timely evidence, and explicit methodological comparison. 

\begin{figure}[t]
\centering
\includegraphics[width=\textwidth]{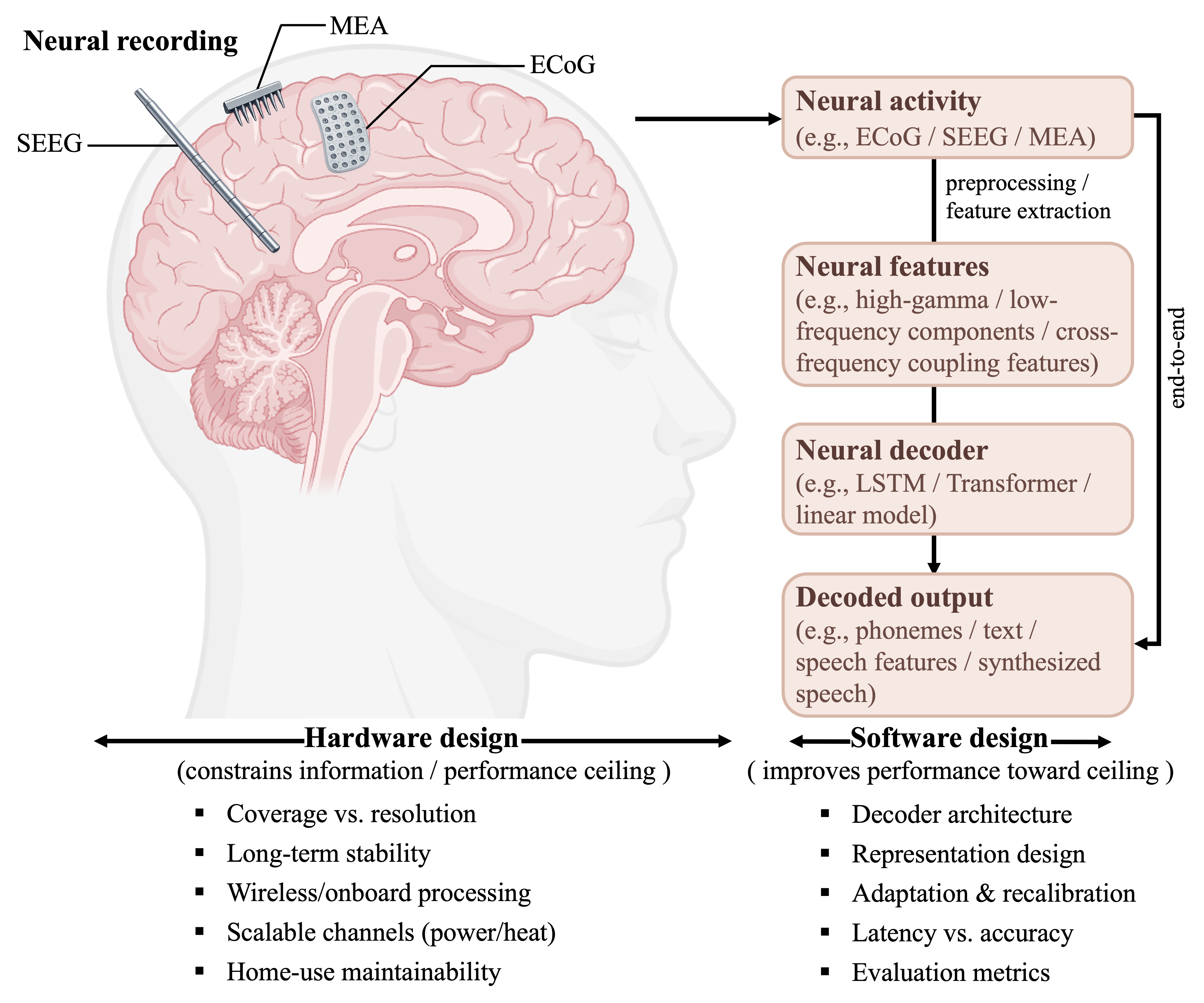}
\caption{Schematic of intracranial speech BCI decoding pipeline with hardware-software co-design. Neural activity from ECoG/SEEG/MEA recordings is transformed into neural features and decoded into phonemes, text, speech features, or synthesized speech; hardware design largely constrains the performance ceiling for neural-information recovery, whereas software design determines how closely it is approached within this signal-derived limit; language priors could additionally improve end-task performance by contributing complementary linguistic information.}
\label{fig:modalities}
\end{figure}

Building on this framework, we make three specific contributions. First, we provide an end-to-end, decision-oriented synthesis that links neural mechanisms, recording methodologies, experiment design, decoding models, evaluation strategies, and translational deployment into a cohesive framework. Fig.~\ref{fig:modalities} previews this end-to-end pipeline, tracing the path from neural recordings (e.g., ECoG, SEEG, and intracortical MEAs) through feature extraction and neural decoding to outputs including phonemes, text, speech features, and synthesized speech, while making explicit how hardware and software choices jointly shape each stage. Second, we organize the field around five interconnected design questions that span the entire BCI development pathway: (i) \emph{What} neural representations best capture speech intention across overt, mimed, and imagined speech? (ii) \emph{Where and how} should neural activity be recorded to balance coverage, resolution, chronic stability, and surgical feasibility? (iii) \emph{How} should experimental datasets and annotations be designed to enable cross-subject generalization, multilingual applicability, and reproducible benchmarking? (iv) \emph{Which} decoding architectures and intermediate representations best trade off accuracy, latency, robustness, and interpretability? and (v) \emph{How} should systems be evaluated, governed, and translated while preserving user agency, identity, and safety? Third, we synthesize concrete translational guidance, including a unified evaluation framework and practical design priorities for user-centered clinical deployment. Consistent with these contributions, the following sections progress from a foundational framework of neural mechanisms to the trade-offs inherent in recording modalities and experimental design, followed by algorithmic advances, a proposed unified evaluation framework, and pathways for translational and ethical development. We then conclude with future directions and a synthesis of key priorities. More importantly, we use this synthesis to call for the field's next phase: larger multilingual and multi-center datasets; longitudinal and standardized benchmarks; adaptive yet interpretable decoders; and prospective user-centered clinical validation with transparent reporting and data sharing.

\begin{table}
\centering
\caption{Comparative summary of representative early-stage intracranial speech-decoding and neural speech-synthesis pipelines. Participant count, vocabulary or sentence-set constraint, dataset, and evaluation task are reported alongside performance so that each result can be interpreted within its original evidence base.}
\label{tab:decoding-method-review-early}
\tiny
\setlength{\tabcolsep}{2.3pt}
\begin{tabularx}{\linewidth}{>{\raggedright\arraybackslash}p{0.085\linewidth} >{\raggedright\arraybackslash}p{0.09\linewidth} >{\raggedright\arraybackslash}p{0.115\linewidth} >{\raggedright\arraybackslash}X >{\raggedright\arraybackslash}p{0.09\linewidth} >{\raggedright\arraybackslash}p{0.115\linewidth} >{\raggedright\arraybackslash}p{0.13\linewidth} >{\raggedright\arraybackslash}X}
\toprule
\textbf{Team (Year)} & \textbf{Participants} & \textbf{Method / Framework} & \textbf{Core Techniques} & \textbf{Vocabulary / sentence set} & \textbf{Dataset} & \textbf{Task / evaluation} & \textbf{Representative Performance} \\
\midrule

Herff et al.~\cite{herff2015brain} (2015) & 6 analyzed (7 enrolled); best WER from one participant/session & Brain-to-Text (continuous phrase decoding) & ECoG phone-level acoustic modeling + n-gram LM + pronunciation dictionary; Viterbi decoding of continuous phrases & 10-word test dictionary for best WER; 20 phones + silence for framewise phone error & Read excerpts; 21--49 phrases/session & Overt continuous reading; leave-one-phrase-out; offline & Best participant/session: LM-assisted WER $\sim$25\% for the 10-word condition; framewise phone-class error $<$50\% \\\\

Akbari et al.~\cite{akbari2019towards} (2019) & 5 neural participants + 11 listeners & Auditory-cortex speech reconstruction (DNN-vocoder) & Auditory-cortex decoding to vocoder parameters using DNN regression; comparison against linear spectrogram reconstruction; low/high-frequency neural features + autoencoder bottleneck & 10 digits; 40 test stimuli from 4 speakers & $\sim$30-min continuous stories for neural-model training; TI-46 digits for testing & Passive listening and speech reconstruction; 10-choice listener identification & Listener digit identification: 75\% for DNN-vocoder versus 45\% for the linear baseline \\\\

Anumanchipalli et al.~\cite{anumanchipalli2019speech} (2019) & 5 neural participants; listener WER from participant 1 & Two-stage articulatory$\rightarrow$acoustic neural speech synthesis & ECoG-to-articulatory kinematics (BiLSTM) followed by articulatory-to-acoustic synthesis; explicit intermediate articulatory representation & Closed 25- or 50-word response pools & MOCHA-TIMIT; 101 held-out sentences for participant 1 listener testing & Overt ECoG-to-speech synthesis; closed-pool human transcription & Participant 1 median listener WER: 31\% (25-word pool) and 53\% (50-word pool) \\\\

Makin et al.~\cite{makin2020machine} (2020) & 4; $\sim$3\% WER from the best single participant & Encoder--decoder cortical activity $\rightarrow$ text translation & End-to-end Seq2Seq encoder--decoder (ECoG$\rightarrow$text), auxiliary losses, transfer learning with target-participant training for low-data settings & 30 sentences/$\sim$125 words or 50 sentences/$\sim$250 words; $\sim$3\% on the latter & Picture descriptions; MOCHA-TIMIT (MOCHA-1) & Overt reading; held-out repeated instances of the same closed sentence types; offline & Best-participant mean WER $\sim$3\% (no external LM) on the closed 50-sentence MOCHA-1 set; transfer reduced WER by up to 36\% in selected participant pairs \\\\

Duraivel et al.~\cite{duraivel2023high} (2023) & 4 $\mu$ECoG + 4 ECoG + 7 SEEG comparator participants & High-resolution $\mu$ECoG speech decoding (seq2seq + non-linear decoders) & Intraoperative high-density $\mu$ECoG; SVD-LDA + seq2seq RNN decoding; density/spacing ablation & 52 CVC/VCV nonwords; 9 phonemes & Study-specific nonword set & Auditory-cued overt repetition; phoneme-position classification & Four-vowel accuracy: 71.0\% ($\mu$ECoG), 36.7\% (ECoG), and 36.5\% (SEEG); no WER/PER reported \\

\bottomrule
\end{tabularx}
\end{table}

\begin{table}
\centering
\caption{Comparative summary of representative high-performance and clinical or real-time intracranial speech neuroprosthesis systems. Participant count, vocabulary or sentence-set constraint, dataset, and evaluation task are reported alongside performance so that each result can be interpreted within its original evidence base.}
\label{tab:decoding-method-review-clinical}
\tiny
\setlength{\tabcolsep}{2.3pt}
\begin{tabularx}{\linewidth}{>{\raggedright\arraybackslash}p{0.085\linewidth} >{\raggedright\arraybackslash}p{0.09\linewidth} >{\raggedright\arraybackslash}p{0.115\linewidth} >{\raggedright\arraybackslash}X >{\raggedright\arraybackslash}p{0.09\linewidth} >{\raggedright\arraybackslash}p{0.115\linewidth} >{\raggedright\arraybackslash}p{0.13\linewidth} >{\raggedright\arraybackslash}X}
\toprule
\textbf{Team (Year)} & \textbf{Participants} & \textbf{Method / Framework} & \textbf{Core Techniques} & \textbf{Vocabulary / sentence set} & \textbf{Dataset} & \textbf{Task / evaluation} & \textbf{Representative Performance} \\
\midrule
Willett et al.~\cite{willett2023high} (2023) & 1 (T12); 8 within-participant evaluation days & High-performance intracortical speech neuroprosthesis & MEA recordings (area 6v/44) + RNN phoneme/text decoder + language model; large-vocabulary decoding analysis & 50 and 125,000 words & Moses et al. 50-word set; held-out Switchboard sentences & Real-time cued attempted speech; vocalized and silent/mimed conditions & Vocalized-condition mean: 62 wpm; LM-assisted WER 9.1\% (50-word Moses set) and 23.8\% (125,000-word held-out Switchboard); pre-LM PER 19.7\% vocalized/20.9\% silent \\\\

Metzger et al.~\cite{metzger2023high} (2023) & 1 (Bravo-3) & Multimodal neuroprosthesis (text + audio + avatar) & High-density ECoG + deep multimodal decoding; CTC for unaligned silent attempted-speech training; LM rescoring; personalized voice/avatar outputs & 1,024 words; 9,655-sentence General corpus & Twitter/movie-derived General sentences; 249 novel real-time test sentences & Prompted silent attempted speech; real-time text decoding & Median 78.3 wpm, 25.5\% LM-assisted text-decoding WER, and 18.5\% direct phoneme-sequence PER \\\\

Luo et al.~\cite{luo2023stable} (2023) & 1 (CC01); 35 longitudinal sessions & Stable chronic speech BCI control without recalibration & Chronic ECoG speech-command BCI; CNN-based decoding of speech-related control commands; long-horizon deployment emphasis & 6 commands & Study-specific command trials & Self-paced overt speech for online device control & Median accuracy $\sim$90.59\% for 3 months without recalibration (6-command control paradigm) \\\\

Angrick et al.~\cite{angrick2024online} (2024) & 1 neural participant + 21 listeners; same participant as Luo et al. & Online chronic BCI speech synthesis (ALS) & Chronic ECoG + multi-stage RNN pipeline for online decoding/synthesis; high-gamma feature extraction and spectral reconstruction & 6 keywords & Study-specific command set; 1,570 training trials & Online overt keyword synthesis; six-choice listener identification & 80\% listener identification accuracy across three test days; not WER \\\\

Card et al.~\cite{card2024accurate} (2024) & 1 (T15); later reused by Wairagkar et al. & Rapidly calibrating speech neuroprosthesis (ALS) & Intracortical MEA + phoneme prediction (80 ms step) + LM-assisted text decoding + TTS voice restoration; rapid recalibration workflow & 50 and 125,000 words & BrainGate2 study-specific copy prompts and self-generated conversation & Real-time attempted-vocalized speech; instructed copy and participant-paced conversation & Initial LM-assisted text WER 0.44\% (50-word) and 9.8\% (125,000-word); later mean 2.5\% in copy and 3.7\% in labeled conversation \\\\

Littlejohn et al.~\cite{littlejohn2025streaming} (2025) & 1 (Bravo-3); same participant as Metzger et al. & Streaming brain-to-voice neuroprosthesis (bimodal) & High-density ECoG + RNN-T streaming decoder; simultaneous speech synthesis + text decoding; implicit speech detection; 80 ms chunks & 50 phrases/119 unique words; 1,024 words & AAC phrases; 12,379 unique Twitter/movie-derived General sentences & Online silent attempted speech; General test sentences appeared 6.94 times on average in training & 90.9 wpm (50-phrase AAC), 47.5 wpm (1,024-word); 1,024-word speech/text latency 1.12/1.01 s (detected attempt to first output) \\\\

Wairagkar et al.~\cite{wairagkar2025instantaneous} (2025) & 1 (T15); same participant as Card et al. & Instantaneous expressive voice-synthesis neuroprosthesis & 256-channel intracortical MEA + causal Transformer acoustic decoding + real-time vocoder; expressive/prosodic control decoding & No fixed lexicon for the main evaluation; separate 50-word first-day demo & Study-specific unique cued sentences; 956-sentence main evaluation & Real-time attempted vocalized/mimed speech; direct synthesis; six-choice and open listener tests & Mean six-choice sentence identification 94.34\%; median open-listener WER 43.75\% and PER 34.00\%; neural processing $<$10 ms per 10-ms frame \\

\bottomrule
\end{tabularx}
\end{table}

\begin{table}
\centering
\caption{Comparative summary of representative intracranial speech-decoding frameworks and an auxiliary speech-domain framework for generalization, electrode-robust modeling, and interpretable or structured representations. Participant count, vocabulary or sentence-set constraint, dataset, and evaluation task are reported alongside performance so that each result can be interpreted within its original evidence base.}
\label{tab:decoding-method-review-generalization}
\tiny
\setlength{\tabcolsep}{2.3pt}
\begin{tabularx}{\linewidth}{>{\raggedright\arraybackslash}p{0.085\linewidth} >{\raggedright\arraybackslash}p{0.09\linewidth} >{\raggedright\arraybackslash}p{0.115\linewidth} >{\raggedright\arraybackslash}X >{\raggedright\arraybackslash}p{0.09\linewidth} >{\raggedright\arraybackslash}p{0.115\linewidth} >{\raggedright\arraybackslash}p{0.13\linewidth} >{\raggedright\arraybackslash}X}
\toprule
\textbf{Team (Year)} & \textbf{Participants} & \textbf{Method / Framework} & \textbf{Core Techniques} & \textbf{Vocabulary / sentence set} & \textbf{Dataset} & \textbf{Task / evaluation} & \textbf{Representative Performance} \\
\midrule
Chen et al.~\cite{chen2024neural} (2024) & 48 & Deep-learning + differentiable synthesis neural speech framework & Unified ECoG decoder + differentiable speech synthesizer; ResNet/Swin/LSTM comparisons; causal and low-density settings; left/right hemisphere evaluation & 50 target words; 400 trials/participant & NYU epilepsy ECoG cohort & Five overt word-production tasks; participant-specific 350/50 trial split & Mean Pearson correlation coefficient (PCC) 0.806 for the noncausal ResNet; no WER/PER reported \\\\

Wu et al.~\cite{wu2024speech} (2024) & 10 & SEEG speech decoding with advanced deep learning & Linear regression vs RNN Seq2Seq vs Transformer for mel-spectrogram reconstruction; electrode contribution analysis & Unique-word count not reported; 100 presented word trials/participant & Public SingleWord\allowbreak Production\allowbreak Dutch-\allowbreak iBIDS; Dutch IFA words + digits & Overt single-word reading; offline SEEG-to-mel reconstruction & RNN/Transformer significantly outperformed linear regression; comparable performance achievable with few electrodes in some participants; strong location dependence \\\\

Chen et al.~\cite{chen2025transformer} (2025) & 52 (43 ECoG + 9 SEEG); ECoG cohort overlaps the earlier team study & SwinTW (topology-agnostic transformer for ECoG/SEEG) & Transformer with electrode-wise tokens + temporal window attention + MNI/brain-region positional bias; multi-subject training across arbitrary layouts & 50 target words; 400 trials/participant & NYU 50-word cohort/protocol, extended to SEEG & Five overt word-production tasks; within-participant and neural-decoder participant-unseen analyses; synthesizer trained on target-participant speech & Participant-specific PCC $\sim$0.82 (grid ECoG), 0.838 (extra electrodes), 0.798 (SEEG), and 0.765 (neural-decoder participant-unseen CV) \\\\

Singh et al.~\cite{singh2025transfer} (2025) & 25 & Cross-subject transfer learning via distributed recordings & Seq2Seq (temporal conv + RNN + linear readout), population latent manifold pretraining, frozen transferred recurrent layers, target-trained convolution/readout, regional electrode occlusion analysis & 128 stimuli; 218 unique words; 36 phonemes & Study-specific tongue-twister corpus & Overt articulation and pre-articulatory phoneme-sequence decoding; separate unseen-phrase and target-adapted participant-transfer evaluations & Direct sequence PER (no external LM): 27\%/34\% for fixed-length articulatory/pre-articulatory decoding and 44\%/56\% for variable-length decoding; recurrent-layer transfer PER 43\% versus 49\% within-participant \\\\

Cho et al.~\cite{cho2024coding} (2024) & Not applicable (speech/EMA corpus study) & SPARC articulatory coding framework (structured representation) & Articulatory kinematics coding (SPARC), self-supervised speech modeling + acoustic-to-articulatory inversion + speaker-identity disentanglement & No fixed vocabulary & MNGU0, MOCHA, HPRC, LibriTTS-R, and VCTK speech/EMA corpora & Speech-only articulatory coding, resynthesis, and unseen-speaker zero-shot voice conversion & Multi-corpus unseen-speaker and zero-shot voice-conversion validation; an auxiliary articulatory-representation benchmark without a neural BCI WER/PER endpoint \\\\

Li et al.~\cite{li2025high} (2026) & 9 & Acoustic--linguistic dual-pathway neural speech reconstruction & Dual path: acoustic biLSTM + HiFi-GAN and neural-driven linguistic Transformer adaptor + TTS (Parler-TTS), fused via voice cloning (CosyVoice) & No fixed vocabulary; unique-word count not reported & TIMIT sentences; $\sim$20 min passive listening/participant & Offline perceived-speech ECoG-to-speech reconstruction; Whisper/Wav2Vec2 transcription of synthesized audio & Mel-spectrogram coefficient of determination ($R^2=0.824\pm0.029$); ASR-based WER $18.9\pm3.3\%$ and PER $12.0\pm2.5\%$ \\

\bottomrule
\end{tabularx}
\end{table}

\section{Neural Mechanism Framework for Intracranial Speech Decoding}

\begin{figure}[t]
\centering
\includegraphics[width=\textwidth]{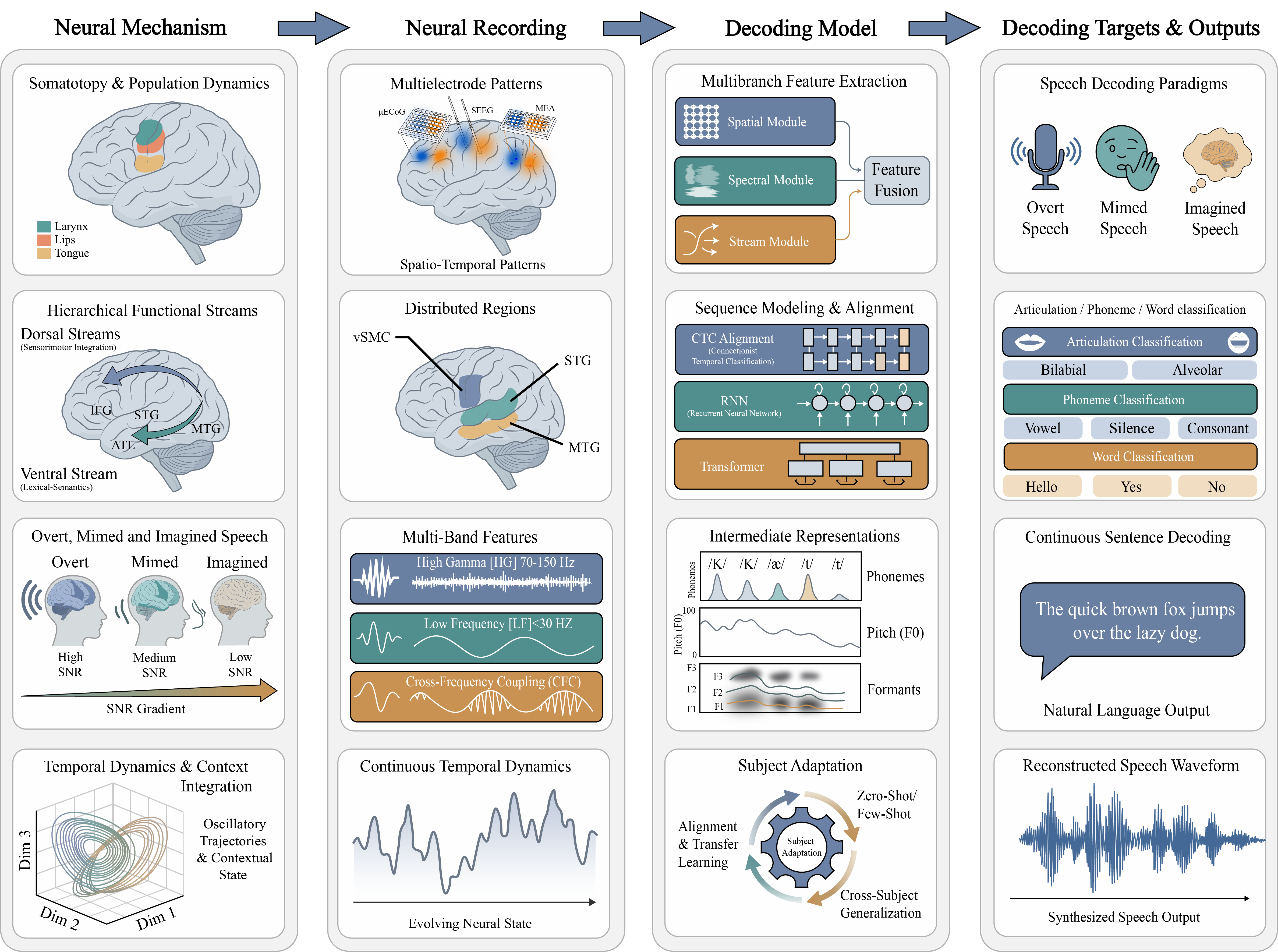}
\caption{Mechanism–Recording–Model–Task framework for neural speech decoding. Neural mechanisms of speech, including somatotopic organization, dorsal–ventral processing streams, differences among overt, mimed, and imagined speech, and temporal population dynamics, shape the neural signals recorded with ECoG, SEEG, and MEAs. These recordings capture spatially distributed activity, multiband spectral features, and continuous temporal dynamics. Decoding models integrate multibranch feature extraction, sequence architectures such as RNNs and Transformers, alignment objectives such as CTC, intermediate speech representations including phonemes, pitch, and formants, and subject-adaptation strategies. The framework supports decoding across overt, mimed, and imagined speech, with outputs spanning articulation, phoneme, and word classification, continuous sentence generation as natural-language text, and reconstruction of speech waveforms for synthesized speech.}
\label{fig:neural}
\end{figure}

As summarized in Fig.~\ref{fig:neural}, this section links four neural-mechanistic themes to a common mechanism--recording--model--task framework, highlighting how biological constraints shape measurable signals and, in turn, influence choices in decoder architecture.

\subsection{Somatotopic Organization and Articulatory Kinematics in Sensorimotor Cortex}
Within the sensorimotor cortex, the ventral sensorimotor cortex (vSMC) and middle precentral gyrus (midPrCG) exhibit spatially intermixed but somatotopically organized tuning to vocal tract articulators, including the lips, tongue, jaw, and larynx~\cite{musch2020transformation,wang2023distributed,hullett2025parallel}. This intermixing extends to the single-electrode and single-neuron level: in the ventral premotor cortex (Brodmann area 6), even a tiny region of just $3.2 \times 3.2 mm^2$ exhibits a dense and overlapping representation of multiple speech articulators. This pattern is not confined to ventral speech areas: even the dorsal 'hand knob' of the precentral gyrus shows broad tuning across phonemes, with many electrodes modulated for multiple phonemes, further evidence that neural activity related to speech production is spatially intermixed~\cite{wilson2020decoding}. Similarly, single-neuron recordings reveal heterogeneous speech-feature tuning during perception and structured phonetic encoding during production~\cite{leonard2024large,khanna2024single}, and electrodes in vSMC encode coordinated trajectories involving multiple articulators rather than single articulator movements. Despite this intermixing, population-level somatotopic organization is evident. The vSMC and midPrCG are somatotopically organized along the ventral–dorsal axis, representing different muscle groups that control the speech articulators~\cite{bouchard2013functional,chartier2018encoding,silva2024speech}. This results in an overall dorsoventral arrangement of articulator representations (e.g., lips, jaw, tongue, larynx) on the ventral pre- and post-central gyri. For example, the lips are represented more dorsally, while the tongue is distributed more broadly across ventral areas. These representations cluster into four main groups, corresponding to coronal, labial, dorsal, and vocalic articulatory configurations. The somatotopic organization is further supported by two lines of evidence. First, during passive listening to speech sounds, differential activation occurs in the precentral motor cortex that mirrors the somatotopic representation of the articulators used to produce those phonemes~\cite{pulvermuller2006motor}. Second, spatial patterns of gamma band activity over the sensorimotor cortex show distinct organizations for consonants versus vowels, confirming the coordination of distinct articulatory representations~\cite{chakrabarti2015progress}. The cortical representation of vocal tract articulators shows both somatotopic organization and intermixing. Complementary speech-domain evidence comes from linear transformations between speakers' articulatory coordinates~\cite{cho2024coding} and speech synthesis from peripheral muscle activity~\cite{gowda2026emg2speech}.

Emergent population-level representations and harmonic oscillator dynamics are key features of speech encoding in the sensorimotor cortex. In a five-participant ECoG cohort, Anumanchipalli et al.~\cite{anumanchipalli2019speech} found that neural population activity yielded low-dimensional articulatory trajectories with biphasic vowel--consonant dynamics consistent with harmonic oscillation. Individual electrodes in vSMC encode diverse, functionally stereotyped articulatory kinematic trajectories (AKTs). These out-and-back trajectories exhibit damped harmonic oscillator dynamics and correspond to the vocal tract constrictions that define phonemes~\cite{chartier2018encoding}. Principal component analysis and clustering reveal that the cortical state-space is organized into clusters corresponding to major oral articulators and vowels, with consonants and vowels occupying distinct regions consistent across subjects~\cite{bouchard2013functional}. This study also revealed harmonic oscillator-like dynamics underlying the temporal progression of articulatory representations. Population-level neural dynamics during speech include a large condition-invariant signal (CIS) at movement initiation, followed by rotatory (oscillatory) dynamics in the neural state during articulation~\cite{stavisky2019neural}. The Directions Into Velocities of Articulators (DIVA) model models feedforward and sensory-feedback interactions among cortical areas during speech motor control~\cite{golfinopoulos2010integration}. In the later intracortical study of one BrainGate2 participant (T12), Willett et al.~\cite{willett2023high} used naive Bayes classifiers on 1 s of neural activity to decode 33 orofacial movements with 92\% accuracy, 39 phonemes with 62\% accuracy, and 50 words with 94\% accuracy across repeated within-participant sessions. Together, the earlier five-participant ECoG cohort and the later T12 study provide independent cross-platform agreement on mixed, low-dimensional articulatory organization, while the quoted classifier values remain specific to T12. Decoding performance improves with the number of electrodes in a log-linear relationship.

At the cellular scale, single-neuron recordings complement field-potential evidence by resolving feature selectivity that is obscured in aggregate signals. In the STG, neurons across cortical depths show heterogeneous tuning to phonetic, prosodic, envelope, and sequence features during speech perception~\cite{leonard2024large}. During speech production, language-dominant prefrontal neurons encode planned phonetic sequences in a temporally ordered cascade~\cite{khanna2024single}. STG population spiking also predicts high-frequency ECoG responses, directly linking cellular activity to mesoscopic field potentials~\cite{leonard2024large}. In speech BCIs, intracortical arrays use spikes and threshold crossings to access local phonetic and articulatory population states at high resolution~\cite{willett2023high,card2024accurate}. These signals complement the broader field-potential coverage of ECoG and SEEG, integrating cellular and mesoscopic evidence within the mechanism--recording--model framework.

Controversies persist regarding the consistency of somatotopic patterns across individuals. Intracranial studies provide evidence for emergent population-level representations, but there is individual variability in spatial patterns. Speaker-dependent variations in sensorimotor cortex responses captured by SEEG are attributed mostly to differences in electrode placement~\cite{arthur2024speech}, with interindividual variability in neural activations and electrode placement contributing to inconsistencies in spatial activation patterns~\cite{meng2023continuous}. Variability is also observed in the extent and consistency of population dynamics and neural modulation between participants, with some showing stronger and more stable modulation than others. Additionally, while population-level articulatory dynamics are consistent across speakers~\cite{anumanchipalli2019speech}, somatotopic articulatory representations in motor cortex are not consistently active or spatially organized during speech perception across individuals~\cite{cheung2016auditory}, and some consistencies in somatotopic tuning across subjects are observed alongside ongoing variability and inconsistencies. The generalizability of multi-subject models is supported by accommodating inter-subject variability in electrode placement, though performance variability remains~\cite{chen2025transformer}. Variability across individuals in somatotopic patterns and tuning also contributes to controversies regarding the consistency of these representations~\cite{panachakel2021decoding}.

To distill the mechanistic implications of evidence from somatotopy and population-level analyses, we summarize our synthesis in the ``Somatotopy and Population Dynamics'' column of Table~\ref{tab:neural-critical-synthesis}. Specifically, mixed local tuning and low-dimensional trajectories jointly constrain feature design and decoding targets; intermixed-yet-somatotopic encoding coexists with robust CIS/oscillatory motifs, while substantial inter-individual variability persists. This synthesis corresponds to the somatotopy/population-dynamics branch in Fig.~\ref{fig:neural}.

\subsection{Hierarchical Functional Streams in Speech Processing}
Speech processing is widely recognized to involve two hierarchical functional streams: a dorsal form-to-articulation stream and a ventral form-to-meaning stream~\cite{fridriksson2016revealing,martin2019use}. The dorsal stream, connecting temporal and frontal regions via the arcuate fasciculus, supports speech production and phonological processing, while the ventral stream, coursing through the extreme capsule fiber system, is responsible for comprehension and semantic processing~\cite{qiu2025review}. This dual-stream architecture is substantiated by lesion studies, where dorsal stream damage leads to motor and phonological deficits, and ventral stream damage results in comprehension and lexical-semantic impairments. Distributed networks for sensorimotor integration, including primary motor/premotor cortices and temporal areas, support dorsal stream form-to-articulation processing, with distinct topographies for local motor potential (LMP) and frequency-based features indicating different neural origins~\cite{schalk2007decoding}.

During speech perception, intracranial recordings reveal frequency- and region-dependent tracking of the speech envelope. The posteromedial Heschl's gyrus (HG), the presumed core auditory cortex, resolves the temporal envelope of speech stimuli through mechanisms across a wide range of ECoG frequencies. High-frequency event-related band power (ERBP) reflects fine-grained temporal encoding in this region, even under conditions of compromised intelligibility~\cite{nourski2009temporal}. High-gamma activity (70–150 Hz) tracking of speech is concentrated in the superior temporal gyrus (STG), while low-frequency phase tracking (1–7 Hz) is more widespread, encompassing the STG, anterior temporal cortex, and inferior frontal cortex~\cite{golumbic2013mechanisms}.

Phonetic representations are encoded in distributed neural populations. The STG exhibits a caudal-anterior functional gradient, with posterior "onset" zones sensitive to acoustic onsets and phrase boundaries, and anterior "sustained" zones encoding ongoing phonetic and lexical-semantic features~\cite{hamilton2018spatial}. Within STG, phonetic feature selectivity is organized by manner and place of articulation, with distinct clusters for obstruents (plosives, fricatives) and sonorants (vowels, nasals), forming a multidimensional acoustic feature space~\cite{mesgarani2014phonetic}. During speech production, phonetic features can be decoded from neural activity in the vSMC and STG, with confusion patterns clustering according to articulatory characteristics~\cite{moses2019real}. In an overt speech-production task with concurrent auditory feedback, broadband-gamma SEEG signals from bilateral temporal and motor cortices reconstructed the acoustic speech envelope. The strongest temporal contributions arose from the left STG and right MTG~\cite{ivucic2025speech}. The face motor cortex (FMC) shows locally distributed neural assemblies encoding articulatory movements, enabling higher word classification accuracy than Wernicke's area, which exhibits more abstract semantic encoding~\cite{kellis2010decoding}. Phonetic discriminability during continuous speech production involves dynamic activation of Broca's area (supporting articulation planning), sensorimotor areas (supporting articulation), and STG (supporting speech processing), with peak activation timing aligned to phone production~\cite{herff2015brain}. Speech decoding leverages both articulatory information from the sensorimotor cortex and perceptual or semantic information from the temporal lobe. The vSMC often provides the most informative signals, with increased electrode coverage further improving decoding accuracy~\cite{stavisky2025restoring}.

Lexical-semantic processing engages ventral stream regions. Word prediction primarily activates lower-order regions like the STG and MTG, while sentence-level prediction recruits default mode network (DMN) regions, such as the temporoparietal junction (TPJ), medial prefrontal cortex (mPFC), and precuneus~\cite{zhou2025hierarchical,heilbron2022hierarchy}. Lexical activation is associated with the left posterior MTG (pMTG), while lexical selection engages the left inferior frontal gyrus (LIFG) and medial frontal areas. Lesions to the left pMTG result in persistent word retrieval deficits~\cite{ries2016choosing}. More specifically, ECoG decoding of abstract versus concrete semantic attributes was supported by delayed IFG activity, whereas early STG activity did not support significant semantic classification in that analysis~\cite{nagata2022spatiotemporal}. During natural speech comprehension, IFG activity tracked higher-order contextual transformations, while mid-STG activity was associated primarily with auditory-phonological processing, consistent with complementary processing across functional streams~\cite{goldstein2025temporal}.

Cross-stream interaction is evident in overlapping functional networks and dynamic information flow. The STG's onset and sustained zones interact to support speech parsing, with onset activity providing temporal landmarks and sustained activity enabling continuous phonetic processing. Functional interactions between word and sentence prediction levels occur sparsely at sentence boundaries, supporting the sparse updating hypothesis. Conduction aphasia, arising from impaired integration between dorsal and ventral streams, underscores the critical role of cross-stream communication for fluent speech. Recurrent neural computations within the STG may enable the binding of form to articulation and meaning, rendering stream reliance context-sensitive~\cite{yi2019encoding}.

Individual differences in stream localization are observed across spatial, spectral, and functional domains. In HG, envelope following shows considerable intersubject variability, with some subjects maintaining high-frequency ERBP envelope tracking even for unintelligible speech. The frequency profile of gamma reactivity (low- versus high-gamma activity) and its spatial topography vary across subjects, reflecting individual differences in functional anatomy~\cite{crone2001induced}. STG phonetic feature clusters show systematic individual variation but share common organizational principles. Speech-responsive electrode distributions, including tone- and syllable-discriminative electrodes in vSMC and widespread speech-responsive regions, exhibit intersubject variance influenced by language behaviors and functional anatomy~\cite{zhang2024brain}. Sensorimotor cortex activation patterns during speech perception overlap with those observed during production but show individual-specific differences, potentially leading to false-positive activations in speech BCIs trained on production data~\cite{schippers2024don}.

Task demands modulate the relative engagement of the streams, with high-gamma and low-frequency responses providing complementary temporal information across sampled cortical regions~\cite{goldstein2025temporal,golumbic2013mechanisms}. Phonological production tasks rely more on the dorsal stream, while lexical-semantic comprehension tasks depend on the ventral stream. At the system level, decoding latency and speed vary with speech type and vocabulary complexity~\cite{littlejohn2025streaming}. Different speech stimuli, such as phonemes, words, and sentences, engage distinct hierarchically organized networks~\cite{gwilliams2025hierarchical}, with acoustic-phonetic features driving dorsal stream reliance and lexical-semantic features enhancing ventral stream engagement~\cite{wang2025progress}.

To frame hierarchical stream organization in a comparative way, the ``Hierarchical Functional Streams'' column of Table~\ref{tab:neural-critical-synthesis} summarizes the central pattern, where convergent dorsal--ventral evidence is accompanied by unresolved questions about stream overlap and context-dependent dominance; this aligns with the dorsal--ventral pathway component of Fig.~\ref{fig:neural}.

\subsection{Neural Signatures of Overt, Mimed, and Imagined Speech}
Overt, mimed, and imagined speech exhibit distinct neural correlates across frequency bands and cortical regions. For instance, power spectrum changes for overt and imagined speech are spatially comparable but less pronounced for imagined speech, with fewer cortical sites showing significant changes, particularly in broadband high-frequency activity (BHA), also termed high-gamma activity. In auditory (superior temporal) cortex, high-gamma activity increases during overt speech but decreases during imagined speech, consistent with the absence of auditory feedback, whereas in sensorimotor regions it increases during both conditions~\cite{proix2022imagined,de2024imagined}. ECoG studies have demonstrated distinct spatio-temporal cortical activation patterns during continuous overt and silently repeated speech production: fronto-motor regions (ventral primary motor cortex, ventral premotor cortex, inferior frontal gyrus) activate prior to speech output during overt speech, while temporal regions (superior and middle temporal gyri) activate primarily during and after speech output, reflecting auditory feedback processing~\cite{brumberg2016spatio,khalilian2024corollary,ozker2024speech}. Direct comparison of overt and silently repeated speech ECoG signals reveals common activations in auditory cortex, premotor and primary motor cortex, frontal eye fields, and dorsolateral prefrontal cortex, but silent-repetition signals do not reliably encode acoustic features like the speech intensity envelope. MEG studies further show distinct decoding performances across stages (pre-stimuli, perception, preparation/imagination, production/articulation) for overt and imagined speech, with the highest accuracy during production, followed by imagination~\cite{dash2020decoding}. Additionally, imagined speech involves distinct neural mechanisms engaging Broca's and Wernicke's areas with differential activation patterns~\cite{su2025systematic,wandelt2024representation}.

Evidence of hierarchical channel nesting and differential frequency contributions to decoding is substantial. While high-gamma activity provides the best signal for overt speech decoding, both low- and higher-frequency power and local cross-frequency-coupling (CFC) contribute importantly to imagined speech decoding, particularly in phonetic and vocalic spaces. Low-frequency power (e.g., theta, low-beta) and cross-frequency dynamics (e.g., phase-amplitude coupling between low-beta and high-gamma activity) contain key information for imagined speech decoding. Spatial-spectral-temporal convolutional neural network (CNN) approaches, which integrate spatial information from multiple sensors with spectral and temporal features, enhance decoding accuracy by modeling spatial--spectral--temporal structure in the sampled signals. Intracranial electroencephalography (iEEG) studies identify distinct frequency bands: high-gamma activity linked to phoneme articulation and low-frequency theta band activity associated with syllabic and prosodic modulation, with cross-frequency coupling (e.g., theta phase-gamma amplitude coupling) reflecting hierarchical neural dynamics~\cite{al2025mistr}. Thalamocortical studies further reveal hierarchical processing, where thalamic neurons prefer higher temporal modulation rates, and cortical neurons show slower preferences, indicating differential frequency contributions across regions~\cite{miller2002spectrotemporal}. For overt speech, high-gamma activity (70--170 Hz) is highly task-related, with spectrogram reconstruction showing correlations peaking around the first formant frequency (300 Hz)~\cite{herff2016towards}. Combining low-frequency components (0–50 Hz) with the high-gamma envelope (70–150 Hz) yields higher speech reconstruction accuracy than single bands, implicating complementary hierarchical encoding~\cite{akbari2019towards}.

Controversies persist regarding how far imagined-speech decoding can scale beyond constrained paradigms. Supportive intracranial evidence comes from three independent research programs: Proix et al.~\cite{proix2022imagined} observed above-chance low- and cross-frequency decoding in 13 participants across three task cohorts; Martin et al.~\cite{martin2016word} studied five participants and obtained significant imagined-word classification in four; and Kunz et al.~\cite{kunz2025inner} later characterized attempted and imagined speech in four implanted BrainGate2 participants. Their agreement establishes qualitative replication of decodable imagined-speech information across distinct tasks. Non-invasive phrase decoding during imagination stages provides complementary evidence~\cite{dash2020decoding}. Counter-evidence includes weak acoustic-envelope recoverability during silent repetition~\cite{brumberg2016spatio}, declines in decoding performance in earlier continuous speech pipelines~\cite{herff2015brain}, and systematic reviews emphasizing heterogeneity in tasks, compliance checks, and preprocessing choices~\cite{su2025systematic,zhang2026imagined}. Methodologically, limited paired neural and speech data for supervision and temporal misalignment persist as practical bottlenecks~\cite{chen2024neural,verwoert2022dataset,he2025vocalmind,zhao2025open}. Overall, current studies indicate feasibility for proof-of-concept tasks but suggest that clinical-grade reliability remains to be established.

Within-study comparisons report task- and band-dependent differences among overt, mimed (mouthed without vocalization), and imagined speech, with overt conditions typically exhibiting stronger and more spatially extensive high-gamma activity than imagined conditions~\cite{proix2022imagined,brumberg2016spatio,soroush2023nested}. This task-dependent pattern both motivates and constrains imagined-speech decoding: positive feasibility evidence shows above-chance decoding of imagined speech in constrained word and phrase paradigms using low-frequency and CFC features~\cite{proix2022imagined,martin2016word,dash2020decoding}, with recent implanted-participant work reporting real-time decoding of attempted and imagined speech~\cite{kunz2025inner}. At the same time, negative findings indicate that signals from silent repetition and imagined speech encode acoustic detail less reliably and exhibit substantial variability across participants and tasks~\cite{brumberg2016spatio,herff2015brain,su2025systematic}, suggesting that robust large-vocabulary communication is not yet established across users and settings for imagined speech. Crucially, task- and band-dependent signal quality guides the choice of intermediate representations for decoding. Overt speech-related motor signals are stronger and more directly linked to acoustic features, supporting decoding strategies that exploit high-gamma articulatory correlates~\cite{martin2014decoding,chartier2018encoding}. In contrast, imagined speech exhibits weaker, more spatially diffuse activity and poorer discriminability of articulatory representations in high-gamma activity; instead, it more reliably engages perceptual (phonetic and vocalic) representations indexed by low-frequency bands and CFC~\cite{proix2022imagined,livezey2019deep,de2024imagined}, consistent with the flexible abstraction hypothesis~\cite{cooney2018neurolinguistics}. Given the task-dependent signal quality and typical data scarcity in implanted settings, this motivates targeting low-frequency and CFC features~\cite{proix2022imagined,livezey2019deep}. Reported data-efficiency gains in production-based decoding~\cite{chen2024neural,angrick2024online} further motivate comparisons of explicit articulatory–acoustic intermediate parameters (e.g., pitch, formant frequencies)~\cite{bouchard2014control} with direct spectrograms or latent vectors to assess their data efficiency in imagined-speech decoding. Thus, imagined-speech variability demands adaptation toward perceptual and low-frequency structure~\cite{proix2022imagined,cooney2018neurolinguistics,de2024imagined}, while overt speech can exploit higher-SNR, high-gamma articulatory signals~\cite{martin2014decoding,chartier2018encoding}.

To contextualize mimed and imagined speech mechanisms for decoding design, the synthesis appears in the ``Overt, Mimed and Imagined Speech'' column of Table~\ref{tab:neural-critical-synthesis}, where the current pattern suggests task-dependent overt--imagined signal differences together with the practical contribution of low-frequency/CFC features; this is the speech-mode pathway highlighted in Fig.~\ref{fig:neural}.

\subsection{Temporal Dynamics and Context Integration in Speech Representation}
Empirical neural data reveal temporal context dependence in phonological representations, motivating temporal decoding models. Temporal context is modeled in the Long Short-Term Memory (LSTM) network introduced by Hochreiter and Schmidhuber in 1997~\cite{hochreiter1997long}, which, through its memory cells featuring constant error carousels and multiplicative gate units, is designed to control information flow. This architecture enables the storage of context-dependent information over long intervals, thereby facilitating context-sensitive sequence prediction. Building on this, Graves et al.~\cite{graves2006connectionist} combined bidirectional LSTM (BiLSTM) networks with Connectionist Temporal Classification (CTC) loss for unsegmented sequence prediction. The encoder supplies context-sensitive label probabilities, and CTC sums their frame-wise products over monotonic paths that collapse to the target sequence after repeated labels are merged and blanks removed. The value of temporal modeling for speech decoding is empirically reinforced by recent studies. Luo et al. (2022)~\cite{luo2022brain} showed that deep learning models like BiLSTMs map temporally structured neural features to sequence-based speech representations, consistent with independently observed context-sensitive cortical representations~\cite{musch2020transformation,wang2023distributed}. Corroborating this, Metzger et al. (2023)~\cite{metzger2023high} provided evidence that the sensorimotor cortex (SMC) maintains temporally structured articulatory representations. These representations are somatotopically organized by place of articulation (POA) and are actively involved in context-dependent phonological encoding.

Evidence from intracranial recordings reveals that dynamic neural population activity during speech production is organized in ways consistent with motor-control-related patterns. Specifically, Metzger et al. (2023)~\cite{metzger2023high} reported strong correlations between the neural encoding of individual phones and attempted orofacial movements, including tongue raising and lip puckering. These findings are consistent with dynamic SMC representations of articulatory gestures, but they do not by themselves establish the cortical computation that generates those representations. Using high-density micro-electrocorticographic ($\mu$ECoG) arrays, Duraivel et al. (2023)~\cite{duraivel2023high} showed that neural population dynamics in the SMC are organized by articulatory features and phonemes. The observation that decoding errors most frequently involve phonemes sharing articulatory features is consistent with a motor-control-based hypothesis of neural encoding, but error structure alone does not establish that mechanism. Further supporting this framework, Martin et al. (2016)~\cite{martin2016word} reported that dynamic high-gamma time courses obtained from ECoG recordings reflect complex temporal neural dynamics during speech production. Their use of dynamic time warping (DTW) to accommodate temporal variability resulted in improved classification accuracy, supporting the value of temporal alignment for decoding. Complementing this, Heelan et al. (2019)~\cite{heelan2019decoding} applied LSTM decoders to reconstruct spectrotemporal and phonetic features from secondary auditory cortex recordings in non-human primates, providing complementary evidence for decoding perceived speech from auditory population activity.

The integration of contextual information through temporal dynamics enables the decoding of continuous, naturalistic speech by allowing models to process unsegmented sequences and leverage distributed neural representations. Graves et al. (2006)~\cite{graves2006connectionist} showed that CTC eliminates the need for pre-segmented data, allowing BiLSTM networks to train directly on continuous input sequences and handle temporal variability, a capability critical for naturalistic speech processing. In one participant, Metzger et al. (2023)~\cite{metzger2023high} demonstrated that a bidirectional RNN with CTC loss could decode prompted, silently attempted sentences from ECoG at a median 78 words per minute, with the recurrent network itself proving critical for performance. This reflects the network's ability to integrate neural activity over time. Supporting this, Luo et al. (2022)~\cite{luo2022brain} reported that sentence-level neural activity encodes richer temporal information than isolated phonemes, and that multi-stage deep neural networks, bridging neural features to audio synthesis via intermediary representations, can capture the dynamics of speech motor patterns, enhancing continuous speech decoding. Duraivel et al. (2023)~\cite{duraivel2023high} independently demonstrated ordered-phoneme sequence decoding without explicit onset labels in four $\mu$ECoG participants, with four ECoG and seven SEEG participants as comparators; Martin et al.'s separate five-participant ECoG cohort~\cite{martin2016word} likewise supported temporal sequence structure. These cohorts provide cross-study convergence on the value of temporal context while using different tasks. Further evidence comes from Soroush et al. (2023)~\cite{soroush2023nested}, who identified multi-frequency temporal features (theta, alpha, beta, high-gamma activity) proximal to speech onset that support continuous speech representation. Broader spatial integration also appears important: Heelan et al. (2019)~\cite{heelan2019decoding} found that decoding performance improves with higher neural channel counts, reflecting the importance of integrating context across spatially distributed cortical populations~\cite{verwoert2025whole}. Collectively, these findings underscore that temporal neural structure and model-based context integration provide a strong basis for decoding fluent, naturalistic speech from intracranial brain recordings.

To integrate neural temporal evidence with model-level results, a concise summary is provided in the ``Temporal Dynamics and Context Integration'' column of Table~\ref{tab:neural-critical-synthesis}, where recurrent context integration improves sequence decoding but attribution to neural mechanism versus model capacity remains unsettled; in Fig.~\ref{fig:neural}, this corresponds to the temporal population-dynamics pathway linked to sequence-model architectures such as RNNs and Transformers, together with alignment objectives such as CTC.

\begin{table}
\centering
\scriptsize
\caption{Critical synthesis of neural mechanism evidences across four dimensions, organized by consensus, controversies, sources of discrepancies, and experiments to resolve them.}
\label{tab:neural-critical-synthesis}
\begin{tabularx}{\textwidth}{>{\raggedright\arraybackslash\bfseries}p{0.11\textwidth} >{\raggedright\arraybackslash}X >{\raggedright\arraybackslash}X >{\raggedright\arraybackslash}X >{\raggedright\arraybackslash}X}
\toprule
 & \textbf{Somatotopy and Population Dynamics} & \textbf{Hierarchical Functional Streams} & \textbf{Overt, Mimed and Imagined Speech} & \textbf{Temporal Dynamics and Context Integration} \\
\midrule
Consensus & Spatially intermixed but population-somatotopic speech representations are repeatedly observed in vSMC/midPrCG, with robust CIS and oscillatory motifs during initiation and articulation. & A dorsal form-to-articulation stream and a ventral form-to-meaning stream are consistently supported by lesion, stimulation, and intracranial decoding evidence. & Overt--imagined signal differences vary with task and frequency band; overt speech relies more on high-gamma, while imagined speech can carry useful low-frequency and CFC information. & Temporal modeling and context integration consistently improve continuous-speech decoding, especially with unsegmented neural streams and sequence-level objectives. \\
\midrule
Controversy & Subject-level reproducibility of somatotopic layouts and the universality of CIS/oscillatory motifs remain uncertain. & The key dispute is the degree of dorsal-ventral segregation versus overlap and how dominance shifts with task context. & Whether imagined-speech decoding can reach clinically reliable, large-vocabulary performance remains unresolved. & Performance gains may reflect biologically meaningful temporal mechanisms or simply larger model capacity and dataset scale. \\
\midrule
What causes discrepancies & Coverage and localization heterogeneity, inter-subject pathology differences, and task differences (phoneme/word/sentence; overt/mimed/imagined). & Uneven sampling of temporal/frontal/subcortical regions, inconsistent temporal alignment strategies, and non-uniform spectral feature definitions. & Low SNR, missing acoustic alignment labels, uncertain behavioral compliance, variable electrode sampling, and inconsistent CFC definitions. & Different sequence lengths, alignment assumptions (explicit vs. CTC-like), vocabulary complexity, and inconsistent handling of cross-session non-stationarity. \\
\midrule
Experiments to resolve & Standardize cross-center electrode localization, map contacts to Montreal Neurological Institute (MNI)/atlas common space, and learn shared representations with uncertainty estimates. & Use harmonized multi-task protocols within participants and combine causal perturbation with recordings to test directional information flow. & Combine behavioral verification tasks, mimed- and imagined-speech neural-alignment surrogates, self-supervised/contrastive pretraining, and explicit multi-band feature fusion. & Run matched-capacity model comparisons with temporal ablations and longitudinal benchmarks to separate short-term fitting from stable context representations. \\
\bottomrule
\end{tabularx}

\end{table}

\section{Decision-Oriented Hardware Protocol of Intracranial Recording Modalities}
As summarized in Fig.~\ref{fig:modalities}, the choice of recording modality and implant design directly determines which neural features can be captured (e.g., spikes versus field potentials), which anatomical targets can be accessed (surface, depth, or penetrating cortex), and the long-term maintenance burden in chronic use. These factors largely define the attainable information content and establish the fundamental upper bound on performance attainable from the recorded neural signal alone, while decoder design primarily governs how closely a given implementation approaches this bound. Language priors could additionally improve end-task performance by contributing complementary linguistic information beyond the recording. Consequently, intracranial speech BCIs should treat electrode selection as a primary design variable. In this section, we review intracranial electrode families, representative suppliers, and development trajectories. Quantitative comparisons emphasize hardware-level constraints that shape downstream decoding potential before considering software-side optimization.

\subsection{Intracranial Electrode Families, Representative Suppliers, and Development Trajectories}

\begin{table}[t]
\centering
\scriptsize
\caption{Summary of intracranial electrode families and representative suppliers relevant to speech BCI research and clinical translation. Supplier examples are non-exhaustive; availability and regulatory status vary by region, indication, and system compatibility.}
\label{tab:electrode-families-suppliers}
\begin{tabularx}{\textwidth}{>{\raggedright\arraybackslash\bfseries}p{0.18\textwidth} >{\raggedright\arraybackslash}X >{\raggedright\arraybackslash}X >{\raggedright\arraybackslash}X}
\toprule
\textbf{Electrode family} & \textbf{Typical geometry / signal scale} & \textbf{Typical role in speech BCI} & \textbf{Representative suppliers or platform examples}\\
\midrule
Macro-ECoG grids/strips (subdural or epidural cortical surface) &
Millimeter-scale contacts on flexible strips/grids; captures cortical field potentials and broadband high-gamma over broad peri-Sylvian cortex. &
Clinical mapping-compatible pathway; candidate for reliability-first communication interfaces and broad cortical coverage studies. &
\href{https://adtechmedical.com/epilepsy}{Ad-Tech Medical (Subdural grids/strips)};  
\href{https://investor.integralife.com/news-releases/news-release-details/integra-epilepsy153-products-receive-ce-mark-distribution}{Integra LifeSciences (Epilepsy products/brain mapping lines)};
\href{https://www.neuracle.cn/}{Neuracle (NEO; implantable wireless semi-invasive epidural system)};
\href{https://sinovationmed.com/?lang=en}{Sinovation | HKHS (Intracranial Cortical Electrode: strip/grid models)}\\
\midrule
High-density surface ECoG / $\mu$ECoG (thin-film cortical arrays) &
Sub-millimeter to millimeter contact pitch with flexible/thin-film substrates; improves spatial sampling relative to macro-ECoG while preserving surface coverage. &
Higher-fidelity articulatory and phonemic decoding, speech synthesis feature extraction, and research on fine cortical organization. &
\href{https://cortec-neuro.com/products-components/grid-and-strip/}{CorTec (\textdegree AirRay grid/strip electrodes)}; 
\href{https://nmtc1.com/evo-cortical}{NeuroOne (Evo\textsuperscript{\textregistered} Cortical thin-film electrode)}; 
\href{https://precisionneuro.io/product}{Precision Neuroscience (Layer 7 cortical interface)};
\href{https://www.neuroxess.com/}{NeuroXess (Surftrode; flexible cortical electrode)}.\\
\midrule
SEEG depth electrodes (stereotactic depth shafts) &
Linear contact arrays along depth trajectories; samples distributed cortical and subcortical structures in 3D with clinically routine stereotactic workflows. &
Network-level language/speech studies, deep target access, and clinically practical implantation in epilepsy-style pathways. &
\href{https://diximedical.com/solutions/microdeep}{DIXI Medical (MICRODEEP\textsuperscript{\textregistered} SEEG)}; 
\href{https://adtechmedical.com/epilepsy}{Ad-Tech Medical (Depth electrodes/SEEG)};
\href{https://www.rishena.com/?l=en}{Rishena (Single-use Depth Electrode (SEEG))};
\href{https://sinovationmed.com/?lang=en}{Sinovation | HKHS (Intracranial Depth Electrode)};
\href{https://www.neuroxess.com/}{NeuroXess (Silktrode; flexible deep electrode)}.\\
\midrule
Penetrating intracortical MEA (Utah-style or shank-based arrays) &
Dense penetrating microelectrodes with highest local specificity; captures spikes/threshold crossings, spike-band power, and local field potentials from small cortical territories (often ``up to \textasciitilde 100'' electrodes per array depending on design). &
Fine-scale speech-motor decoding and low-latency/high-rate decoding demonstrations, especially when local spiking activity is the target signal. &
\href{https://blackrockneurotech.com/products/utah-array/}{Blackrock Neurotech (Utah Array)}; 
\href{https://blackrockneurotech.com/products/slant-array/}{Blackrock Neurotech (Utah Slant Array)}; 
\href{https://microprobes.com/products/multichannel-arrays/mea}{MicroProbes for Life Science (penetrating MEAs)};
\href{https://www.stairmed.com/en/}{StairMed (HNE ultra-flexible micro-nano electrode; shank-based penetrating arrays)}.\\
\midrule
Emerging flexible penetrating/threaded arrays (investigational) &
Flexible ultrathin penetrating elements and high channel counts; aims to improve channel scalability and chronic tissue interface properties. &
Potential future route for combining high channel count with improved chronic packaging and fully implanted wireless systems. &
\href{https://neuralink.com/updates/building-safe-implantable-devices/}{Neuralink (threaded arrays; investigational)}; 
\href{https://www.paradromics.com/product}{Paradromics (Connexus BCI; investigational)};
\href{https://www.neuroxess.com/}{NeuroXess (Silktrode / Plextrode; investigational flexible electrodes)};
\href{https://www.stairmed.com/en/}{StairMed (ultra-flexible penetrating electrodes; platform ecosystem)}.\\
\midrule
Hybrid recording-stimulation cortical/depth systems &
Electrode systems designed for both neural recording and stimulation/functional mapping; actual capability depends on electrode specs + lead/connector + stim/record hardware ecosystem. &
Useful for combined mapping-decoding workflows, causal perturbation studies, and translational systems that may require stimulation-assisted functions. &
\href{https://cortec-neuro.com/products-components/grid-and-strip/}{CorTec (\textdegree AirRay: record \& stimulate stated)}; 
\href{https://diximedical.com/solutions/microdeep}{DIXI (SEEG + functional mapping / thermocoagulation context)}; 
\href{https://adtechmedical.com/epilepsy}{Ad-Tech (mapping/stimulation ecosystem dependent)};
\href{https://sinovationmed.com/?lang=en}{Sinovation | HKHS (Intracranial Depth Thermal Coagulation Electrode; mapping/ablation ecosystem)};
\href{https://www.rishena.com/?l=en}{Rishena (SEEG ecosystem; thermocoagulation-oriented product lines described)};
\href{https://www.neuroxess.com/}{NeuroXess (Plextrode; combined deep/cortical electrode)}.\\
\bottomrule
\end{tabularx}
\begin{tablenotes}[flushleft]
\footnotesize
\item Notes: this table emphasizes engineering categories used in review synthesis; exact electrode dimensions, contact counts, magnetic resonance (MR)-conditional labeling, stimulation rating, connectors, and approved indications vary within and across suppliers.
\end{tablenotes}

\end{table}

Electrode choice is a foundational design decision that shapes the decoding target (e.g., text vs.\ speech synthesis), surgical workflow, calibration burden, packaging complexity, and the feasibility of eventual home deployment. The relevant electrode families used in speech BCIs can be broadly categorized into surface cortical arrays (macro-ECoG and high-density/$\mu$ECoG), SEEG depth electrodes, and penetrating intracortical MEAs. In parallel, emerging hybrid systems are seeking to integrate recording and stimulation modalities within chronically stable packaging. Across these electrode families, the primary design consideration involves a nuanced trade-off that balances coverage, spatial specificity, surgical accessibility, and long-term maintainability, factors that extend well beyond signal quality alone. Table~\ref{tab:electrode-families-suppliers} summarizes the major intracranial electrode families used in or adjacent to speech BCI development, together with representative commercial suppliers and platform examples. Note that the supplier list is illustrative and non-exhaustive; verification against local regulatory approvals, regional distribution, and current product portfolios is essential prior to procurement and clinical planning.

The trajectory of intracranial electrodes in speech BCI development reflects a progression from clinically available mapping hardware toward purpose-optimized neuroprosthetic interfaces. Early proof-of-concept decoding studies leveraged electrodes originally designed for epilepsy monitoring, beginning with macro-ECoG grids and strips, and later incorporating SEEG depth electrodes, thereby capitalizing on established neurosurgical workflows and clinical monitoring infrastructure~\cite{leuthardt2004brain,leuthardt2011using,herff2015brain}. This foundational work demonstrated that cortical field potentials, especially in the high-gamma band, contain rich speech information, and that intracranial recordings are capable of supporting continuous or near-continuous decoding in controlled settings.

The next phase focused on improving spatial resolution, task complexity, and decoding fidelity. High-density cortical recordings and improved modeling pipelines enabled stronger articulation-level decoding and speech synthesis, while broader datasets clarified how electrode placement and coverage heterogeneity shape performance~\cite{anumanchipalli2019speech,makin2020machine,moses2021neuroprosthesis,luo2022brain,duraivel2023high}. In parallel, SEEG expanded the accessible anatomical search space beyond cortical surface coverage, supporting distributed language-network sampling and more flexible trajectory planning under clinical constraints~\cite{chen2024neural,wu2024speech,pescatore2025decoding}. Recent work represents a third phase in which electrode hardware is increasingly evaluated by chronic utility rather than acute signal quality alone. Intracortical MEAs enabled high-rate and high-fidelity speech neuroprosthesis demonstrations~\cite{willett2023high,card2024accurate,wairagkar2025instantaneous}, while chronic ECoG systems demonstrated stronger home-use feasibility in selected users with ALS~\cite{luo2023stable,angrick2024online}. This shift reframes the notion of the ``best electrode'' as a use-case-dependent decision -- one that balances coverage, resolution, packaging, calibration burden, and care logistics, rather than simply maximizing any single signal metric~\cite{stavisky2025restoring,singer2025speech,wyse2024stability}.

Looking forward, the most important electrode-development directions for speech neuroprostheses are likely to include: (i) coverage--resolution co-optimization (e.g., hybrid macro+$\mu$ECoG or surface+depth strategies), (ii) chronic biocompatibility improvements and lower-impedance stable interfaces, (iii) fully implanted wireless systems with onboard preprocessing to reduce infection risk and setup burden, (iv) high-channel-count active multiplexed arrays that preserve bandwidth within thermal/power limits, and (v) electrode designs that support both decoding and causal stimulation/mapping for adaptive closed-loop speech systems~\cite{xu2026two,jhilal2025implantable,wairagkar2025instantaneous,stavisky2025restoring}. These future directions are already visible in emerging recording technologies that advance high-fidelity neural capture for speech BCIs by addressing limitations in spatial-temporal resolution, biocompatibility, and coverage. Among these, novel flexible electrode arrays and high-channel-count microelectrode systems have shown particular promise. Xu et al. (2026)~\cite{xu2026two} developed a flexible molybdenum disulfide (MoS2)-based active array with ultrahigh spatial resolution (up to 51 pixels $mm^{-2}$), a broad bandwidth (0.5 Hz--10 kHz), and fast sampling rates (80 kHz), enabling capture of fine-grained high-frequency multiunit activity with µV-range amplitude. Fabricated on flexible polyimide substrates, this array demonstrated excellent biocompatibility and stable SNRs over 4 weeks of chronic implantation, outperforming commercial graphene-based passive arrays in tuning curve selectivity, latency, and amplitude variance.

Intracortical MEAs have also shown significant advancements in capturing speech-related neural activity. In one BrainGate2 participant (T12), Willett et al. (2023)~\cite{willett2023high} used intracortical spiking activity for real-time, cued attempted-speech decoding, achieving 9.1\% WER on the Moses et al. 50-word sentence set, 23.8\% WER with a 125,000-word language model on held-out Switchboard sentences, and 62 words per minute. They further identified that the ventral premotor cortex (area 6v) contains a dense, intermixed representation of speech articulators within a small (3.2 × 3.2 mm$^2$) region, with phoneme articulation details preserved even years after paralysis. Notably, decoding accuracy improved log-linearly with electrode count, suggesting that denser arrays could further enhance performance. In a second single BrainGate2 participant (T15), Wairagkar et al. (2025)~\cite{wairagkar2025instantaneous} implanted 256 microelectrodes in the precentral gyrus and drove a voice synthesizer with near-instantaneous feedback, including control of intonation and emphasis. The T12 and T15 demonstrations thus provide convergence across two participants within the same clinical program. For ECoG, Luo et al. (2022)~\cite{luo2022brain} highlighted innovations such as 32 recording contacts per electrode, robotic insertion techniques to minimize tissue trauma, and flexible array designs, which address traditional limitations in spatial density and coverage. These advances, combined with deep learning methods mapping ECoG high-gamma activity to vocal tract trajectories, support improved neural capture for speech decoding. While long-term ECoG stability for speech remains under investigation, motor BCI research has demonstrated reliable chronic decoding, and safety in individuals with late-stage ALS has been reported.

These emerging technologies, including MoS2-based active arrays with ultrahigh spatiotemporal resolution, high-channel-count intracortical microelectrodes, and advanced ECoG configurations, overcome traditional constraints by enhancing resolution, biocompatibility, and coverage. They enable the capture of fine-grained speech-related neural activity, from spiking dynamics to paralinguistic features, and hold significant potential for advancing the clinical translation of speech BCIs.

\subsection{Quantitative Trade-offs in Resolution, Signal Quality, and Clinical Feasibility}
Willett et al. (2023)~\cite{willett2023high} conducted their intracortical MEA investigation in one participant (T12), with eight within-participant evaluation days providing repeated-session evidence. In terms of spatial sampling density, the 3.2 x 3.2 mm$^2$ MEAs used in their study contained 64 channels each, exhibiting rich and spatially intermixed tuning to speech articulators, which allows for precise decoding from a small cortical area~\cite{willett2023high}. For signal quality, threshold crossing rates with root-mean-square (RMS) thresholds ranging from -4.5 to -3.5 effectively detected strong spiking activity well above the noise floor. Notably, combining spike band power with threshold crossings was found to improve RNN decoding performance. Regarding coverage depth, the arrays were implanted intracortically, targeting superficial cortical layers in area 6v. Although their cortical coverage was narrow, this was offset by the high neural information content per channel. In terms of decoding performance, intracortical MEAs paired with RNN decoders achieved a high decoding speed of 62 words per minute, which is 3.4 times faster than previous BCI speech decoding approaches~\cite{willett2023high}. Real-time cued attempted-speech decoding yielded 9.1\% WER on the Moses et al. 50-word sentence set and 23.8\% WER with a 125,000-word language model on held-out Switchboard sentences. Error rates decreased log-linearly as the electrode count increased. Specifically, doubling the number of electrodes nearly halved the error rate~\cite{willett2023high}.

In considering the matching of task to recording modality, high-resolution MEAs have supported precise speech reconstruction at the phoneme and word levels, including rapid large-vocabulary decoding~\cite{willett2023high,card2024accurate}. These results motivate selecting MEAs when the target signal is local spiking activity, while ECoG grids provide field-potential sampling over broader cortical areas. Participant characteristics, coverage, task design, and decoder architecture jointly inform interpretation of the reported performance. However, MEAs have notable constraints: they require surgical implantation, have limited spatial coverage, and their long-term stability and clinical viability need further validation~\cite{silva2024speech,jhilal2025implantable}. Additionally, in the Willett et al. paradigm, training data requirements are high, with 260--480 sentences collected per day, though decoding still performs reasonably with reduced retraining~\cite{willett2023high}; the required amount depends on electrode coverage, vocabulary, and task design. The study further reported that, in participant T12, threshold-crossing activity from the sampled Brodmann area~44 sites yielded classification accuracy below 12\% for the tested orofacial movements, phonemes, and words within a 1-s post-cue window. This participant- and task-specific result indicates that electrode targeting and coverage can affect performance in the evaluated configuration.

\begin{figure}
\centering
\includegraphics[width=\textwidth]{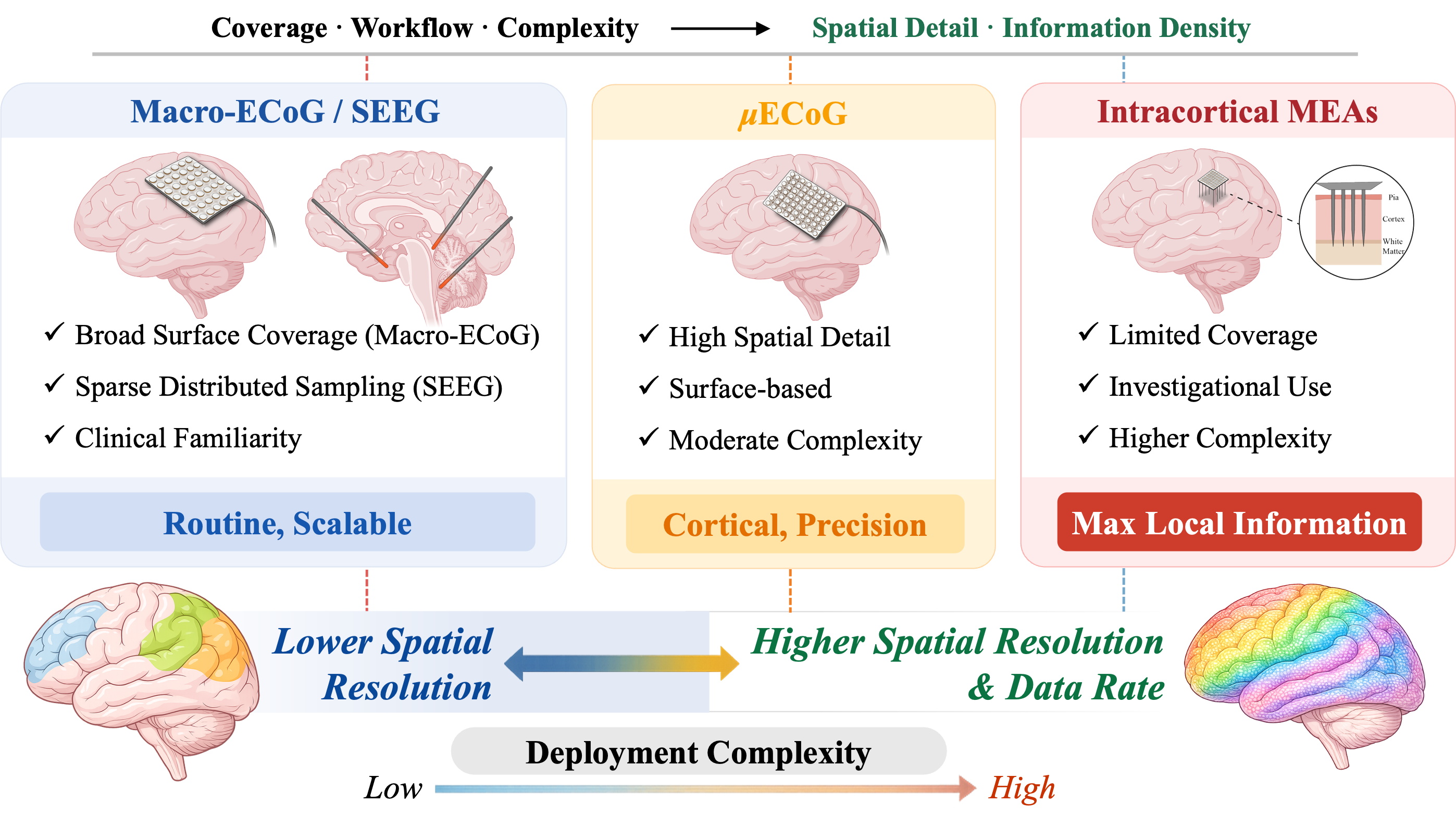}
\caption{Schematic illustration of a practical selection heuristic for intracranial recording modalities. Macro-ECoG prioritizes broad, contiguous cortical surface coverage, whereas SEEG provides sparse, distributed 3D network sampling with deep-target access; both use established clinical implantation workflows. $\mu$ECoG prioritizes cortical spatial detail, and intracortical MEAs prioritize maximal local information density and high-rate decoding at higher implantation/maintenance complexity.}
\label{fig:select}
\end{figure}

A structured and decision-relevant synthesis of these modality trade-offs is provided in Table~\ref{tab:modality-decision-matrix}, which integrates measurement scale, signal type, implantation pathway, chronic evidence, and deployment constraints for real-world speech neuroprostheses. Figure~\ref{fig:select} therefore presents candidate modality--requirement correspondences rather than general allocation rules: macro-ECoG may be considered when broad, contiguous cortical-surface field-potential coverage is required; SEEG when distributed three-dimensional sampling and depth access are required; $\mu$ECoG when cortical spatial detail is a primary constraint; and penetrating intracortical MEAs when the target measurement is local spiking activity from a limited cortical territory. Final selection remains scenario specific and must account for the intended participant, anatomical target and achievable coverage, task protocol, decoder, implantation risk, and long-term maintenance requirements~\cite{leuthardt2004brain,leuthardt2011using,luo2022brain,willett2023high,card2024accurate,wairagkar2025instantaneous,wu2024speech}.

\begin{table}[t]
\centering
\scriptsize
\caption{Quantitative decision matrix for intracranial speech BCI recording modalities. Values represent typical ranges and reported practice patterns rather than absolute limits. Timing entries distinguish processing, update intervals, and event-defined system delays.}
\label{tab:modality-decision-matrix}
\begin{tabularx}{\textwidth}{>{\raggedright\arraybackslash\bfseries}p{0.1\textwidth} >{\raggedright\arraybackslash}X >{\raggedright\arraybackslash}X >{\raggedright\arraybackslash}X}
\toprule
\textbf{Dimension} & \textbf{MEA (intracortical)} & \textbf{ECoG ($\mu$ECoG/macro-ECoG)} & \textbf{SEEG (depth electrodes)} \\
\midrule
Spatial resolution / sampling scale & Sub-millimeter local sampling with highest unit-level specificity in small cortical patches~\cite{willett2023high,wairagkar2025instantaneous}. & Surface sampling from sub-millimeter $\mu$ECoG to millimeter-scale macro contacts; strong cortical field-potential specificity~\cite{duraivel2023high,luo2022brain}. & Millimeter-scale contacts along depth shafts; sparse per contact but distributed 3D sampling across deep and surface targets~\cite{wu2024speech,chen2024neural}. \\
\midrule
Dominant speech-relevant signals & Spikes, threshold crossings, spike-band power, and local field potentials (LFP)~\cite{willett2023high}. & High-gamma activity plus lower-frequency LFP phase/amplitude features~\cite{luo2022brain,golumbic2013mechanisms}. & Depth LFP and high-gamma features with access to deeper structures and distributed language networks~\cite{wu2024speech,pescatore2025decoding}. \\
\midrule
Coverage footprint & Limited cortical territory (local spiking specificity, low areal coverage)~\cite{willett2023high}. & Broad peri-Sylvian cortical coverage for speech motor-auditory mapping~\cite{leuthardt2011using,luo2022brain}. & Sparse, distributed 3D sampling across bilateral and deep targets with clinically constrained but flexible trajectories~\cite{chen2024neural,wu2024speech}. \\
\midrule
Implant pathway (research vs. routine) & Penetrating cortical implantation via craniotomy; currently mostly investigational BCI use~\cite{card2024accurate}. & Subdural/epidural cortical placement with clinically established neurosurgical workflows (primarily epilepsy mapping)~\cite{leuthardt2004brain,leuthardt2011using}. & Stereotactic depth implantation with mature clinical workflow in epilepsy monitoring~\cite{chen2024neural}. \\
\midrule
Chronic stability evidence & Sustained intracortical speech performance has been reported over 8.4 months and independent home use over 19 months; these results incorporated accumulating training data or continuous background decoder updates~\cite{card2024accurate,card2026long}. & ECoG evidence includes 3-month command decoding without model recalibration, session-wise recalibrated synthesis using 5.5-month-old training data, and one-year high-gamma signal analyses~\cite{luo2023stable,angrick2024online,wyse2024stability}. & Speech decoding has been demonstrated with SEEG, whereas speech-specific chronic stability and unattended home use have not yet been evaluated longitudinally~\cite{wu2024speech,chen2025transformer}. \\
\midrule
Home-use feasibility & More than 3,800 h of independent home use over 19 months has been reported in one participant; caregiver support for donning and doffing percutaneous hardware remained part of the workflow~\cite{card2026long}. & Multi-month home operation has been demonstrated in one participant, including a 3-month fixed-decoder command interface~\cite{luo2023stable}. & The clinically established implantation route motivates prospective home-use studies, but no speech-specific chronic home-use series currently supports a stability claim~\cite{chen2024neural,wu2024speech}. \\
\midrule
Training data and calibration burden & Current pipelines range from daily labeled training~\cite{willett2023high} to little-to-no explicit daily speech calibration supported by continuous background fine-tuning~\cite{card2026long}. & The reported burden ranges from no retraining or baseline recalibration during 3-month command use~\cite{luo2023stable} to session-wise recalibration for online synthesis~\cite{angrick2024online}. & Prospective chronic studies are needed to quantify speech-specific calibration burden. \\
\midrule
Reported timing quantities & Neural acquisition-to-synthesized-frame processing $<$10 ms; phoneme prediction updates every 80 ms~\cite{wairagkar2025instantaneous,card2024accurate}. & Median speech-offset-to-decoding-registration delay of 1.24 s has been reported; online word-level synthesis is feasible~\cite{luo2023stable,angrick2024online}. & Latency evidence remains less standardized and often task/pipeline dependent in current speech studies~\cite{wu2024speech}. \\
\midrule
Regulatory and device maturity & Predominantly investigational for communication neuroprostheses (Investigational Device Exemption (IDE)-stage evidence)~\cite{card2024accurate,stavisky2025restoring}. & Electrode modality/procedures are clinically mature for mapping; chronic communication indication remains investigational~\cite{leuthardt2004brain,stavisky2025restoring}. & High procedural clinical maturity for diagnostic use; chronic speech-prosthetic translation still emerging~\cite{chen2024neural,stavisky2025restoring}. \\
\bottomrule
\end{tabularx}

\end{table}

\subsection{Chronic Implant Stability and Clinical Deployment Constraints}
Longitudinal evidence supports two complementary routes to stable operation: sustained signal or end-to-end performance with ongoing data collection, recalibration, or decoder updates; and stable performance with a frozen decoder and no baseline or session-wise recalibration. Adaptive operation shows how chronic utility can be maintained despite neural non-stationarity, while frozen-decoder operation demonstrates the potential to reduce recalibration burden. Table~\ref{tab:longitudinal-stability-evidence} organizes this evidence by recording modality and update status. Longitudinal speech BCI evidence currently comes from ECoG and intracortical arrays, complemented by short-term speech-decoding studies using SEEG.

\begin{table}[t]
\centering
\scriptsize
\caption{Representative longitudinal speech BCI evidence separated by recording modality, duration and use setting, and decoder-update status. Signal persistence and sustained performance under adaptation are distinguished from stable performance without recalibration.}
\label{tab:longitudinal-stability-evidence}
\begin{tabularx}{\textwidth}{>{\raggedright\arraybackslash}p{0.08\textwidth} >{\raggedright\arraybackslash}p{0.10\textwidth} >{\raggedright\arraybackslash}X >{\raggedright\arraybackslash}X >{\raggedright\arraybackslash}X}
\toprule
\textbf{Study} & \textbf{Recording modality} & \textbf{Longitudinal evidence / use setting} & \textbf{Recalibration and data-collection status} & \textbf{Stability claim supported} \\
\midrule
Luo et al.~\cite{luo2023stable} & Chronic ECoG & Self-paced control of computer applications and external devices over 3 months in one participant. & Training data were collected before deployment; the reported 3-month use required neither model retraining nor baseline recalibration. & Direct participant-level evidence for stable fixed-decoder command classification without recalibration. \\
\midrule
Angrick et al.~\cite{angrick2024online} & Chronic ECoG & Three online speech-synthesis sessions on three days, conducted 5.5 months after acquisition of the training data. & Each online session included recalibration but no model retraining; test data were collected in each session. & Persistence of a useful neural--acoustic mapping across months under session-wise recalibration. \\
\midrule
Wyse-Sookoo et al.~\cite{wyse2024stability} & Chronic ECoG & High-gamma speech signals were analyzed longitudinally over one year in the same implanted participant. & Repeated longitudinal recordings supported signal analysis; unattended frozen-decoder operation was not the evaluated endpoint. & Signal-level persistence relevant to decoder design, distinct from end-to-end performance without recalibration. \\
\midrule
Card et al.~\cite{card2024accurate} & Intracortical MEAs & 84 home sessions over 32 weeks (8.4 months), including more than 248 cumulative hours of communication, in one participant. & Data from multiple days were added continuously to calibrate the decoder; during 29 personal-communication-only days, three participant-requested updates each used approximately 7.5 min of labeled data. & Sustained high performance with continuing data accumulation and occasional explicit recalibration, not fixed-decoder stability. \\
\midrule
Card et al.~\cite{card2026long} & Intracortical MEAs & Near-daily independent home use over 19 months, exceeding 3,800 h, in the same participant. & The speech decoder was continuously recalibrated in the background and required little-to-no explicit daily calibration; the cursor decoder used a short daily calibration routine. & Strong evidence for independent chronic deployment with low explicit speech-calibration burden and continuous adaptation. \\
\bottomrule
\end{tabularx}
\end{table}

This evidence establishes two complementary routes to longitudinal utility: a fixed ECoG command decoder can operate for months without recalibration in a selected participant, and adaptive ECoG or intracortical systems can preserve useful performance across longer intervals by incorporating session-wise or continuous updates. Card et al.~\cite{card2026long} extend the intracortical evidence from repeated research sessions to sustained independent home use, while also showing that continuous background fine-tuning is part of the stability mechanism. The present evidence is therefore strongest for modality-specific feasibility in individual users; larger longitudinal cohorts can determine how update burden and performance generalize across implants and clinical settings.

Safety and operational outcomes should be interpreted alongside decoder performance. Chronic ECoG and intracortical reports have documented encouraging safety outcomes in the studied participants~\cite{luo2023stable,card2024accurate,card2026long}, while percutaneous hardware, donning and doffing assistance, labeled-data requirements, and explicit or background decoder updates remain measurable components of deployment burden. Treating these components as primary endpoints enables direct comparison of autonomous use. For deployment decisions, pre-specified surgical, infection, thermal-safety, and essential operational requirements define the set of eligible configurations. A configuration that fails any prerequisite is excluded, and $\mathcal{U}$ is not calculated for it; accuracy, responsiveness, or convenience therefore cannot compensate for ineligibility. Accuracy, responsiveness, and residual deployment burden guide trade-offs only within the eligible set. To formalize these deployment trade-offs, we propose a transparent conceptual framework for scenario-specific multi-criteria assessment among eligible configurations: a \emph{bounded} deployment utility can be written as
\begin{equation}
\mathcal{U} = w_{A}\tilde{A} + w_{R}\tilde{R} - \sum_{k=1}^{5}\lambda_{k}\tilde{C}_{k},
\label{eq:deployment-utility}
\end{equation}
where $\tilde{A}$ denotes a normalized task-accuracy score (larger is better), $\tilde{R}$ denotes a normalized responsiveness score derived from end-to-end latency (larger is better), and $\tilde{C}_k$ are normalized clinical deployment-burden scores (larger is worse). All components are mapped to $[0,1]$ to make $\mathcal{U}$ dimensionless and bounded (avoiding scale dependence on units such as ms vs.\ s), and weights are nonnegative and optionally normalized for interpretability (e.g., $w_A+w_R+\sum_{k=1}^{5}\lambda_k=1$). 

The framework couples primary reporting of individual outcomes with a secondary, scenario-specific utility summary. Prospective validation can establish how this conceptual summary relates to deployment outcomes. In practice, $\mathcal{U}$ can be computed per candidate system configuration (hardware + decoder + workflow) to illustrate trade-offs among eligible alternatives under scenario-specific weight settings. Concretely, $\tilde{A}$ can be instantiated using task-appropriate accuracy metrics mapped to $[0,1]$ (e.g., the WER normalization defined in Section~\ref{sec:clt-benchmark} for text output), and $\tilde{R}$ can be instantiated by mapping latency into $[0,1]$ via a pre-registered monotone transformation over an acceptable latency range. Accordingly, the deployment-burden vector is taken as 
\begin{equation}
\tilde{C}=[\tilde{C}_{\text{infection}},\tilde{C}_{\text{power}},\tilde{C}_{\text{packaging}},\tilde{C}_{\text{care}},\tilde{C}_{\text{home}}],
\end{equation}
with the following operational interpretations:
\begin{itemize}
\item $\tilde{C}_{\text{infection}}$ (percutaneous-interface risk): transcutaneous connectors increase infection risk; quantify using infection events per implant-day together with intervention burden (e.g., clinic visits, antibiotics, revisions), mapped to $[0,1]$.
\item $\tilde{C}_{\text{power}}$ (battery--wireless budget): wireless bandwidth for high-rate neural streams must be balanced against battery life and tissue heating constraints; quantify using average power draw, wireless duty-cycle, and thermal margin, mapped to $[0,1]$.
\item $\tilde{C}_{\text{packaging}}$ (long-term encapsulation): hermetic sealing quality, corrosion resistance, and impedance drift determine effective service lifetime; quantify using impedance drift, failure-rate proxies, or predicted service lifetime, mapped to $[0,1]$.
\item $\tilde{C}_{\text{care}}$ (patient adherence and maintainability): daily setup time, caregiver workload, and recalibration frequency directly affect sustained home adoption; quantify using minutes/day, caregiver involvement, and recalibration events per week, mapped to $[0,1]$.
\item $\tilde{C}_{\text{home}}$ (domestic robustness): system performance under household acoustic noise, motion artifacts, and electromagnetic interference must be prospectively stress-tested; quantify using performance degradation under stress (e.g., $\Delta\mathrm{WER}$, dropout rate, latency inflation), mapped to $[0,1]$.
\end{itemize}

\section{Algorithmic Advances in Generalization and Interpretability}
In the hardware--software co-design view of Fig.~\ref{fig:modalities}, algorithmic choices determine how effectively recorded neural information is converted into robust linguistic outputs under fixed recording constraints. Recording hardware therefore remains fundamental to signal fidelity and neural-information recovery, while language priors could provide complementary external information that further improves end-task performance. Beyond subject-specific laboratory demonstrations, clinically viable speech BCIs require decoders that generalize across participants, heterogeneous electrode layouts, sessions, and speech modes, while remaining interpretable enough to support mechanistic hypothesis testing and user-centered safeguards. This section reviews recent advances spanning topology-agnostic sequence and transformer architectures, biologically grounded articulatory intermediates, language-prior assistance, dual-path and multi-modal decoding frameworks, and interpretable models for testing hypotheses about neural representations.

\begin{figure}
\centering
\includegraphics[width=\textwidth]{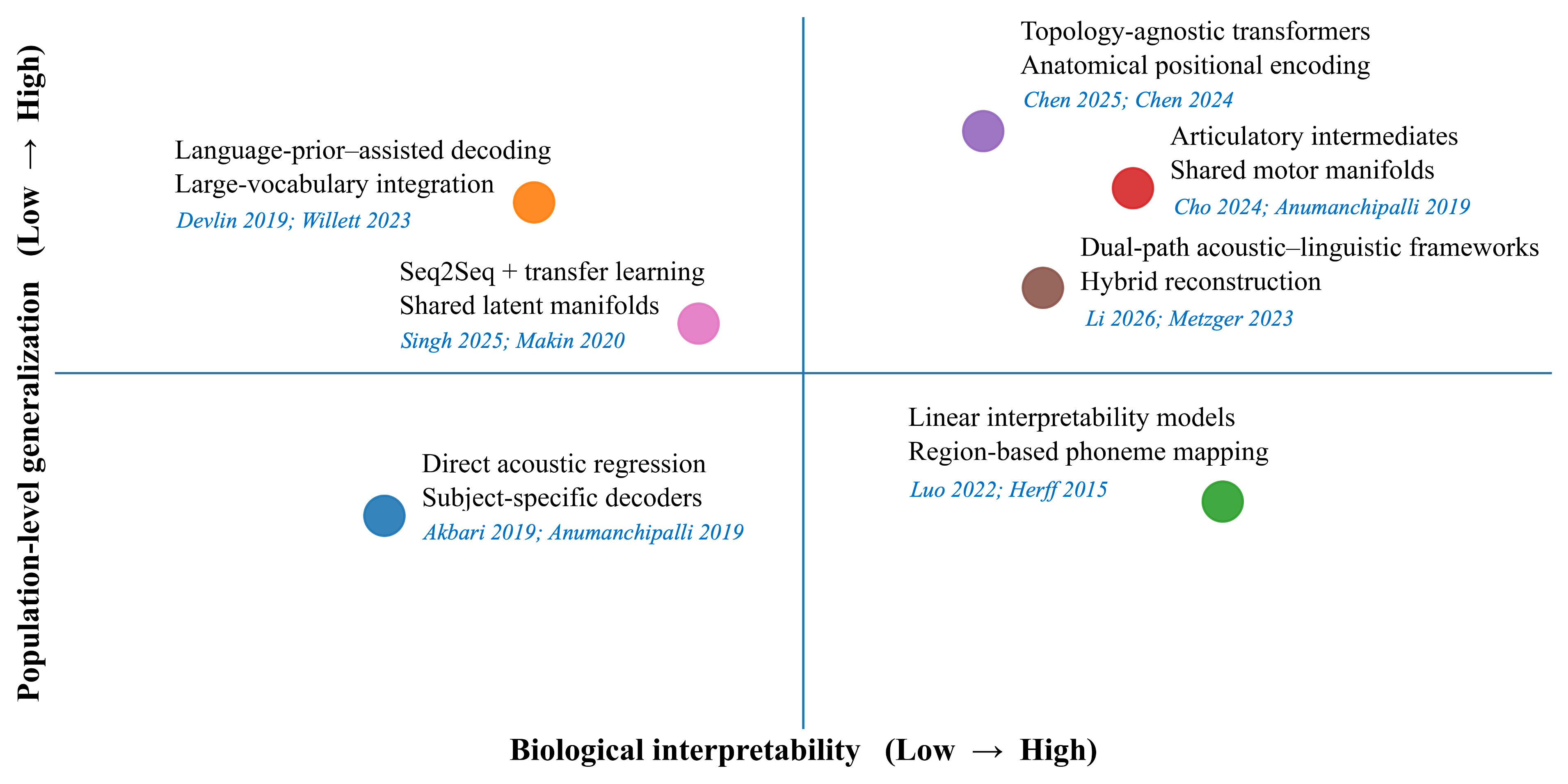}
\caption{Conceptual landscape of speech BCI decoding models. Decoding approaches are positioned qualitatively along two axes: biological interpretability and population-level generalization. Direct acoustic models typically offer limited interpretability and cross-subject robustness, whereas articulatory intermediates and topology-agnostic transformers occupy the high-generalization regime. Dual-path frameworks integrate acoustic and linguistic representations, and language-prior–assisted models improve large-vocabulary performance but may introduce bias. Positions reflect conceptual trends rather than quantitative benchmarking.}
\label{fig:landscape}
\end{figure}

\subsection{Cross-Subject Generalization via Sequence and Transformer Architectures}
This section addresses the software-design side of Fig.~\ref{fig:modalities}: decoder architecture, intermediate representations, and adaptation strategies largely determine how efficiently neural features (e.g., high-gamma, low-frequency, or cross-frequency coupling (CFC) features) are converted into robust outputs under a fixed recording setup. Fig.~\ref{fig:timeline} outlines this hardware--software co-design perspective by organizing representative intracranial speech-decoding frameworks into three thematic lanes, with compact technical tags that make cross-task design choices visually comparable. To synthesize the trade-space discussed in this section, Fig.~\ref{fig:landscape} further provides a conceptual map of major decoding paradigms by representation interpretability and the transfer protocol. This framing helps explain why the following subsections emphasize topology-agnostic transformers and articulatory intermediate targets as promising routes for scalable and interpretable decoding, while also highlighting where language-prior assistance and direct acoustic decoding introduce distinct strengths and risks.

Cross-subject generalization in speech BCIs has been significantly advanced through the application of sequence and transformer architectures, which address the challenges of variable electrode placements and sparse cortical coverage. Sequence-to-sequence models have demonstrated robust performance in decoding speech-related neural signals across participants. Singh et al. (2025)~\cite{singh2025transfer} studied 25 epilepsy participants performing an overt tongue-twister task comprising 128 stimuli, 218 unique words, and 36 phonemes. Their sequence-to-sequence transfer framework isolated shared latent manifolds and achieved a median variable-length PER of 43\%, compared with 49\% for within-participant training. The framework pretrained model components on one participant, then compared transfer configurations; the best-performing configuration froze the recurrent layer while training the remaining participant-specific components, including the input convolution and linear readout, on target-participant neural data to accommodate variable electrode configurations. Makin et al. (2020)~\cite{makin2020machine} studied four participants with an RNN-LSTM encoder--decoder for ECoG-to-text decoding. Their approximately 3\% minimum was the best participant's mean WER on held-out repeated instances from a fixed set of 50 MOCHA-1 sentences (about 250 unique words); within the study's closed overt-reading sentence inventories, cross-participant transfer with target-participant training reduced relative WER by up to 36\%, including across ECoG arrays in different hemispheres. For SEEG data, Wu et al. (2024)~\cite{wu2024speech} studied 10 epilepsy participants, each reading 100 presented Dutch-word trials; the number of unique words was not reported. Both RNN-based and transformer-based sequence-to-sequence models outperformed linear regression baselines in mel-spectrogram reconstruction accuracy, with comparable results achievable using only a few electrodes, highlighting the robustness of these architectures to electrode sparsity.

Transformer-based decoders have further enhanced cross-subject generalization by leveraging anatomical information and shared neural representations. As complementary out-of-modality evidence, Défossez et al. (2023)~\cite{defossez2023decoding} introduced a shared deep convolutional architecture with a participant-specific layer, trained on non-invasive MEG/EEG data using contrastive learning aligned to pretrained speech representations (wav2vec 2.0). This group-derived model outperformed individual-specific decoders, achieving up to 70.7\% top-10 segment-level accuracy in MEG datasets in zero-shot retrieval of held-out perceived speech segments from trained participants. Within the intracranial modality scope, recent transformer studies have evaluated cross-participant transfer using intracortical arrays, ECoG, and SEEG. Zhang et al. (2026)~\cite{zhang2026cross} pretrained the BraIn-to-Text (BIT) transformer encoder on 367 h of Utah-array recordings from humans and monkeys across speech and arm-related motor tasks. After participant-specific fine-tuning, the non-ensembled cascaded model achieved 6.35\% WER on the one-participant Brain-to-Text '24 benchmark (T12; 125,000-word vocabulary; 1,200 held-out Switchboard sentences; cued attempted speech) and 4.06\% WER on the separate one-participant Brain-to-Text '25 benchmark (T15; 125,000-word vocabulary; 1,450 held-out multi-corpus sentences; cued attempted speech). For imagined speech in participants T12 and T15 using the 50-word Kunz et al. inner-speech dataset~\cite{kunz2025inner}, pretrained variants reduced WER by 39--45\% relative to a transformer trained from scratch. These results demonstrate the value of transferable initialization followed by participant-specific fine-tuning. Chen et al. (2025)~\cite{chen2025transformer} evaluated the topology-agnostic SwinTW transformer in 52 participants (43 ECoG and nine SEEG), extending the team's earlier work in an ECoG cohort overlapping its 2024 dataset. The shared model was fitted and tested on held-out trials from the same 15 ECoG participants (PCC 0.837 versus 0.831 for participant-specific models), whereas participant-unseen generalization was evaluated separately by participant-held-out cross-validation across the 43-participant ECoG cohort (mean PCC 0.765); the nine-participant SEEG-only analysis yielded a mean PCC of 0.798. SwinTW handles arbitrarily positioned electrodes by using their 3D anatomical locations (MNI coordinates and brain region embeddings) instead of 2D grid indices, generating tokens from each electrode and applying temporal window attention without participant-specific layers. The synthesis pipeline pairs this shared neural decoder, applied to unseen participants without target-participant neural training, with a synthesizer pretrained on each target participant's speech. This is why topology-agnostic transformer families are highlighted for transfer potential in Fig.~\ref{fig:landscape}: their architectural inductive bias is designed around anatomical variability rather than fixed electrode grids, which addresses an important barrier to cross-subject transfer.

Collectively, these sequence and transformer approaches suggest that explicitly modeling shared structure—via latent articulatory manifolds or anatomy-aware tokenization—is central to cross-subject transfer. Singh et al.~\cite{singh2025transfer} demonstrated that multi-subject models, which concatenate subject-specific features and learn shared recurrent layers, reduced PER and were robust to regional electrode occlusion (REO), particularly showing resilience in sensorimotor and temporal lobe electrode removal, whereas single-subject models were significantly affected by removal of key speech production regions. In the non-invasive MEG/EEG setting, Défossez et al.~\cite{defossez2023decoding} similarly found that decoding accuracy increased with more training participants, indicating the model learned common neural representations across individuals. Chen et al.~\cite{chen2025transformer} showed that SwinTW's use of anatomical positional bias enabled learning of shared latent representations across diverse electrode layouts, improving cross-subject robustness. Despite these advancements, challenges in achieving consistent cross-subject performance remain~\cite{zhao2026rmetnet}. Singh et al.~\cite{singh2025transfer} noted that group model accuracy plateaus at ~3-4 subjects, and transfer learning performance improves with increased similarity of electrode coverage patterns between training and inference subjects. Wu et al.~\cite{wu2024speech} highlighted electrode contribution variability across subjects: some participants exhibit critical electrodes whose removal causes significant performance degradation, while others show distributed information, complicating generalization; additionally, SEEG's sparse, clinically constrained coverage limits consistent performance. Makin et al.~\cite{makin2020machine} pointed to limited training data per participant as a barrier to scaling to larger vocabularies and unconstrained language. Chen et al.~\cite{chen2025transformer} acknowledged that while SwinTW generalizes well to unseen participants, achieving consistently high accuracy for subjects outside training cohorts remains a challenge, motivating future research into larger datasets and subject-specific refinements.

For clinical deployment, population-level generalization and individual personalization should be viewed as complementary rather than competing objectives. Multi-participant models can learn shared neural--linguistic representations across heterogeneous participants and electrode layouts, but studies using participant-specific layers, anatomy-aware representations, and cross-subject transfer learning indicate that some degree of user-specific adaptation is likely to remain necessary~\cite{defossez2023decoding,chen2025transformer,singh2025transfer}. A group-derived model may therefore serve as a robust initialization, while a relatively small calibration dataset is used to adapt selected components, such as the electrode-level input projection, participant-alignment or normalization parameters, and the final decoding head, rather than retraining the entire model from scratch~\cite{makin2020machine,defossez2023decoding,singh2025transfer}. Recent speech-neuroprosthesis studies further demonstrate that subject-specific decoders can be calibrated and progressively refined over repeated use, although the amount of initial training and subsequent recalibration varies substantially across systems and recording modalities~\cite{willett2023high,metzger2023high,card2024accurate,luo2023stable}. Following initial calibration, controlled online or session-wise adaptation may further compensate for neural non-stationarity. Related intracranial BCI studies have demonstrated self-calibration, long-term closed-loop decoder adaptation, and latent-space alignment as potential strategies for reducing repeated supervised recalibration~\cite{jarosiewicz2015virtual,silversmith2021plug,karpowicz2025stabilizing}. Such adaptation should nevertheless be constrained by held-out performance checks, drift-detection criteria, and explicit user confirmation to limit gradual error accumulation and unintended changes in decoded content. Future studies should consequently report not only cross-subject accuracy, but also the amount of individual calibration data, calibration time, adapted model components, update frequency, longitudinal stability, and the incremental performance gained through personalization.

\subsection{Biologically Grounded Articulatory Representations for Cross-Speaker Transfer}
Biologically grounded articulatory representations, which encode the coordinated kinematic trajectories of vocal-tract movements, have emerged as effective intermediate targets in speech BCIs, offering distinct advantages in data efficiency in neural decoding and cross-speaker generalization in speech-domain models. These representations leverage the low-dimensional, somatotopic encoding of articulatory gestures in the sensorimotor cortex, where neural activity reflects the physical shaping of the vocal tract to produce speech sounds. Studies have demonstrated that such articulatory features are highly conserved across speakers, with state-space trajectories of decoded articulation showing high similarity (r > 0.8) between participants, suggesting a shared neural representation of articulatory kinematics that facilitates cross-speaker transfer. Cho et al.'s Speech Articulatory Coding (SPARC) framework~\cite{cho2024coding} was validated on multi-speaker speech and EMA corpora, providing speech-domain evidence for transferable articulatory representations. Anumanchipalli et al.'s independent five-participant ECoG study~\cite{anumanchipalli2019speech} provides complementary neural evidence, together supporting representation-level convergence across speech-corpus and neural studies. SPARC proposes a universal articulatory space in which a single-speaker EMA template can be adapted to multiple speakers via linear affine transformations that compensate for anatomical differences, enabling speaker-agnostic acoustic-to-articulatory inversion with performance comparable to multi-speaker systems.
Accordingly, Fig.~\ref{fig:landscape} highlights the physical interpretability and transfer potential of articulatory-intermediate paradigms: the representation itself remains interpretable in terms of speech motor control.

In terms of data efficiency, articulatory intermediates have shown robust performance with limited training data. Anumanchipalli et al.~\cite{anumanchipalli2019speech} studied five participants; the cross-participant trajectory analysis used the three participants with comparable sentence material, while the 25-min data-efficiency analysis used participant 1. Decoding articulatory kinematics before transforming them to acoustics enabled reliable speech synthesis with limited training data and outperformed direct acoustic decoding. This efficiency is attributed to the low-dimensional manifold of articulatory kinematics, which constrains the high-dimensionality of acoustic signals, making the neural mapping more learnable. Additionally, articulatory coding supports speech-domain zero-shot voice conversion by disentangling speaker-specific voice texture from articulatory features. SPARC, for example, uses a speaker identity encoder to separate articulation from voice characteristics, enabling accent-preserving zero-shot conversion by switching speaker embeddings while maintaining articulatory trajectories.

Compared to acoustic or text-based targets, articulatory frameworks have demonstrated improved synthesized speech quality. Anumanchipalli et al.~\cite{anumanchipalli2019speech,silva2024speech} demonstrated that decoding articulatory features before transforming them into acoustics yields higher-quality synthesis than directly decoding acoustics from neural activity. Furthermore, complementary speech BCI architectures support large-vocabulary decoding at increasingly conversational rates. Metzger et al.~\cite{metzger2023high} achieved a median rate of 78.3 words per minute in one participant with anarthria during real-time text decoding of prompted, silently attempted sentences from the 1,024-word-General task using CTC-based sequence decoding. Clinically, Wairagkar et al.~\cite{wairagkar2025instantaneous,wood2025brain} demonstrated naturalistic speech restoration in a participant with ALS: an intracortical BCI generated speech personalized to the participant's pre-ALS voice with closed-loop feedback and controllable paralinguistic features, including intonation and emphasis.

However, trade-offs exist between biological interpretability and decoding accuracy. While articulatory representations offer physically interpretable targets for studying the neural correlates of speech motor control, some degradation in naturalness has been observed in zero-shot voice conversion, particularly for rare combinations of accent and voice. This suggests that while the low-dimensional, biologically grounded nature of articulatory features enhances representation interpretability and can support transfer under specified adaptation protocols, optimizing for maximum accuracy may require balancing these constraints against the complexity of acoustic outputs.

\subsection{Language Priors: Utility and Risks}
In BCI pipelines, linguistic priors (from n-gram to neural language models) have been associated with lower decoding error or improved output coherence in continuous decoding and large-vocabulary settings~\cite{herff2015brain,moses2021neuroprosthesis,willett2023high,card2024accurate}. In Willett et al.'s 125,000-word task~\cite{willett2023high}, the RNN produced phoneme sequences with raw PERs of 19.7\% for vocalized and 20.9\% for silent attempted speech; the subsequent language-model-assisted word decoder yielded WERs of 23.8\% and 24.7\%, respectively. A dual-path framework further reported improved intelligibility when acoustic and linguistic pathways were fused~\cite{li2025high}. Zhang et al. (2026) integrated the pretrained BIT encoder with a compact audio-LLM on the one-participant Brain-to-Text '24 benchmark (T12; 125,000-word attempted-speech task; 1,200 held-out Switchboard sentences): end-to-end WER was 15.67\% for the single model and 10.22\% with ensembling, compared with 24.69\% for the prior end-to-end entry~\cite{zhang2026cross,willett2023high}. Outside BCI, wav2vec/HuBERT-style fusion and transformer language models report consistent WER reductions, supporting the plausibility of this mechanism~\cite{baevski2020wav2vec,hsu2021hubert,devlin2019bert}. From a computational-neuroscience perspective, LLM-based components also provide a useful bridge between linguistic theory and neural data, and model-brain alignment analyses can help evaluate whether language-prior representations are biologically plausible rather than only performance-improving~\cite{zhang2026linguistics, guo2026generative}. In Fig.~\ref{fig:landscape}, these language-prior--assisted approaches are highlighted as a useful route for improving large-vocabulary decoding by complementing neural predictions with external sequence priors.

However, language-prior assistance also introduces risks. In brain-to-voice settings, generative linguistic pathways can bias outputs toward probable text patterns, potentially shifting away from intended content~\cite{li2025high}. Ethical analyses and design recommendations further indicate agency and ownership risks when model priors dominate user intent unless start/stop, confirmation, and veto controls are explicit~\cite{maslen2021control,sankaran2023recommendations,littlejohn2025streaming, guo2026generative}. At a representation level, dataset-derived bias and incomplete brain-alignment of current LLM features indicate unresolved attribution and reliability issues~\cite{radford2019language,caucheteux2022deep,goldstein2025temporal}, while semantic reconstruction studies further foreground privacy risks around high-level language content extraction from neural signals~\cite{tang2023semantic}. This risk profile is also reflected in Fig.~\ref{fig:landscape}, where the benefit of language-prior assistance is shown alongside a potential bias cost: better lexical coverage and fluency may come at the expense of faithful intent expression if priors are insufficiently constrained. Accordingly, current evidence suggests that language priors can improve fluency and error rates in constrained or assisted settings~\cite{li2025high,littlejohn2025streaming,wang2025progress}, while real-world risk rates (intent drift, bias harms, latency penalties) remain to be established through prospective multi-user evaluations~\cite{silva2024speech,stavisky2025restoring,maslen2021control,sankaran2023recommendations,littlejohn2025streaming}.

\subsection{Dual-Path and Multi-Modal Decoding Frameworks}
Dual-path decoding frameworks are designed to address the traditional trade-off between acoustic naturalness and linguistic intelligibility by concurrently reconstructing acoustic and linguistic features from neural signals. Herff et al.~\cite{herff2015brain} enrolled seven epilepsy-monitoring participants and analyzed sessions from six in their early Brain-to-Text study of overt continuous reading. The approximately 25\% minimum WER came from one best-performing session under leave-one-phrase-out evaluation with a 10-word test dictionary; the phone-classification task used 20 phone classes plus silence and achieved an error below 50\%. The system integrates acoustic and linguistic representations to decode continuously spoken speech from intracranial ECoG recordings. The acoustic pathway models high-gamma activity from ECoG as context-independent Gaussian phone models, capturing neural activity across time intervals with feature stacking to integrate temporal context (±200 ms), while the linguistic pathway employs statistical language models (n-gram) and a pronunciation dictionary. By combining these via Viterbi decoding to find the most likely word sequence, the system balances neural signal decoding with lexically and syntactically informed prediction.
In the conceptual landscape (Fig.~\ref{fig:landscape}), dual-path frameworks are therefore characterized by their paired acoustic and linguistic representations, because they explicitly combine neural-conditioned acoustic reconstruction with linguistic structure rather than relying on either pathway alone.

Li et al.~\cite{li2025high} evaluated a more recent explicit dual-path framework in nine epilepsy-monitoring participants during offline passive-listening reconstruction of TIMIT sentences, using approximately 20 minutes of neural recordings per participant. The framework decodes neural activity through parallel neural-to-acoustic and neural-to-text-to-speech pathways and fuses their outputs. The acoustic pathway utilizes a bidirectional LSTM decoder and a pre-trained high-fidelity generative adversarial network (HiFi-GAN) to reconstruct detailed spectrotemporal speech features, capturing speaker-specific timbre and prosody. The neural-to-text-to-speech pathway uses a Transformer-based adaptor to decode neural activity into discrete word tokens, which are synthesized into speech via a text-to-speech (TTS) generator (Parler-TTS) to capture syntactic and semantic information. Fusion is achieved through voice cloning (CosyVoice 2.0), which uses the acoustic output as a voice reference and the decoded word tokens as text input. The fused system achieved MOS $3.956\pm0.173$, mel-spectrogram coefficient of determination ($R^2=0.824\pm0.029$), Whisper-base transcription WER $18.9\pm3.3\%$, and phoneme-recognizer PER $12.0\pm2.5\%$; the study did not report a fixed or unique-word lexicon size. Ablation studies confirm that integrating both pathways delivers substantial improvements over the neural-to-acoustic or neural-to-text-to-speech pathway alone, with significant reductions in WERs and PERs while maintaining high acoustic fidelity.

Architectural advances, such as the Transformer model introduced by Vaswani et al.~\cite{vaswani2017attention}, provide a foundation for dual-path decoding through multi-head self-attention mechanisms. Multi-head attention allows the model to jointly attend to information from different representation subspaces, providing a flexible computational component for decoders with explicitly defined acoustic and linguistic targets. Encoder-decoder attention layers further enable integration of multiple data modalities, while positional encoding injects sequence order information crucial for preserving prosody and lexical content in the output. Although originally designed for machine translation, these principles are directly applicable to speech decoding frameworks aiming to reconstruct acoustic and linguistic features concurrently.

Multi-modal integration enhances decoding robustness across speech types by combining complementary information from different pathways. The older Herff cohort (six analyzed participants) and the newer Li cohort (nine participants) provide independent design-level convergence that linguistic structure can complement acoustic reconstruction; Herff used overt phrase-to-text decoding, whereas Li used passive-listening speech reconstruction. In Herff et al.'s system~\cite{herff2015brain}, neural decoding and conventional language modeling jointly supported word and phone decoding across participants and sessions with varying electrode placements and task conditions. In Li et al.~\cite{li2025high}, the dual-path framework combined acoustic naturalness with lexical content and outperformed either pathway alone. Extending this complementary-representation strategy, Pescatore et al.~\cite{pescatore2025decoding} decoded word-level lexical-semantic categories with an average accuracy of 21\% across participants (chance level 10\%). In a communication BCI, such signals may constrain error correction or provide a fallback when direct phoneme- or word-level decoding is uncertain.

\subsection{Interpretable Decoding Models for Neural Representation Analysis}
Interpretable decoding models are critical for testing hypotheses about neural representations of speech by characterizing predictive links between neural activity and articulatory or linguistic components, thereby advancing both scientific understanding and clinical BCI optimization. Singh et al. (2025)~\cite{singh2025transfer} tested their sequence-to-sequence model in 25 epilepsy participants performing the same 128-stimulus tongue-twister protocol, providing a multi-participant within-study evaluation of direct phoneme-sequence PER. The model integrates temporal convolutional networks, recurrent neural networks, and a linear readout layer to isolate phoneme identity probabilities, enabling interpretable mapping from neural data to phonemic sequences. It outperformed linear baselines in phoneme prediction and, through REO analysis, characterized model reliance on sampled inputs from ventral sensorimotor cortex, posterior superior temporal gyrus, and superior temporal sulcus during phoneme prediction, supporting predictive region-to-phoneme mappings aligned with known neuroanatomical contributions to speech. Transfer learning with a shared recurrent layer further captured transferable speech-related structure, producing latent feature embeddings that facilitated decoding with variable electrode coverage and supported shared representations across the sampled participants.

Linear models have served as interpretable baselines for analyzing predictive neural patterns, complementing more complex architectures. Forward-pattern transformations account for feature covariance and noise when relating decoder weights to physiological signal patterns~\cite{haufe2014interpretation}. Luo et al. (2022)~\cite{luo2022brain} noted that phoneme decoding studies using linear classifiers achieved accuracies ranging from ~20\% to over 70\% with intracranial recordings, directly mapping neural features to articulatory or linguistic components. Beyond linear approaches, deep learning models such as convolutional neural networks (CNNs) mapped high-gamma activity to speech features like mel-spectrograms, while recurrent neural networks and encoder-decoder frameworks improved word and sentence decoding accuracy. Speech synthesis approaches, which first map neural activity to intermediary representations (e.g., spectrotemporal features or vocal tract trajectories) before acoustic reconstruction via vocoders like WaveNet, enhance interpretability by exposing testable neural-to-representation mappings and supporting clinical BCI optimization.

For imagined speech, visualization-based analyses characterize the spatiotemporal features that support decoding; model/data randomization and stability analyses assess the specificity and reproducibility of the resulting attribution maps~\cite{adebayo2018sanity}. Jiang et al.~\cite{jiang2026decoding} used FAST-generated attribution maps to identify discriminative spatiotemporal patterns in scalp EEG over frontal and temporal regions during imagined word production. Left frontal EEG channels showed pronounced model attribution, providing scalp-level candidate features for studies of speech planning and phonological processing. Per-utterance saliency maps further demonstrated reduced attribution magnitude with repeated imagined-speech rehearsal. Through spatial-temporal tokenization and transformer encoding, the model linked neural signals to imagined-speech labels, informing hypotheses about imagined-speech neural dynamics and guiding targeted decoder optimization.

These interpretable decoding models—ranging from Seq2Seq architectures and linear baselines to visualization-driven frameworks—support neural representation analysis by mapping neural features to articulatory and linguistic components, advancing both scientific knowledge and clinical translation of speech BCIs. Combining these predictive maps with independent perturbation or convergent physiological evidence provides a route to mechanistic hypothesis testing. To synthesize these cross-cutting developments, Fig.~\ref{fig:timeline} places representative frameworks by publication year within three lanes and annotates each with method tags (e.g., sequence/transformer decoders, CTC/LM assistance, streaming synthesis, or transfer-learning strategies), highlighting how design priorities have evolved across the field.

\section{Unified Evaluation Framework for Speech Neuroprostheses}
At present, speech neuroprosthesis research lacks a widely accepted unified evaluation framework that can support consistent comparison across heterogeneous output types, experimental tasks, languages, and deployment settings. Existing studies generally select task-specific measures---such as WER or PER for discrete text decoding, acoustic and perceptual measures for speech synthesis, and latency or communication rate for interactive use---and often apply different parameter ranges, aggregation rules, and testing conditions. Consequently, results reported for different systems are difficult to compare directly, and improvements in laboratory decoding accuracy cannot readily be interpreted in terms of perceptual quality, conversational utility, or long-term clinical performance.

To address this gap, we synthesize the evaluation practices and metrics identified across the reviewed literature and propose a unified, multidimensional evaluation framework that integrates objective accuracy, perceptual quality, naturalistic expressivity, conversational utility, and longitudinal stability. The framework further provides a cross-language, cross-task structure that organizes task-specific metrics within a common scoring logic and makes metric selection, normalization, and weighting assumptions explicit. In this way, the proposed framework connects technical decoding performance with perceptual and deployment-relevant outcomes and provides a common basis for designing evaluation protocols and comparing systems across research settings.

\subsection{Experiment Design Protocol for Generalizable Speech BCI Development}

\begin{figure}
\centering
\includegraphics[width=\textwidth]{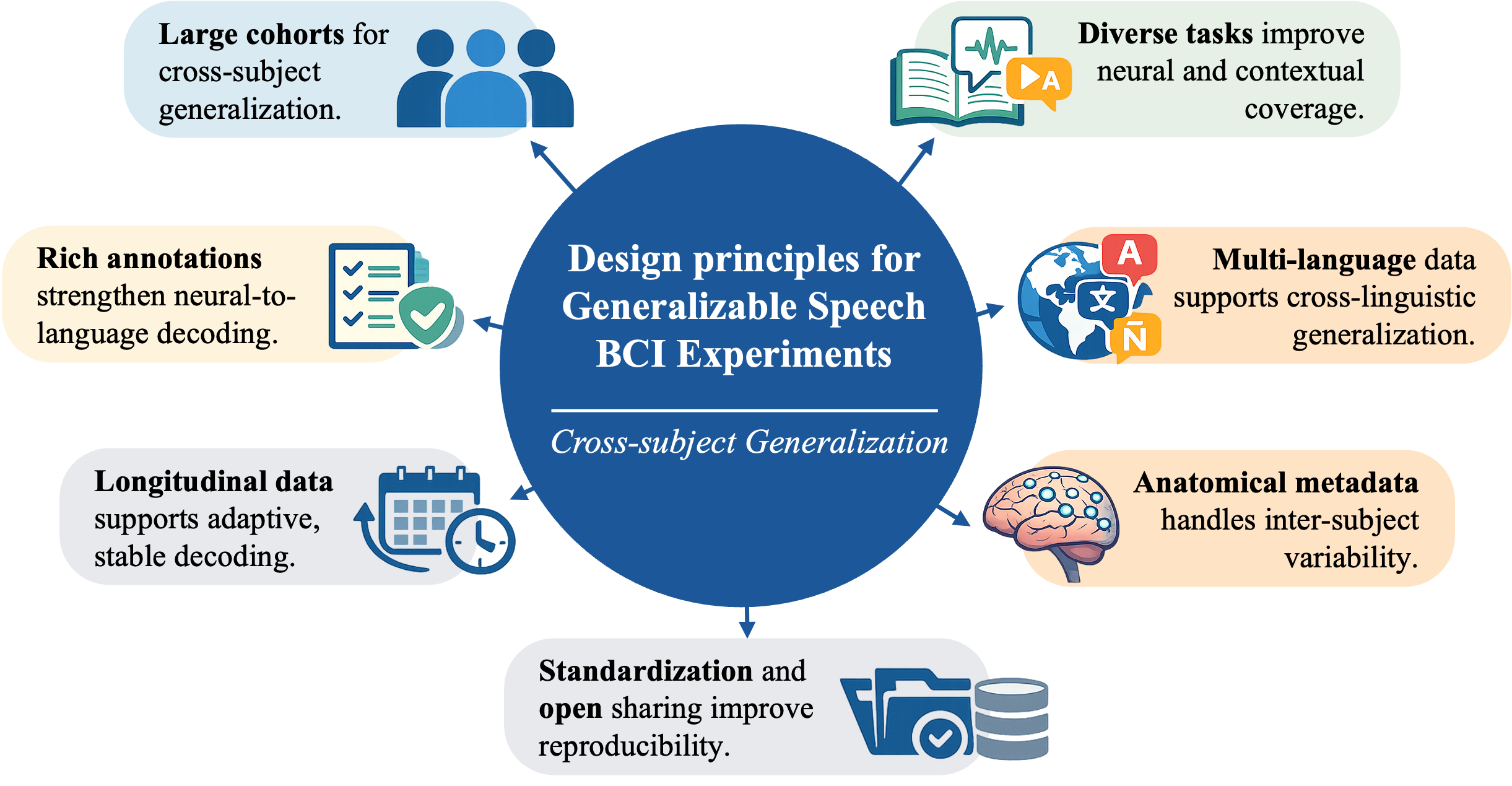}
\caption{Design principles for generalizable speech BCI experiments. Large cohorts, diverse tasks, multi-language coverage, rich annotations, anatomical metadata, standardization/open sharing, and longitudinal collection jointly support cross-subject generalization and more robust performance of speech-related neural decoding.}
\label{fig:dataDesign}
\end{figure}

As illustrated in Fig.~\ref{fig:dataDesign}, when developing generalizable speech BCIs, dataset design must prioritize critical characteristics that address the bottleneck of poor cross-subject generalization. A large cohort size is foundational. Chen et al.'s SwinTW study~\cite{chen2025transformer} included 52 participants (43 ECoG and nine SEEG): its shared model was trained and evaluated on held-out trials from the same 15 ECoG participants (PCC 0.837 versus 0.831 for participant-specific models), while participant-unseen performance was tested separately using a shared neural decoder and a synthesizer pretrained on each test participant's own speech, with participant-held-out cross-validation across the 43-participant ECoG cohort (mean PCC between the reconstructed and ground-truth mel-spectrograms of 0.765, paired with a mean Short-Time Objective Intelligibility (STOI+) score of approximately 0.18). The 10-participant Brain Treebank likewise supports multi-participant model development~\cite{wang2024brain}. These designs show how large cohorts can support both shared-model fitting and participant-held-out validation. Even smaller cohorts, such as the 4 participants in a Mandarin Chinese BCI study~\cite{feng2025acoustic}, contribute by focusing on language-specific linguistic structures, though larger cohorts remain critical for broader generalization. Beyond cohort size, clinical representativeness requires etiological diversity across ALS, brainstem stroke, locked-in syndrome, and spinal cord injury, with residual vocal and articulatory function annotated separately.

Diverse speech tasks are essential to capture varied neural responses across different speech contexts. Datasets have included tasks like Auditory Repetition, Auditory Naming, Sentence Completion, Visual Reading, and Picture Naming, with 400 trials per participant~\cite{chen2025transformer}, as well as reading monosyllabic characters (covering all Pinyin syllables) and varied-length sentences. Naturalistic paradigms, such as watching Hollywood movies to capture over 38,000 sentences~\cite{wang2024brain}, further enhance diversity by reflecting real-world language perception, supporting representation learning beyond controlled production tasks.

Current intracranial speech and language datasets provide limited linguistic coverage. The publicly available resources summarized in Table~\ref{tab:public-intracranial-language-datasets} comprise English, Mandarin Chinese, and Dutch data, including a single Mandarin--English bilingual dataset. Although an English--Spanish speech neuroprosthesis demonstrated cross-language transfer through shared cortical articulatory representations, the evidence was derived from a single participant~\cite{silva2024bilingual}. Direct cortical recordings further indicate that shared articulatory representations coexist with language-specific encoding of phonological sequences, lexical information, and lexical tone~\cite{silva2024bilingual,bhaya2026shared,li2021human}. Systematic evaluation of cross-language generalizability therefore calls for linguistically diverse cohorts and held-out-language validation.

Rich linguistic annotations provide structured supervised signals to handle data variability. For example, 18 speech parameters (e.g., pitch, formants, loudness) extracted from spectrograms guide decoder training, while Mandarin BCI systems model initials, tones, and finals to decode syllables. The Brain Treebank enhances annotations with part-of-speech tags, dependency parses, word onsets/offsets, and scene labels~\cite{wang2024brain}, enabling fine-grained neural analysis of language processing.

To account for individual differences, datasets incorporate anatomical electrode metadata, such as MNI coordinates and brain region indices, allowing suitable architectures to accommodate heterogeneous electrode layouts without subject-specific layers. Studies also report performance metrics per participant and annotate electrode locations using common brain atlases, enabling assessment of robustness to variability in electrode placement and neural signal characteristics. Collectively, these dataset features—large cohorts, diverse tasks, multi-language coverage, rich annotations, anatomical metadata, etiological diversity, and residual-function labels—directly address generalization bottlenecks, paving the way for clinical translation of speech BCIs.

In addition to these design principles, standardized data formats and open sharing practices are crucial for fostering collaboration and accelerating progress in the field. Adopting common data structures (e.g., Brain Imaging Data Structure (BIDS) for neuroimaging) and providing comprehensive documentation facilitate reproducibility and enable researchers to build upon existing datasets effectively. Moreover, incorporating longitudinal data collection can help capture neural signal variability over time, informing the development of adaptive decoding algorithms that maintain performance in real-world settings. Table~\ref{tab:public-intracranial-language-datasets} inventories representative public intracranial speech and language datasets. Production-oriented resources range from controlled isolated-word datasets, such as SingleWordProductionDutch-iBIDS~\cite{verwoert2022dataset} and Du-IN~\cite{zheng2024discrete}, to multimode word-and-sentence recordings in VocalMind~\cite{he2025vocalmind}. The Brain Treebank~\cite{wang2024brain} and Podcast ECoG~\cite{zada2025podcast} instead provide naturalistic perception data with extensive linguistic annotations. These perception datasets support acoustic and linguistic representation learning from stimulus-linked neural responses. The inventory also exposes persistent reproducibility constraints. Production datasets remain limited in cohort size or vocabulary, language coverage is concentrated in English and Mandarin Chinese, and heterogeneous tasks, modalities, annotations, and data structures impede direct comparison. Public availability therefore needs to be coupled with harmonized metadata, shared task definitions, standardized data splits, and common evaluation protocols to support reproducible reuse and cross-dataset validation.

\begin{table}[t]
\centering
\scriptsize
\caption{Representative publicly available intracranial speech and language datasets relevant to speech BCI development. Datasets are grouped into speech-production paradigms that can support output decoding and language-perception paradigms suited to representation learning. Availability was verified from the cited dataset papers and repositories.}
\label{tab:public-intracranial-language-datasets}
\setlength{\tabcolsep}{3.2pt}
\renewcommand{\arraystretch}{1.08}

\begin{tabularx}{\textwidth}{>{\raggedright\arraybackslash}p{0.105\textwidth} >{\raggedright\arraybackslash}p{0.10\textwidth} >{\raggedright\arraybackslash}p{0.23\textwidth} >{\raggedright\arraybackslash}p{0.10\textwidth} >{\raggedright\arraybackslash}X >{\raggedright\arraybackslash}p{0.08\textwidth} >{\raggedright\arraybackslash}p{0.08\textwidth}}
\toprule
\textbf{Authors (year)} & \textbf{Dataset} & \textbf{Size} & \textbf{Language} & \textbf{Task} & \textbf{Recording modality} & \textbf{Public availability} \\
\midrule
\multicolumn{7}{l}{\textbf{\textit{Speech-production and output decoding}}} \\
\midrule
Verwoert et al. (2022) & SWPD-iBIDS~\cite{verwoert2022dataset} & 10 participants; 1,000 word trials; 1,103 contacts & Dutch & Overt reading of 100 isolated words & SEEG & \href{https://doi.org/10.17605/OSF.IO/NRGX6}{OSF (NRGX6)} \\
\midrule
Bouchard et al. (2013) & CV-syllable ECoG~\cite{bouchard2013functional} & 4 participants; 256-channel recordings & English & Overt production of consonant--vowel syllables & High-density ECoG & \href{https://doi.org/10.48324/dandi.000019/0.220126.2148}{DANDI (000019)} \\
\midrule
He et al. (2025) & VocalMind~\cite{he2025vocalmind} & 1 participant; 67.85 min; 20 words and 100 sentences & Mandarin Chinese & Vocalized, mimed, and imagined production of words and sentences & SEEG & \href{https://zenodo.org/records/14696348}{Zenodo (14696348)} \\
\midrule
Zheng et al. (2024) & Du-IN~\cite{zheng2024discrete} & 12 participants; $\sim$3 h task plus $\sim$12 h non-task data per participant & Mandarin Chinese & Overt reading of 61 words (50 repetitions each) & SEEG & \href{https://huggingface.co/datasets/liulab-repository/Du-IN}{Hugging Face (Du-IN)} \\
\midrule
NeuroVox (2025) & NeuroVox~\cite{neurovox_2025_13332808} & 2 public participants; 3,069 retained utterances; $\sim$149 min & Mandarin Chinese and English & Overt reading of Mandarin and English words and Arabic digits & SEEG & \href{https://zenodo.org/records/13332808}{Zenodo (13332808)} \\
\midrule
\multicolumn{7}{l}{\textbf{\textit{Language-perception and representation learning}}} \\
\midrule
Wang et al. (2024) & Brain Treebank~\cite{wang2024brain} & 10 participants; 43.5 h; 1,688 electrodes; 223,068 word instances & English & Naturalistic viewing of Hollywood movies with spoken dialogue & SEEG & \href{https://ww2.nemar.org/dataset/nm000253}{NEMAR (nm000253)} \\
\midrule
Zada et al. (2025) & Podcast ECoG~\cite{zada2025podcast} & 9 participants; 1,330 electrodes; 30 min and $>$5,000 words per participant & English & Listening to a 30-min English podcast narrative & ECoG & \href{https://doi.org/10.18112/openneuro.ds005574.v1.0.2}{OpenNeuro (ds005574)} \\
\midrule
Zhang et al. (2026) & ECoG Tonal Speech Perception~\cite{zhang2026fine} & 4 participants; 20 sessions; $\sim$76.3 min total; 128--256 channels & Mandarin Chinese & Listening to continuous sentence-level Mandarin speech from ASCCD & High-density ECoG & \href{https://doi.org/10.57760/sciencedb.27618}{Science Data Bank (27618)} \\
\bottomrule
\end{tabularx}
\end{table}

\subsection{Task-Protocol Heterogeneity and Its Implications for BCI Co-Design}

Reported decoding performance cannot be interpreted independently of the task protocol used to elicit neural activity. Intracranial speech BCI protocols differ in speech mode (perception, overt, attempted, mimed, or imagined speech), stimulus--response structure (reading, repetition, naming, or spontaneous generation), linguistic unit, vocabulary constraint, external cueing, temporal alignment, feedback, and online versus offline operation. These choices determine the available ground truth, signal-to-noise ratio (SNR), timing precision, training-data requirements, and ecological validity. Consequently, similar metrics obtained under different protocols may represent substantially different levels of technical and clinical difficulty.

Overt reading, repetition, and naming provide synchronized audio and comparatively precise phonetic or acoustic labels, supporting supervised decoding of spectrograms, articulatory trajectories, phonemes, and text with regression, sequence-to-sequence, or differentiable speech-synthesis models~\cite{anumanchipalli2019speech,makin2020machine,chen2024neural}. ECoG coverage over ventral sensorimotor cortex (vSMC) and superior temporal gyrus (STG) captures articulation- and auditory-feedback-related high-gamma activity, whereas SEEG can sample distributed and deeper components of the speech network. These protocols are valuable for mechanistic investigation, model pretraining, and electrode-selection studies, but performance may depend on acoustic feedback and actual articulator movement. Moreover, temporary epilepsy-monitoring cohorts provide evidence under clinically determined electrode coverage; translation to chronic therapeutic users with anarthria or severe paralysis requires evaluation with coverage and task protocols suited to those users.

Attempted or mimed speech is more closely aligned with communication restoration in people who cannot produce intelligible speech but retain speech-motor intention. Although both conditions may lack audible output, mimed speech usually retains visible articulation, whereas attempted speech in paralysis may involve weak or absent peripheral movement; residual kinematics should therefore be reported. Accurate vocalization provides aligned acoustic targets and auditory feedback, inaccurate or dysarthric vocalization provides a distorted acoustic trace, and miming retains articulation without voiced output. These regimes alter signal availability, label quality, and task difficulty, so decoder performance should be compared within matched residual-function strata. These protocols generally motivate coverage of vSMC, ventral premotor cortex, and related speech-motor regions: high-density ECoG samples spatially distributed high-gamma patterns, whereas penetrating microelectrode arrays (MEAs) provide high-resolution spiking or spike-band activity from smaller cortical territories~\cite{willett2023high,metzger2023high,card2024accurate}. Because a reliable acoustic target may be absent, decoders often use phonemes, articulatory features, text, or intended words as intermediate targets. Sequence architectures such as recurrent networks, transducers, and causal Transformers model temporal dependencies, while objectives such as CTC help accommodate uncertain alignment. Such protocols are directly relevant to people with amyotrophic lateral sclerosis (ALS), brainstem stroke, locked-in syndrome (LIS), or severe dysarthria, although their feasibility may still depend on residual articulatory control and reliable training prompts.

Imagined speech removes the requirement for overt movement and may be attractive for individuals with complete anarthria, severe fatigue, or minimal residual motor output. However, imagined speech can produce weaker, less spatially focal, and less precisely time-locked activity than overt or attempted speech~\cite{proix2022imagined,de2024imagined,kunz2025inner}. This profile motivates evaluating broader frontotemporal ECoG or SEEG coverage alongside localized motor-cortex sampling, together with low-frequency activity, cross-frequency coupling, and distributed network features. Given the lower SNR and limited trial counts, current imagined-speech studies more often use constrained classification, regularized models, transfer learning, or representation learning than direct large-vocabulary synthesis. Thus, imagined speech has substantial long-term clinical relevance but less mature evidence for rapid, open-vocabulary communication.

Task structure introduces an additional co-design constraint. Cue-based, closed-set command or spelling protocols provide repeated, well-aligned trials and can support reliability-first communication with relatively simple classifiers and limited but stable coverage~\cite{luo2023stable}. In contrast, open-vocabulary sentence production and interactive conversation require self-paced speech-onset detection, causal sequence modeling, streaming output, language-prior integration, and explicit latency and agency controls~\cite{littlejohn2025streaming,wairagkar2025instantaneous}. Speech-perception and naturalistic-listening paradigms can provide large datasets for acoustic, phonological, or semantic representation learning, particularly with STG and higher-order language-area coverage~\cite{akbari2019towards,wang2024brain}. Table~\ref{tab:task-protocol-codesign} summarizes these protocol-dependent design choices.

\begin{table}[t]
\centering
\scriptsize
\caption{Task-protocol differences and their implications for intracranial recording design, neural modeling, target populations, and translational aims. Protocol categories are not mutually exclusive, but they represent different levels of signal availability, alignment, ecological validity, and clinical difficulty.}
\label{tab:task-protocol-codesign}
\begin{tabularx}{\textwidth}{>{\raggedright\arraybackslash}p{0.14\textwidth} >{\raggedright\arraybackslash}X >{\raggedright\arraybackslash}X >{\raggedright\arraybackslash}X >{\raggedright\arraybackslash}X}
\toprule
\textbf{Task protocol} & \textbf{Typical iEEG type and coverage} & \textbf{Neural features and models} & \textbf{Target population or aim} & \textbf{Key limitation} \\
\midrule
Overt reading, repetition, or naming & ECoG over vSMC/STG; SEEG across distributed language networks & High-gamma and acoustic features; regression, Seq2Seq, or speech reconstruction & Mechanistic studies, model pretraining, and electrode selection in research-access cohorts & Acoustic feedback and actual movement limit direct transfer to paralysis \\
\midrule
Attempted or mimed speech & vSMC, ventral premotor cortex, and area 6v; high-density ECoG or MEA & High-gamma or spikes; phoneme/articulatory targets; CTC, RNN, transducer, or Transformer & ALS, severe dysarthria, brainstem stroke, or LIS & Residual motor ability varies and acoustic labels may be unavailable \\
\midrule
Imagined speech & Broader frontotemporal ECoG or SEEG coverage & Low-frequency, cross-frequency, and distributed features; classification, transfer, or representation learning & Complete anarthria, severe fatigue, or minimal residual motor output & Low SNR, uncertain timing, and predominantly constrained vocabularies \\
\midrule
Closed-set commands or spelling & Limited but stable motor or language-area coverage & CNN or conventional classifiers & Reliability-first home communication and control & Restricted communication rate and expressive range \\
\midrule
Open-vocabulary continuous speech & High-density or strategically distributed multi-area coverage & CTC, RNN-T, causal Transformer, and language-prior-assisted decoding & Natural conversation and fidelity-first speech restoration & High data and calibration demands; latency and intent-attribution risks \\
\midrule
Listening or naturalistic perception & ECoG/SEEG over STG and higher-order language areas & Acoustic, semantic, or contextual representation models & Mechanistic analysis and representation pretraining & Stimulus-linked targets; production-task validation needed for communication use \\
\bottomrule
\end{tabularx}
\end{table}

Taken together, task protocol, recording modality and coverage, neural representation, decoder architecture, and clinical aim should be treated as coupled rather than independent choices. Studies should therefore report speech mode, cueing and timing procedures, vocabulary size, repetition structure, online feedback, residual speech or movement, electrode type and anatomical coverage, neural features, decoder target, and intended-use population. Cross-study comparisons should be stratified by protocol class to preserve the interpretation of results within their speech-mode, cueing, and vocabulary conditions.

\subsection{Limitations of Conventional Metrics and Role of Perceptual Evaluation}
In the co-design view of Fig.~\ref{fig:modalities}, evaluation strategy is part of the software stack rather than a post hoc reporting detail: it determines whether improvements in decoder architecture or adaptation are detectable, comparable, and clinically meaningful.

Conventional objective metrics in speech neuroprosthesis evaluation, such as Pearson correlation coefficient (PCC) and mel cepstral distortion (MCD), suffer from significant limitations and inconsistent usage, hindering meaningful cross-task comparisons~\cite{wu2025improved}. PCC, widely relied upon in many studies, is sensitive to analysis parameters such as window size and focuses on co-variation rather than absolute signal differences, thereby failing to capture perceptual intelligibility or naturalness. Similarly, MCD, which operates on mel-frequency cepstral coefficients (MFCCs) and gives greater weight to perceptually relevant spectral features than correlation-based metrics, does not inherently account for temporal misalignment and may penalize perceptually acceptable reconstructions with systematic spectral offsets. It therefore does not fully reflect speech intelligibility or perceived quality. The inconsistent application of these metrics, together with variations in parameter settings and experimental conditions, further limits the comparability of results across studies and datasets.

To address these limitations, perceptual evaluation should be treated as user-facing evidence rather than as an objective ground truth. Listener-derived word error rate (WER) and character error rate (CER) quantify the word- or character-sequence accuracy of transcripts produced by human listeners, whereas phoneme error rate (PER) measures phonemic accuracy when phoneme-level reference annotations or transcriptions are available. For a reference sequence of $N>0$ words, characters, or phonemes, the corresponding error rate is $(S+D+I)/N$, where $S$, $D$, and $I$ count substitutions, deletions, and insertions, respectively. Mean opinion scores (MOS) capture perceived naturalness and, depending on the rating protocol, may also assess intelligibility, voice quality, or speaker resemblance. These measures remain dependent on the listener population, speech materials, presentation conditions, and evaluation protocol. Crowd-sourced transcription is nevertheless valuable because it directly tests intelligibility beyond acoustic-feature similarity~\cite{metzger2023high}. In aphasic speech assessment, non-expert ratings of clarity, fluidity, and prosody have approached expert-level performance, suggesting that appropriately designed listener studies can provide scalable and clinically relevant evidence~\cite{le2016automatic}. Human transcription results and MOS should therefore be reported together with listener demographics, sample size, rating instructions, playback conditions, and aggregation procedures~\cite{wu2025improved}.

Perceptual evaluation can improve comparability by linking technical results to human experience. Human-listener judgments, including closed-set transcript matching, open transcription, and Likert-scale ratings, should be used to contextualize objective measures such as MCD rather than being treated as interchangeable with them~\cite{wairagkar2025instantaneous,silva2024speech}. For large-scale, open-vocabulary, or multilingual evaluations, automatic speech recognition (ASR)-based transcription may provide a scalable proxy or complement to human scoring, but it should not be interpreted as a direct measure of human perception. Studies using ASR-based evaluation should identify the ASR system and model version, specify decoding and language-model settings, assess language-, accent-, and pathology-specific biases, and measure agreement with human transcriptions on a representative subset. An integrated evaluation strategy combining objective acoustic measures, human perceptual judgments, and carefully validated ASR-based proxies can therefore support more reproducible, user-centered, and clinically meaningful comparisons across speech neuroprosthesis systems.

\subsection{Proposed Benchmark for Cross-Task and Cross-Language Comparison}
\label{sec:clt-benchmark}
The variability and lack of standardization in evaluation metrics across speech neuroprosthesis studies, particularly for continuous speech decoding, have hindered meaningful comparisons across tasks, languages, and system architectures~\cite{stavisky2025restoring}. While discrete speech decoding benefits from well-established metrics such as WER and PER that measure word- and phoneme-sequence accuracy, continuous voice synthesis lacks standardized metrics, with objective measures (e.g., mel-cepstral distortion, frequency band power correlation) often poorly corresponding to speech intelligibility, and subjective perceptual scores varying across studies. Moreover, differences in participant population, speech mode, task cueing, vocabulary constraints, and online versus offline operation can substantially alter task difficulty and the interpretation of reported performance. This heterogeneity raises a central methodological question: \emph{How can speech-neuroprosthesis systems be compared fairly across evaluation regimes without collapsing fundamentally different tasks and clinical use cases into a single, potentially misleading performance ranking?} To address this question, benchmarking should distinguish direct numerical comparability from broader evidence synthesis. Because protocol difficulty differs substantially, benchmark results should first be reported within predefined strata defined by speech mode, cueing condition, vocabulary constraint, online/offline operation, and participant population. Direct WER/PER benchmarking is most informative when participant population, vocabulary size, sentence-inventory openness, dataset, speech mode, task cueing, online/offline operation, and scoring procedure are aligned. When these conditions differ, studies should instead be compared in terms of shared mechanisms, robustness trends, user-facing capabilities, and system feasibility, rather than ranked solely by raw error rates. This distinction motivates a stratified benchmark that preserves task-specific interpretation while providing a common structure for cross-study synthesis.

Accordingly, we propose a two-tier, representation-agnostic framework. The mandatory tier reports user-facing outcomes separately: task-appropriate accuracy or intelligibility, end-to-end latency, perceived naturalness, and resemblance to the user's intended or pre-morbid voice when synthesized voice is evaluated. The optional tier contains task- and architecture-specific diagnostic metrics, including PER, STOI, MCD, mel-spectrogram coefficient of determination ($R^2$), prosodic correlations, and timing synchrony. Individual outcomes and diagnostic metrics are reported in their original units, with a candidate direction-normalized weighted composite providing a secondary within-stratum summary for prospective validation. Let $x_{\ell,t,i,m}$ denote the raw value of metric $m$ for language $\ell$, task $t$, and sample/session/participant index $i$; let $[a_m,b_m]$ be the pre-registered normalization bounds for metric $m$, and let $d_m\in\{+1,-1\}$ indicate whether larger values are better ($+1$) or worse ($-1$). For the composite, each raw value is mapped to a bounded score $z_{\ell,t,i,m}\in[0,1]$ by

\begin{equation}
z_{\ell,t,i,m}=
\begin{cases}
\min\left\{1,\max\left\{0,\dfrac{x_{\ell,t,i,m}-a_m}{b_m-a_m}\right\}\right\}, & d_m=+1, \\[6pt]
\min\left\{1,\max\left\{0,\dfrac{b_m-x_{\ell,t,i,m}}{b_m-a_m}\right\}\right\}, & d_m=-1.
\end{cases}
\end{equation}

For metrics that must be mapped to perceptual quality (e.g., objective acoustic features to MOS), a separately calibrated model $f_{\theta}$ would be required to generate a perceptual proxy $\hat{p}_{\ell,t,i}=f_{\boldsymbol{\theta}}(\mathbf{x}_{\ell,t,i,\mathcal{M}_{\mathrm{obj}}})$, which is then normalized using the same rule to obtain $z^{(p)}_{\ell,t,i}$. The task-language benchmark score is defined as

\begin{equation}
\bar{z}_{\ell,t,m}=\frac{1}{N_{\ell,t}}\sum_{i=1}^{N_{\ell,t}} z_{\ell,t,i,m}, \qquad
\bar{z}^{(p)}_{\ell,t}=\frac{1}{N_{\ell,t}}\sum_{i=1}^{N_{\ell,t}} z^{(p)}_{\ell,t,i}, \qquad
B_{\ell,t}=\sum_{m\in\mathcal{M}_{t}} w_m \bar{z}_{\ell,t,m} + w_p \bar{z}^{(p)}_{\ell,t},
\end{equation}
where $N_{\ell,t}$ is the number of evaluated samples/sessions, $\bar{z}^{(p)}_{\ell,t}$ is the averaged normalized perceptual proxy (or observed MOS when available), and $\sum_{m\in\mathcal{M}_{t}} w_m + w_p = 1$. The descriptive cross-language, cross-task (CLT) benchmark for a fixed, predefined task--language mixture is then

\begin{equation}
B_{\mathrm{CLT}}=\sum_{\ell\in\mathcal{L}} \alpha_{\ell} \sum_{t\in\mathcal{T}} \beta_{t} B_{\ell,t},
\end{equation}
where $\sum_{\ell\in\mathcal{L}}\alpha_{\ell}=1$ and $\sum_{t\in\mathcal{T}}\beta_{t}=1$. This summary accompanies the individual outcomes within each stratum and supports descriptive comparisons for the same predefined task--language mixture. In practice, standardized data collection and task design support the core outcome layer, whereas pre-registered ranges, weights, and perceptual mapping models define the secondary composite.

Table~\ref{tab:clt-benchmark-defaults} gives an illustrative parameterization of the secondary composite across representative task classes. The metric families cover content fidelity, usability, perceptual quality, and expressivity, while the displayed ranges provide candidate clipping bounds. Short-Time Objective Intelligibility (STOI) is included in the suggested synthesis-task set. The numerical weights were derived using a structured swing-weighting~\cite{parnell20092} rationale: for each task $t$, a ``swing'' for metric $m$ is defined as improving it from the worst admissible level to the best admissible level within the pre-registered range $[a_m,b_m]$ (accounting for direction $d_m$), while holding all other metrics at their worst levels. We then assign each metric an a priori swing-importance score $s_m$ that reflects its relative contribution to the task goal (content fidelity vs.\ usability vs.\ perceptual quality), and obtain the reported weights by normalization. These values provide transparent starting points for sensitivity testing.

\begin{table}
\caption{Illustrative parameterization template for secondary composite scores in the proposed cross-language, cross-task benchmark.}
\label{tab:clt-benchmark-defaults}
\scriptsize
\begin{tabularx}{\linewidth}{@{\extracolsep{\fill}} l l l c c c c}
\toprule
\textbf{Task $t$} & \textbf{Metric set $\mathcal{M}_t$} & \textbf{Metric $m$} & \textbf{\shortstack{Normalization\\bounds $[a_m,b_m]$}} & \textbf{$d_m$} & \textbf{Weight} \\
\midrule
$t_{\mathrm{disc}}$ (copy/spelling) & \shortstack[l]{$\{\mathrm{WER},\mathrm{PER},L_{\mathrm{on}},R_{\mathrm{comm}}\}$} & $\mathrm{WER}$ & $[0,1]$ & $-1$ & $0.35$ \\
 &  & $\mathrm{PER}$ & $[0,1]$ & $-1$ & $0.20$ \\
 &  & $L_{\mathrm{on}}$ & $[0,3]$ s & $-1$ & $0.15$ \\
 &  & $R_{\mathrm{comm}}$ & $[0,30]$ wpm & $+1$ & $0.10$ \\
 &  & $p$ & $[1,5]$ & $+1$ & $w_p=0.20$ \\
\midrule
$t_{\mathrm{synth}}$ (cont. synth.) & \shortstack[l]{$\{\mathrm{STOI},\mathrm{MCD},R^2_{\mathrm{mel}},L_{\mathrm{on}}\}$} & $\mathrm{STOI}$ & $[0,1]$ & $+1$ & $0.25$ \\
 &  & $\mathrm{MCD}$ & $[0,10]$ dB & $-1$ & $0.20$ \\
 &  & $R^2_{\mathrm{mel}}$ & $[0,1]$ & $+1$ & $0.15$ \\
 &  & $L_{\mathrm{on}}$ & $[0,3]$ s & $-1$ & $0.10$ \\
 &  & $p$ (MOS) & $[1,5]$ & $+1$ & $w_p=0.30$ \\
\midrule
$t_{\mathrm{expr}}$ (prosody) & \shortstack[l]{$\{R_{f0},R_{\mathrm{eng}},F1_{\mathrm{inton}},\Delta t_{\mathrm{sync}}\}$} & $R_{f0}$ & $[0,1]$ & $+1$ & $0.20$ \\
 &  & $R_{\mathrm{eng}}$ & $[0,1]$ & $+1$ & $0.10$ \\
 &  & $F1_{\mathrm{inton}}$ & $[0,1]$ & $+1$ & $0.20$ \\
 &  & $\Delta t_{\mathrm{sync}}$ & $[0,0.5]$ s & $-1$ & $0.15$ \\
 &  & $p$ (MOS) & $[1,5]$ & $+1$ & $w_p=0.35$ \\
\midrule
$t_{\mathrm{conv}}$ (interactive) & \shortstack[l]{$\{\mathrm{WER},L_{\mathrm{on}},S_{\mathrm{turn}},U_{\mathrm{stable}}\}$} & $\mathrm{WER}$ & $[0,1]$ & $-1$ & $0.20$ \\
 &  & $L_{\mathrm{on}}$ & $[0,3]$ s & $-1$ & $0.20$ \\
 &  & $S_{\mathrm{turn}}$ & $[0,1]$ & $+1$ & $0.25$ \\
 &  & $U_{\mathrm{stable}}$ & $[0,1]$ & $+1$ & $0.15$ \\
 &  & $p$ (utility) & $[1,5]$ & $+1$ & $w_p=0.20$ \\
\bottomrule
\end{tabularx}
\par\vspace{2pt}
{\footnotesize\textbf{Notes:} $t_{\mathrm{disc}}$ denotes discrete text-output tasks such as copy-typing, spelling, or prompted word/sentence transcription (closed or open vocabulary); $t_{\mathrm{synth}}$ denotes continuous speech synthesis/reconstruction tasks where the output is a waveform or acoustic features (e.g., mel-spectrogram) rendered to audio; $t_{\mathrm{expr}}$ denotes expressivity/prosody tasks focusing on pitch, energy, intonation category, emphasis, and timing synchrony; $t_{\mathrm{conv}}$ denotes interactive conversational tasks evaluated in turn-taking settings (participant-paced dialogue), emphasizing latency, turn success, and stability over time. For each task $t$, the listed weights satisfy $\sum_{m\in\mathcal{M}_t} w_m + w_p = 1$. The entries provide author-proposed starting values for protocol development and comparative testing. For prospective comparison, ranges and weights should be co-designed and maintained by a multi-center steering group comprising researchers, clinicians, end users, and methodologists. The resulting public specification should be versioned and preregistered by each study, and direct comparisons should use the same version.}
\end{table}

To solve the problem of non-comparable metrics, the benchmark integrates objective and subjective measures. Objective metrics such as Pearson correlation coefficients for spectrogram reconstruction and STOI scores can quantify technical performance, as shown by Angrick et al.~\cite{angrick2019speech} where these metrics demonstrated significantly better-than-chance performance across participants. For discrete outputs, PER and WER remain valuable measures of phoneme- and word-sequence accuracy, although their interpretation depends on whether the evaluated transcript is generated by human listeners or ASR; neither metric directly measures semantic preservation or successful functional communication. In one participant with ALS using a 125,000-word vocabulary, Card et al. reported a mean text-decoding WER of 2.5\% in the final five instructed-delay copy-task evaluations and 3.7\% over 925 participant-paced conversational sentences with known intended labels; the latter is distinct from the participant's sentence-level correct/mostly-correct/incorrect ratings~\cite{card2024accurate}. Importantly, such integrated approaches, which relate objective decoding metrics to perceptual intelligibility and communication efficacy, have been validated in speech synthesis and automatic speech recognition research, supporting their feasibility in speech neuroprosthetics.

Li et al.~\cite{li2025high} evaluated their dual-pathway framework in nine epilepsy-monitoring participants during offline passive-listening reconstruction of TIMIT sentences. The fused system achieved mel-spectrogram coefficient of determination ($R^2=0.824\pm0.029$), MOS $3.956\pm0.173$, Whisper-base transcription WER $18.9\pm3.3\%$, and phoneme-recognizer PER $12.0\pm2.5\%$; the evaluated unique-word lexicon size was not reported.

The benchmark process could harmonize task design, including both controlled tasks (e.g., instructed-delay copy tasks) and ecologically valid scenarios (e.g., participant-paced conversational modes), as implemented in Card et al.~\cite{card2024accurate} to enable cross-task evaluation. For open-vocabulary and multilingual benchmarks, ASR-based transcription calibrated against a representative human-scored subset can support reproducible large-scale scoring.

\subsection{Metrics for Naturalistic Expressivity and Conversational Utility}
Naturalistic expressivity and conversational utility in speech neuroprostheses demand metrics that extend beyond basic intelligibility, capturing prosodic nuances, temporal dynamics, and expressive features critical for real-world communication.

Prosodic accuracy, which includes elements like pitch, loudness, intonation, and timbre, is vital for conveying lexical contrasts, sentence intonation, emotional tone and speaker identity. For tonal-language tasks, tone-label or tone-bearing-syllable accuracy measures lexical contrasts, while f0 correlation characterizes pitch-contour agreement~\cite{li2021human}. PER provides an additional tone-sensitive diagnostic when its label inventory includes tone. Source features such as pitch (f0) and loudness, encoded within frameworks such as SPARC, directly represent prosodic elements critical for expressive speech~\cite{cho2024coding}. In one participant, Wairagkar et al.~\cite{wairagkar2025instantaneous} demonstrated real-time neural control of statement-versus-question intonation and word emphasis, enabling naturalistic prosodic variation in synthesized speech.

Conversational latency, defined as the interval from an observable reference event to a specified output endpoint, is crucial for maintaining natural conversation flow, as delays longer than a few seconds can disrupt interaction and cause frustration. For onset delay ($L_{\mathrm{on}}$), the protocol preregisters a start event---an external GO cue, measurable articulatory onset, a neural-detector trigger, or explicit user initiation---and the first text or non-silent audio output. Utterance-completion and software-registration delays use their respective output events. Comparisons use matched event pairs, with detector-triggered timing characterizing the output pipeline from detection onward. Processing chunk size and inference time additionally characterize the computational demands of streaming output. For example, a system using recurrent neural network transducer (RNN-T) models achieved median delays from detector-defined attempted-speech onset to output onset of ~1.12 s for speech synthesis and ~1.01 s for text decoding on the 1,024-word-General set, with 80-ms processing chunks and median per-chunk inference latency of ~12 ms, enabling incremental speech production~\cite{littlejohn2025streaming}. Wairagkar et al.~\cite{wairagkar2025instantaneous} reported near-instantaneous closed-loop synthesis with less than 10 ms from neural acquisition to synthesized samples per frame, with additional audio-driver playback delay, highlighting progress toward real-time conversational dynamics. However, some frameworks still have a real-time factor of approximately 2--3, requiring approximately 2--3 s of total processing time per second of input, necessitating engineering focus to reduce latency for improved conversational utility.

Expressive feature decoding encompasses the extraction of prosodic, paralinguistic, and speaker-specific features enabling nuanced communication. Bimodal decoding models, such as those enabling simultaneous streaming speech synthesis and text decoding, have shown tight temporal synchrony (median absolute timing difference ~170-185 ms between text and speech onset), enhancing expressivity through synchronized multimodal outputs~\cite{littlejohn2025streaming}. Frameworks like SPARC incorporate speaker identity embedding, disentangled from articulatory features, allowing control of voice texture and accent preservation, vital for naturalistic expressivity. Accordingly, accuracy, end-to-end latency, naturalness, and voice resemblance should be reported as distinct user-facing outcomes. Prosodic correlations, feature-level scores, and synchrony measures should serve as optional diagnostics. This two-tier structure prevents optimization of representation-specific proxies from displacing user-relevant goals, while preserving technical measures for diagnosing system behavior and guiding improvement.

\subsection{Long-Term Performance and Stability Evaluation Protocols}
\label{sec:long-term-stability-evaluation}
Longitudinal evaluation should report three endpoints separately: (i) signal stability, quantified by channel retention and feature-distribution drift; (ii) frozen-decoder stability, quantified by performance drift before any update; and (iii) adaptive operational stability, quantified after scheduled or background updates. The modality-specific evidence underlying these endpoints is consolidated in Table~\ref{tab:longitudinal-stability-evidence}; this subsection therefore focuses on evaluation design rather than repeating individual cases.

Adaptive stabilization can update feature baselines, decoder biases, or model weights during use, using inferred targets or closed-loop data to reduce disruptive supervised recalibration~\cite{jarosiewicz2015virtual,silversmith2021plug}. Self-supervised or unsupervised representation stabilization instead uses unlabeled neural activity and reconstruction or temporal-dynamics objectives to align each new session with a reference latent space. NoMAD implements this label-free principle by updating the neural-to-latent mapping without new behavioral labels~\cite{karpowicz2025stabilizing}. Speech BCI evaluations should compare these approaches with frozen and supervised-recalibration baselines while reporting update triggers, data volume, latency, and error accumulation.

A prospective protocol should pre-specify the recording modality, evaluation interval, frozen-model test windows, update triggers, update frequency, labeled-data volume, and whether an update is baseline normalization, input-layer alignment, full retraining, or continuous fine-tuning. Each longitudinal time point should report performance immediately before and after updating, together with explicit calibration time, autonomous use hours, caregiver or researcher support, hardware downtime, and the amount of newly collected data incorporated into the model. Reporting both unadapted and adapted trajectories permits stable neural representations, stable fixed-decoder performance, and successful compensation for non-stationarity to be evaluated as distinct scientific outcomes.

For home-use studies, these measures should be paired with task-matched accuracy, latency, communication rate, failure and dropout events, and user-initiated recalibration. Modality-stratified reporting is essential: ECoG and intracortical findings can define current reference trajectories, whereas SEEG requires its own chronic speech-specific longitudinal evaluation before cross-modality stability comparisons are made.

\section{Translational Pathways and Ethical Design for Clinical Speech Neuroprostheses}

\begin{figure}
\centering
\includegraphics[width=\textwidth]{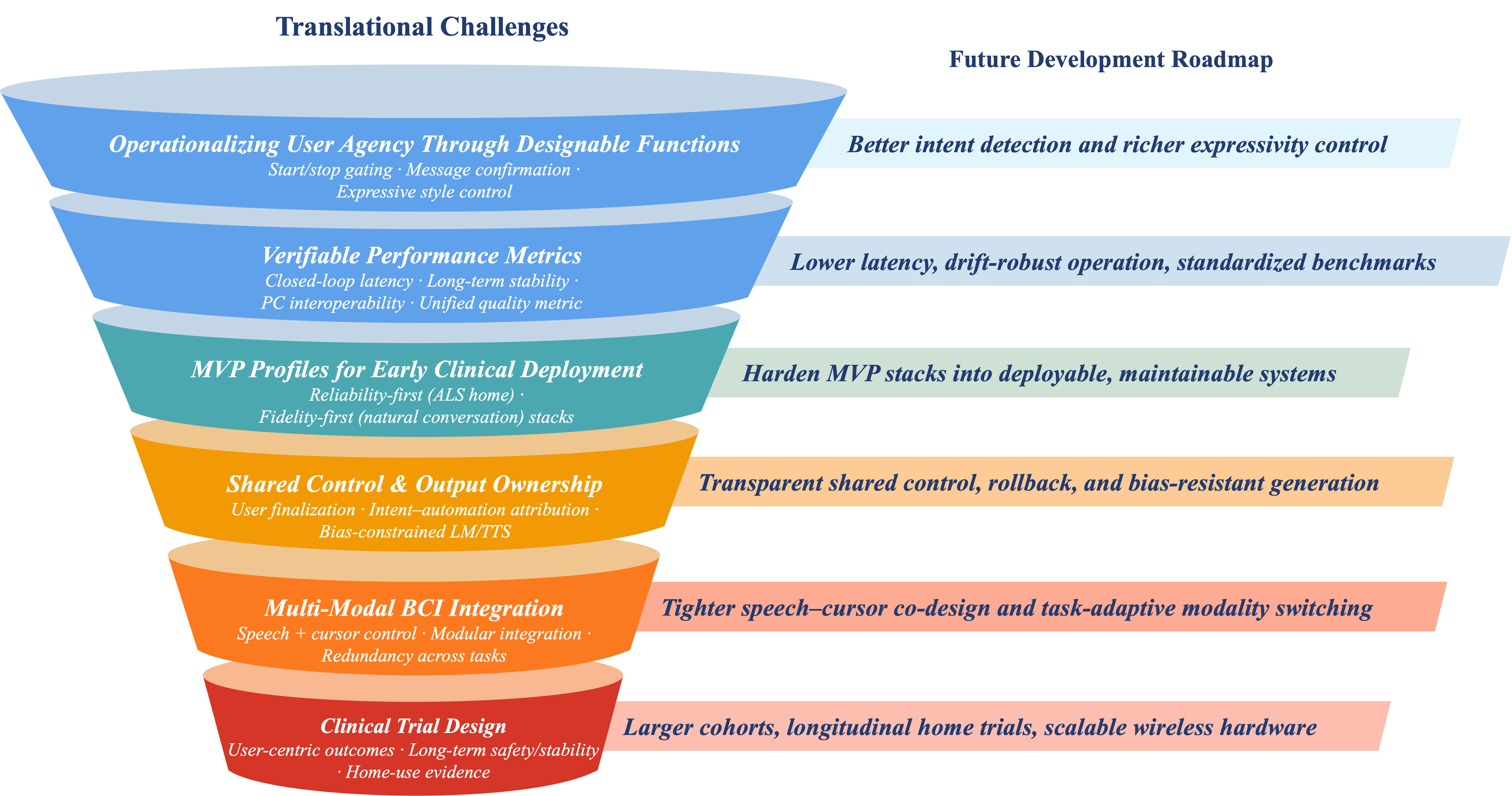}
\caption{Translational gate framework of speech BCIs for clinical speech neuroprostheses. The funnel depicts sequential gates from laboratory innovation to clinical deployment, spanning user agency, verifiable real-world metrics, MVP readiness, ethical shared control, multimodal utility, and regulatory validation. The roadmap highlights aligned future directions, emphasizing that clinical adoption requires progress across all gates, not optimization of a single metric.}
\label{fig:translational}
\end{figure}

As summarized in Fig.~\ref{fig:translational}, this section frames clinical translation as a gated process in which user agency, verifiable metrics, Minimum Viable Product (MVP) definition, ethical shared control, multimodal utility, and regulatory validation must advance together for deployment readiness.

\subsection{Operationalizing User Agency Through Designable Functions}
User agency in clinical speech neuroprostheses can be operationalized through specific designable functions that address ethical concerns about automated output shaping by centering user volitional control. As the first translational gate in Fig.~\ref{fig:translational}, these functions include decoder start-stop control, message confirmation, and expressive style modulation, each contributing to ensuring users retain authority over their communication.

Decoder start-stop control supports user control over when decoding initiates and terminates, reducing the risk of unintended output. Speech-detection mechanisms identify speech-related task states through explicit modules trained to identify volitional speech attempts~\cite{sankaran2023recommendations} or implicit detection within decoding models, such as the RNN-T architecture, which enables continuous operation with implicit start-stop detection from neural signals. For example, Littlejohn et al. demonstrated implicit speech detection in offline continuous decoding with an experimental ``GO'' cue specifying task timing~\cite{littlejohn2025streaming}. Detection under this instructed protocol does not establish that a system can distinguish authorized communication from spontaneous private inner speech. In addition, decoder gating can be augmented with peripheral biosignals such as electromyography (EMG), for example using a deliberate EMG switch or push-to-talk gesture as an explicit user-controlled start signal and a rapid veto/kill-switch to halt synthesis~\cite{judge2025exploring,lobo2014non}. Such hybrid gating or switching schemes can reduce false positives by requiring concordant volitional evidence (e.g., detected speech attempts plus deliberate EMG activation) before allowing output, and by supporting reliable transitions between control modes~\cite{kreilinger2012switching}. Advances in EMG classification further support robust detection of user-triggered control signals for start/stop gating~\cite{he2026dsranet,he2026multi}. Explicit user-controlled gating can provide executory control (go-command) and veto control to avoid unintentional speech and halt production if needed~\cite{maslen2021control}. Inner-speech findings further inform gating design: Kunz et al. found overlapping inner- and attempted-speech representations alongside a differentiating motor-intent dimension and demonstrated strategies to reduce unintended inner-speech decoding, including keyword-based unlocking~\cite{kunz2025inner}. These approaches are candidate safeguards rather than validated solutions for protecting private inner speech in spontaneous use.

Message confirmation mechanisms allow users to review and explicitly accept decoded content before speech generation, mitigating errors and enhancing ownership. Real-time visual display of decoded text enables users to read and check accuracy before accepting an output, while simultaneous brain-to-text display supports monitoring during synthesis.

Expressive style modulation empowers users to shape communicative nuances, aligning outputs with personal identity. Wairagkar et al.~\cite{wairagkar2025instantaneous} personalized the synthesizer using the participant's pre-ALS voice and enabled real-time modulation of question intonation, word emphasis, and short melodies. For instance, the brain-to-voice neuroprosthesis implemented parallel binary classifiers to decode user intent for question intonation or word-emphasis modulation, enabling closed-loop control over expressive features~\cite{wairagkar2025instantaneous}. Beyond vocal features, facial animation can accompany speech synthesis to express non-verbal gestures, further enhancing expressive capacity~\cite{metzger2023high,silva2024speech}. Customization of avatars or communication parameters (e.g., language model adaptivity) also contributes to user agency by aligning the system with individual preferences.

\subsection{Translational Priorities with Verifiable Performance Metrics}
Translational priorities for clinical speech neuroprostheses necessitate verifiable performance metrics to address bottlenecks of insufficient real-world utility, with key focuses on closed-loop audio feedback latency, long-term stability, and interoperability with personal computer systems. This corresponds to the verifiable real-world metrics gate in Fig.~\ref{fig:translational}. For translation, long-term stability should be verified using the endpoint definitions in Section~\ref{sec:long-term-stability-evaluation} and the modality-stratified evidence in Table~\ref{tab:longitudinal-stability-evidence}, alongside study duration. Deployment reports should quantify independent-use days and hours, explicit calibration minutes per day, background-update frequency and data volume, caregiver or researcher assistance, downtime, and performance drift before and after updates. Current chronic speech BCI results support ECoG and intracortical deployment pathways in selected participants under different update regimes. A SEEG deployment pathway requires prospective characterization using the same calibration, autonomy, and performance-retention endpoints, extending the evidence base to chronic depth recordings.

Closed-loop audio feedback latency is essential for natural communication, and advancements in this area are notable. In the same CortiCom ECoG participant, Luo et al.~\cite{luo2023stable} reported a median 1.24-s interval from speech offset to decoding registration, while Angrick et al.~\cite{angrick2024online} later implemented online synthesis and playback with word-delayed auditory feedback. In the separate BrainGate2 participant T15, Card et al.~\cite{card2024accurate} operated an intracortical neuroprosthesis in real time with phoneme-probability updates every 80 ms. These two participant-specific systems document progress toward conversational feedback under different latency definitions and output tasks.

Interoperability with personal computer control systems is vital for integrating speech BCIs into daily life. In 2023, Luo et al.~\cite{luo2023stable} showed that their ECoG-based system could control various computer applications and external devices in real time, while in 2015, Herff et al.~\cite{herff2015brain} noted that their Brain-to-Text approach, though evaluated offline, is well-suited for real-time online application on desktop computers. In 2024, Angrick et al.~\cite{angrick2024online} demonstrated interoperability through communication board control, and in 2024, Card et al.~\cite{card2024accurate} integrated their neuroprosthesis with personal computers via Bluetooth keyboard functionality, enabling activities like email writing and integration with eye tracking for user confirmation.

To ensure consistent evaluation of these metrics, in 2025, Wu et al.~\cite{wu2025improved} proposed a unified metric using a Random Forest regressor that combines STOI and MCD to predict human-rated MOS, addressing the limitations of individual objective metrics. This cross-validated model provides a verifiable performance metric that correlates with subjective perception in the evaluated datasets, supporting waveform-quality assessment.

\subsection{Minimum Viable Product (MVP) Profiles for Early Clinical Deployment}
Because communication reliability and speech naturalness place different demands on hardware and algorithms, an actionable translational strategy is to define use-case-specific MVP stacks rather than seeking a single universal architecture. This operationalizes the MVP readiness gate in Fig.~\ref{fig:translational} and is also consistent with the organizing logic of Fig.~\ref{fig:timeline}, whose lane structure emphasizes that clinically relevant progress depends on jointly improving decoding performance, robustness, calibration burden, and deployment feasibility rather than optimizing a single benchmark metric.

\textbf{ALS home communication (reliability-first MVP):} prioritize recording modalities on the basis of direct chronic speech BCI evidence and quantified update burden. ECoG currently provides participant-level evidence for 3-month fixed-decoder command use, whereas intracortical arrays provide extensive independent home-use evidence with continuous background adaptation~\cite{luo2023stable,card2026long}. SEEG is a candidate for prospective home-use evaluation when its anatomical access is advantageous, with progression based on speech-specific longitudinal calibration and autonomy endpoints. Use explicit start/stop gating, confirmation or veto control, and a simplified high-frequency lexicon or phrase bank to reduce unintended output and caregiver load~\cite{sankaran2023recommendations,wairagkar2025instantaneous}. Practical endpoints should include stable multi-week operation, predictable latency, and low maintenance overhead.

\textbf{High-naturalness conversational speech (fidelity-first MVP):} consider MEA or high-density $\mu$ECoG as candidate recording approaches when the target is rich prosody, voice identity preservation, and low-latency synthesis. Articulatory-intermediate, direct acoustic, and text-based decoding pathways offer candidate routes to speech synthesis, with channel density and calibration requirements assessed for the intended user~\cite{willett2023high,wairagkar2025instantaneous,anumanchipalli2019speech,cho2024coding,li2025high}. Practical endpoints should include near-real-time feedback, timed from a specified observable event to first audible output, and robust control of expressive features (intonation/emphasis).

This two-profile framing clarifies that clinical deployment decisions should be benchmarked against scenario-specific success criteria, not only pooled average decoding metrics.

\subsection{Ethical Considerations in Shared Control and Output Ownership}
Shared control between users and automated decoding systems in speech BCIs introduces nuanced ethical dynamics, as demonstrated by systems where automated suggestions are paired with user oversight. This directly maps to the ethical shared-control gate in Fig.~\ref{fig:translational}. In one BrainGate2 participant (T12), Willett et al.~\cite{willett2023high} displayed the language model's decoded words in real time and allowed the user to end the decoding trial by button press. Independently, in one ECoG participant (Bravo-3), Metzger et al.~\cite{metzger2023high} required active silent-speech attempts to drive multimodal text, speech, and avatar output. These separate single-participant programs provide convergence on active user participation at the system-design level. Such designs support the efficiency of neural and language-model decoding and user control over task execution.

User ownership of synthetic speech and moral responsibility for its outputs remain open ethical questions; technical separation of articulatory and speaker representations does not resolve them. Their assessment requires considering the neural evidence supporting the content, the contribution of generative processing, and users' meaningful opportunities to confirm or veto the output~\cite{maslen2021control,sankaran2023recommendations}. In 2024, Cho et al.~\cite{cho2024coding} developed the SPARC framework, which disentangles speaker-agnostic articulatory features in speech/EMA corpora from speaker-specific voice texture (handled by automated encoders). This separation offers a design principle for BCI transparency by distinguishing the representation of articulatory content from voice rendering, but it does not by itself establish ownership or responsibility. In 2023, Metzger et al.~\cite{metzger2023high} supported personal expression by personalizing synthetic speech to the participant's pre-injury voice and animating facial avatars to reflect intended emotional expressions, supporting continuity with the participant's pre-injury voice and chosen expressive presentation without resolving authorship of a generated utterance.

Design strategies to balance efficiency and user control must address both technical and ethical dimensions. In 2023, Willett et al.~\cite{willett2023high} noted that retraining decoders daily improves adaptation to neural changes, but unsupervised adaptation could reduce user burden, suggesting that adaptive algorithms should prioritize user convenience without sacrificing control. In 2023, Metzger et al.~\cite{metzger2023high} introduced complementary output modalities (text, speech, avatar) and early stopping for silence detection, providing a basis for interfaces that offer context-appropriate output selection and control over communication flow. We therefore recommend combining interpretable articulatory representations~\cite{cho2024coding} with explicit user correction and shared-control functions~\cite{maslen2021control,sankaran2023recommendations}. Additionally, in 2026, Li et al.~\cite{li2025high} highlighted risks of generative model biases (e.g., Parler-TTS favoring linguistically probable outputs), advocating for dual-pathway architectures that integrate acoustic and linguistic decoding to mitigate such biases while preserving user-specific vocal identity.

Alongside technical advances such as Feng et al.'s Mandarin brain-to-sentence decoder~\cite{feng2025acoustic}, these examples underscore that ethical design in shared control requires user agency, transparent decoding, and personalized output. These safeguards can support personal expression, but synthetic speech should not be assumed automatically to represent the user's intent or identity when generative processing contributes to the output.

\subsection{Multi-Modal BCI Integration for Comprehensive Communication}
Multi-modal integration can align communication restoration with computer access for users whose functional priorities span both domains. In Fig.~\ref{fig:translational}, this combination corresponds to the multimodal utility gate for broader clinical communication tasks. P300-, motor-imagery-, and steady-state visually evoked potential-based BCIs provide reliable selection and device-control functions, whereas speech BCIs enable open-ended linguistic communication through text or synthesized audio. Cursor control complements this expressive channel by supporting precise navigation, selection, and device manipulation. Multi-modal integration can therefore be configured around each user's communication goals, digital-access requirements, and preferred interaction strategies.

Direct multi-modal evidence currently centers on intracortical arrays implanted in speech motor cortex. Singer-Clark et al.~\cite{singer2025speech} demonstrated two-dimensional cursor movement and click control from this cortical target in one participant. Card et al.~\cite{card2026long} subsequently extended the same participant's system to long-term independent use with four intracortical arrays and parallel brain-to-text, cursor-movement, and click decoders. Over 19 months, the participant accumulated more than 3,800 h of home use. Prompted speech decoding achieved 99.2\% word accuracy with a 125,000-word vocabulary, and cursor control reached $2.90 \pm 0.16$ bits s$^{-1}$ in structured grid-task assessments. Complementary evidence shows that spoken English can be decoded from intracortical arrays in dorsal precentral regions commonly targeted for arm- and hand-movement BCIs~\cite{wilson2020decoding}. Together, these findings support systematic evaluation of both speech-primary and motor-primary cortical targets using function-specific performance as a joint design criterion.

At the system level, modular architectures can route a common neural recording stream into parallel, task-specific decoding pathways. Speech decoders can model neural activity associated with attempted articulation, while cursor decoders independently map attempted hand or arm movement to pointer velocity and click commands. The resulting outputs can be coordinated within a shared interface for communication, navigation, and device control. This architecture couples shared recording hardware with independently trained modality-specific decoders, preserving the distinct neural--behavioral mapping required for each function~\cite{singer2025speech,card2026long}.

Clinical value is maximized when modality selection and implant placement are jointly optimized around the user's functional priorities and anatomy. A shared cortical target offers a route to supporting both speech and cursor control within one implantation strategy when it reaches the required performance for both functions~\cite{singer2025speech,card2026long}. When the optimal targets differ, target selection can prioritize the function that carries the greatest individual benefit, while multi-site strategies can be evaluated where clinically appropriate. Because the clinical populations for speech restoration and hand-function restoration overlap only partially, candidate selection can distinguish speech-primary, cursor-primary, and integrated functional profiles~\cite{sankaran2023recommendations,stavisky2025restoring}. Prospective comparisons across participants, implant targets, and home-use tasks can define performance thresholds for each profile and guide individualized multi-modal BCI design.

\subsection{Target Populations and Clinical Trial Design for Speech Neuroprosthesis Validation}
Participant selection criteria for speech neuroprosthesis clinical trials should distinguish between research-access cohorts and the intended therapeutic populations for chronic implantation, while balancing clinical relevance, feasibility, safety, and the likelihood of meaningful functional benefit. This subsection addresses the regulatory validation gate in Fig.~\ref{fig:translational}, where study design determines whether promising prototypes can generate approval-relevant evidence. The most realistic near-term candidates for invasive speech neuroprostheses are individuals with severe, chronic, and otherwise irreversible loss of intelligible speech or reliable motor-based communication, particularly those with late-stage ALS accompanied by anarthria or profound dysarthria, brainstem stroke or locked-in syndrome, and selected individuals with tetraplegia whose speech or conventional access channels are inadequate. Appropriate candidates should generally retain sufficient cognition, language comprehension, communicative intent, and the ability to formulate intended linguistic content; they should also derive limited benefit from conventional augmentative and alternative communication (AAC) and have an acceptable surgical risk--benefit profile.

Early feasibility studies may nevertheless prioritize participants with some residual speech-related motor activity because such signals facilitate decoder development and validation. For example, one study included a person with bulbar-onset ALS who retained limited orofacial movement and vocalization ability despite being unable to speak intelligibly~\cite{willett2023high}. Similarly, the BrainGate2 pilot clinical trial enrolled participants with tetraplegia and residual speech ability to safely investigate neural decoding performance~\cite{wilson2020decoding}, while another BrainGate2 participant with tetraplegia and dysarthria demonstrated feasibility in a more severely impaired population~\cite{singer2025speech}. However, residual vocal or articulatory function should not be treated as a universal eligibility requirement, because individuals with complete anarthria may ultimately have the greatest unmet need. Surveys of potential users with ALS and spinal cord injury further indicate that willingness to undergo implantation depends on whether anticipated communication benefits, including sufficiently high accuracy (e.g., $>90\%$), justify the surgical and long-term maintenance burdens~\cite{stavisky2025restoring}. For individuals with milder or potentially reversible impairments, effective access to conventional AAC, or an unfavorable surgical risk profile, conventional approaches---including eye-tracking or switch-based AAC---should remain first-line or complementary options for communication and system fallback where usable. Scalp EEG, fNIRS, and other non-invasive or peripheral interfaces have modality- and task-specific roles in research, rehabilitation, candidate screening, or communication support.

Outcome measures in such trials prioritize user experience by focusing on metrics that reflect real-world communication efficacy and user preferences. Performance evaluations have included both vocalized and silent (mouthing) speech attempts, with participants reporting preferences for silent speech due to reduced fatigue. Wilson et al.~\cite{wilson2020decoding} recorded two BrainGate2 participants, each overtly speaking a 420-word corpus; the reported 33.9\% offline RNN accuracy across 39 phonemes was the best-participant result, with lower performance in the second participant. WER, PER, and communication rate characterize sequence accuracy and speed, while turn success, user ratings, and communication utility assess functional communication (Section~5.5).

Long-term safety monitoring and decoder stability are essential for regulatory approval, addressing neural non-stationarities and device longevity. Strategies include using RNN architectures with day-specific input layers and rolling feature adaptation to handle across-day and within-day neural changes. Implanted sensors like Utah arrays and ECoG grids have demonstrated stable recordings for months to years (e.g., up to 7 years for Utah arrays, 3 years for ECoG), supported by adaptive algorithms that maintain performance without frequent recalibration. Clinical trials such as BrainGate2 operate under U.S. Food and Drug Administration (FDA) IDE and institutional review board (IRB) approvals, with monitoring extending to home environments to assess real-world reliability.

These design elements directly address barriers to regulatory approval by demonstrating safety, efficacy, and user-centric outcomes. For example, the use of IDE-approved protocols ensures compliance with regulatory standards, while evidence of stable decoding performance, implant safety, and meaningful communication output helps meet regulatory expectations of efficacy. Ongoing trials, including BrainGate2, implement these principles by exploring suboptimal cortical areas to de-risk future trials targeting ventral speech cortex, demonstrating multi-modal control (speech and cursor) from a single implant, and planning commercial trials with fully wireless sensors. Despite promising results, further validation in larger cohorts remains necessary to establish clinical viability.

\section{Future Directions}

\begin{figure}
\centering
\includegraphics[width=\textwidth]{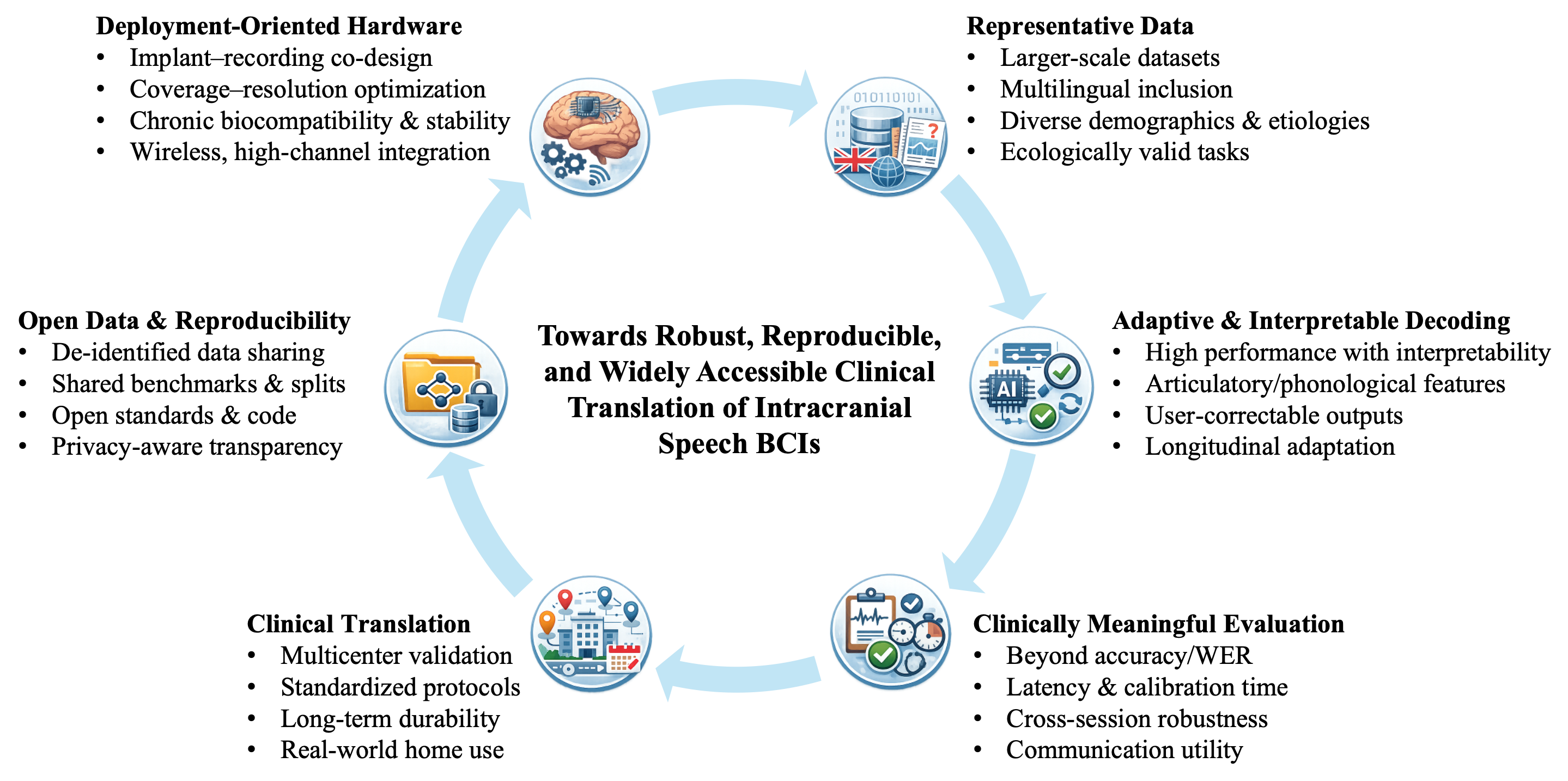}
\caption{Future directions for intracranial speech BCIs. A coordinated roadmap highlighting six interdependent priorities: hardware and implant co-design, representative multilingual datasets, adaptive/interpretable decoding, standardized clinical evaluation, multicenter validation and deployment, and open data/standards for reproducible progress.}
\label{fig:direction}
\end{figure}

As summarized in Fig.~\ref{fig:direction}, future research in intracranial speech BCIs should prioritize a coordinated roadmap that links hardware and algorithmic progress to real-world clinical deployment. A foundational systems priority is recording-hardware and implant co-design. Beyond improving acute signal quality, future electrode development should target coverage--resolution co-optimization (e.g., hybrid macro+$\mu$ECoG and surface+depth strategies), chronic biocompatibility and low-impedance interface stability, fully implanted wireless architectures with onboard preprocessing, and high-channel-count active multiplexed arrays that preserve bandwidth within thermal and power limits. Platforms that support both decoding and causal stimulation/functional mapping may also enable adaptive closed-loop speech systems. These hardware advances should be evaluated with deployment-relevant endpoints such as infection risk, packaging durability, setup burden, and home-environment robustness.

On the data side, a foundational priority is the development of larger, multi-language, and clinically representative datasets to improve generalization across populations and linguistic contexts~\cite{yang2023biot, levin2026cross}. Current evidence is dominated by English-speaking participants and highly controlled tasks, which limits external validity~\cite{tang2025semantic}. Expanding data collection to include tonal and morphologically diverse languages, broader age ranges, and heterogeneous etiologies of speech impairment will enhance both scientific rigor and translational relevance~\cite{jamali2024semantic}.

At the model-development level, adaptive and interpretable decoding approaches remain a central priority. Although deep learning approaches have substantially improved decoding performance, black-box behavior can undermine user trust, calibration transparency, and ethical accountability. Future systems should therefore combine high performance with interpretable intermediate representations (e.g., articulatory or phonological features), user-correctable outputs, richer control of lexical tone where relevant, expressive prosody, and paralinguistic cues, and longitudinal adaptation strategies that reduce daily recalibration burden while preserving user agency~\cite{pereira2018toward}. In this setting, population-trained models may serve as transferable priors, supporting data-efficient adaptation of selected subject-specific components and subsequent carefully constrained longitudinal updates. Improving reliable imagined-speech decoding under task-dependent signal strength and limited temporal alignment is another important frontier for clinically flexible communication~\cite{kunz2025inner}. An important extension of adaptive decoding is the closed-loop co-adaptation of the user, neural decoder, and speech synthesizer. Continuous and accurate real-time feedback about decoded neural activity has been shown to be necessary for users to learn effective control of an imagined-speech BCI, suggesting that the combination of machine and human learning can improve BCI controllability~\cite{bhadra2025learning}. Building on this principle, recent demonstrations of online and near-instantaneous speech synthesis provide a technical foundation for investigating whether low-latency auditory feedback could similarly support user adaptation and online refinement of the neural-to-speech mapping~\cite{angrick2024online,wairagkar2025instantaneous}. Future studies should determine whether the neural decoder and speech synthesizer can be jointly adapted during real-time use while minimizing interference from feedback latency and preserving the user's intended content, vocal identity, and agency.

Beyond model optimization, the field would benefit from broader adoption of standardized evaluation frameworks that capture clinically meaningful outcomes rather than isolated model metrics. In addition to accuracy or WER, studies should report latency, calibration time, robustness across sessions, failure modes, user workload, and communication utility in ecologically valid tasks~\cite{levin2026cross}. Cross-language and cross-task benchmarks are an important starting point, but they must be harmonized across centers to support fair comparison and cumulative progress~\cite{tang2025semantic}.

The unified evaluation framework and Table~\ref{tab:clt-benchmark-defaults} parameterization proposed in this review, together with the proposed deployment-utility formulation in Eq.~\ref{eq:deployment-utility}, form a conceptual approach requiring prospective validation. The reporting hierarchy prioritizes individual outcomes and uses composite scores as secondary summaries within prespecified task--language and use-case conditions, following safety and feasibility screening. Future work should evaluate the stability of the composite score across datasets, languages, task designs, and participant populations; quantify its sensitivity to alternative metric ranges and weights; calibrate perceptual mappings against human judgments; and compare the proposed settings prospectively across multiple centers. External validation and consensus involving researchers, clinicians, end users, and regulatory stakeholders will be essential for determining whether the framework can evolve from a structured recommendation into a broadly applicable evaluation standard.

At the translational stage, rigorous multi-center clinical validation is necessary to establish safety, efficacy, and durability. Many influential studies remain single-participant or single-center demonstrations. Larger prospective trials with standardized protocols, long-term follow-up, and diverse participant demographics will be essential for clarifying which combinations of recording modality, decoding architecture, and shared-control strategy are clinically viable and scalable for regulatory pathways~\cite{levin2026cross,kunz2025inner}. These trials should also evaluate multimodal use cases (e.g., speech plus cursor control or semantic assistance) and home-use performance, because real-world utility may depend on redundancy and interoperability rather than copy-task accuracy alone.

Across all of these priorities, open sharing of data is not only beneficial but necessary for the field's progress. Intracranial speech BCI research is highly interdisciplinary and resource-intensive, requiring collaboration across neuroscience, engineering, clinical neurology, rehabilitation, linguistics, and ethics. Timely sharing of de-identified datasets, metadata, annotations, and benchmark splits can improve cross-task comparability and help investigators, clinicians, and institutions build on prior work more efficiently~\cite{donoso2026new}, including in lower-resource settings. In parallel with appropriate protections for participant privacy and sensitive neural data, greater openness in data standards, protocols, evaluation criteria, code, and reporting practices will improve reproducibility, reduce duplication of effort, and accelerate equitable clinical translation~\cite{donoso2026new,yang2023biot}.

By advancing these priorities in parallel, intracranial speech BCI research can move beyond isolated technical milestones toward robust, reproducible, and widely accessible communication neuroprostheses for individuals with severe speech impairments.

\section{Conclusion}
This review summarizes the current trajectory of intracranial speech BCIs for speech restoration across neural mechanisms, recording technologies, decoding algorithms, evaluation strategies, and translational design. Overall, the field is progressing from proof-of-concept demonstrations toward more clinically meaningful communication systems, but key bottlenecks remain. At the neural level, intracranial evidence supports a mixed organization with partially somatotopic articulatory coding, dorsal--ventral stream specialization, and frequency-specific dynamics across overt, mimed, and imagined speech. Silent attempted speech, including mimed articulation, has enabled large-vocabulary decoding in implanted participants; imagined-speech decoding has also advanced, while reliable, unconstrained imagined communication remains to be established.~\cite{willett2023high,kunz2025inner} At the systems level, recording modalities continue to present trade-offs in coverage, resolution, invasiveness, and long-term robustness. Intracranial systems are therefore most realistically positioned as higher-risk, potentially high-benefit interventions for carefully selected individuals with profound unmet communication needs, while conventional AAC remains an essential communication option where usable, complemented by non-invasive interfaces according to their demonstrated function and deployment requirements. Realizing meaningful benefit in these intended users will require reliable long-term operation with limited recalibration and manageable setup demands. Although multi-month stability has been demonstrated for both ECoG and intracortical approaches, recalibration burden and setup dependence remain important barriers to routine longitudinal use. In parallel, advances in sequence models, transformers, and articulatory intermediates have improved decoding performance, while language priors (including LLM-based components) may enhance fluency and error correction; however, intent preservation, latency, bias, and privacy risks require continued safeguards and prospective monitoring. 

Importantly, the clinical meaning of decoder performance depends on the task protocol under which it is obtained. Overt, cue-aligned studies in temporary epilepsy-monitoring cohorts provide controlled tests of speech decoding, while translation to self-paced attempted or imagined communication in individuals with severe paralysis depends on validation under those intended-use conditions. Protocol, iEEG modality and anatomical coverage, neural modeling strategy, and intended-use population must therefore be considered jointly when assessing translational readiness. Future progress will depend on standardized and longitudinal evaluation, harmonized benchmarks, explicit reporting of recalibration burden, and larger multi-center clinical cohorts. User agency, shared-control transparency, and multimodal integration should remain central translational priorities, together with prospective safety analyses for language-prior-assisted decoding. In summary, clinically viable intracranial speech BCIs are increasingly plausible, but broader translation will require larger and more diverse datasets, adaptive and interpretable models, rigorous cross-center validation, and transparent data-sharing and reporting practices to support reproducibility and equitable deployment. Sustained open sharing of de-identified datasets, annotations, benchmark definitions, code, and negative results will further strengthen cross-center learning and accelerate responsible clinical translation.

\section*{Acknowledgments}
This work was supported by The Hong Kong Polytechnic University Start-up Fund (Project ID: P0053210), The Hong Kong Polytechnic University Faculty Reserve Fund (Project ID: P0053738), an internal grant from The Hong Kong Polytechnic University (Project ID: P0048377), The Hong Kong Polytechnic University Departmental Collaborative Research Fund (Project ID: P0056428), The Hong Kong Polytechnic University Collaborative Research with World-leading Research Groups Fund (Project ID: P0058097) and Research Grants Council Collaborative Research Fund (Ref: C5033-24G).

\section*{Author Contributions}
D.H. and N.W. contributed to the conceptualization of the study. D.H., W.T.S., and N.W. performed the investigation. D.H. and N.W. drafted the original manuscript. All authors (D.H., W.T.S., and N.W.) reviewed and edited the manuscript. W.T.S. and N.W. acquired funding for the study and supervised the research.

\bibliographystyle{unsrt}
\bibliography{references}

\end{document}